\documentclass[fleqn,usenatbib]{mnras}

\usepackage{newtxtext,newtxmath}
\usepackage[T1]{fontenc}
\usepackage[utf8]{inputenc}

\DeclareRobustCommand{\VAN}[3]{#2}
\let\VANthebibliography\thebibliography
\def\thebibliography{\DeclareRobustCommand{\VAN}[3]{##3}\VANthebibliography}
\DeclareUnicodeCharacter{2212}{-}

\usepackage{graphicx}	% Including figure files
\usepackage{multirow}
\usepackage{rotating}
\usepackage{pdflscape}
\defcitealias{Huitson2017}{H17}

\title[Gemini/GMOS view of the warm Neptune HAT-P-26b]{A new method to measure the spectra of transiting exoplanet atmospheres using multi-object spectroscopy} 
\author[Panwar et al.]{
Vatsal Panwar,$^{1}$\thanks{E-mail: v.panwar@uva.nl}
Jean-Michel D\'esert,$^{1}$
Kamen O. Todorov,$^{1}$
Jacob L. Bean,$^{2}$
Kevin B. Stevenson,$^{3}$
\newauthor
C. M. Huitson,$^{4}$
Jonathan J. Fortney,$^{5}$
and Marcel Bergmann$^{6}$
\\
$^{1}$Anton Pannekoek Institute for Astronomy,
University of Amsterdam, P.O. Box 94249, 1090GE Amsterdam, Noord Holland, NL\\
$^{2}$Department of Astronomy and Astrophysics, University of Chicago, Chicago, IL 60637, USA\\
$^{3}$JHU Applied Physics Laboratory, 11100 Johns Hopkins Rd, Laurel, MD 20723, USA \\
$^{4}$CASA, University of Colorado, 389 UCB, Boulder, CO, 80309-0389, USA \\
$^{5}$Department of Astronomy and Astrophysics, University of California, Santa Cruz, CA 95064, USA \\
$^{6}$NOAO, Gemini Observatory, present address Palo Alto, CA, USA
}

\date{Accepted 2021 December 9. Received 2021 December 9; in original form 2021 June 15}
\pubyear{2021}

\begin{document}
\label{firstpage}
\pagerange{\pageref{firstpage}--\pageref{lastpage}}
\maketitle

\begin{abstract}
% This is a simple template for authors to write new MNRAS papers.
% The abstract should briefly describe the aims, methods, and main results of the paper.
% It should be a single paragraph not more than 250 words (200 words for Letters).
% No references should appear in the abstract.
Traditionally, ground-based spectrophotometric observations probing transiting exoplanet atmospheres have employed a linear map between comparison and target star light curves (e.g. via differential spectrophotometry) to correct for systematics contaminating the transit signal. As an alternative to this conventional method, we introduce a new Gaussian Processes (GP) regression-based method to analyse ground-based spectrophotometric data. Our new method allows for a generalised non-linear mapping between the target transit light curves and the time series used to detrend them. This represents an improvement compared to previous studies because the target and comparison star fluxes are affected by different telluric and instrumental systematics, which are complex and non-linear. We apply our method to six Gemini/GMOS transits of the warm (T$_{\rm eq}$ = 990 K) Neptune HAT-P-26b. We obtain on average $\sim$20 \% better transit depth precision and residual scatter on the white light curve compared to the conventional method when using the comparison star light curve as a GP regressor and $\sim$20 \% worse when explicitly not using the comparison star. Ultimately, with only a cost of 30\% precision on the transmission spectra, our method overcomes the necessity of using comparison stars in the instrument field of view, which has been one of the limiting factors for ground-based observations of the atmospheres of exoplanets transiting bright stars. We obtain a flat transmission spectrum for HAT-P-26b in the range of 490-900 nm that can be explained by the presence of a grey opacity cloud deck, and indications of transit timing variations, both of which are consistent with previous measurements.

\end{abstract}


%%%%%%%%%%%%%%%%%%%%%%%%%%%%%%%%%%%%%%%%%%%%%%%%%%

%%%%%%%%%%%%%%%%% BODY OF PAPER %%%%%%%%%%%%%%%%%%
\section{Introduction}
\label{sec_intro}

Low-resolution transmission spectroscopic observations of transiting gas giant exoplanets have been extensively used to probe their atmospheric compositions. The multi-object spectroscopy (MOS) technique (\citealt{Bean2010}, \citeyear{Bean2011}) has produced spectrophotometric measurements of exoplanet atmospheres at low-resolution (R $\sim$ 10-100) with various ground-based observatories from optical to near infrared (Gemini/GMOS: see e.g. \cite{Crossfield2013}, \cite{Gibson2013}, \cite{Stevenson2014}, \cite{Huitson2017}, \cite{Todorov2019}, \cite{Wilson2021}; VLT/FORS2: see e.g. \cite{Bean2010}, \cite{Sedaghati2017}, \cite{Nikolov2016}, \citeyear{Nikolov2018}, \cite{Carter2019}, \cite{Wilson2020}; Magellan/MMIRS and IMACS: see e.g. \citeauthor{Bean2011} (\citeyear{Bean2011}, \citeyear{Bean2013}), \cite{Rackham2017GJ1214}, \cite{Espinoza2019}, \cite{Weaver2020}, \cite{McGruder2020}; LBT/MODS: see e.g. \cite{Mallonn2016}, \cite{Yan2020}) and long-slit spectrographs at low-resolution (GTC/OSIRIS: see e.g. \cite{Sing2012}, \citeauthor{Murgas2014} (\citeyear{Murgas2014}, \citeyear{Murgas2019}), \cite{Nortmann2016}, \citeauthor{Chen2017} (\citeyear{Chen2017}, \citeyear{Chen2018}, \citeyear{Chen2020}, \citeyear{Chen2021})) which have resulted in the detection of spectral features due to Rayleigh scattering, atomic and molecular absorption, and/or grey opacity clouds (in the form of flat or featureless spectra). The detection of pressure broadened profile of Na I doublet weakly in the atmosphere of WASP-4b (e.g. \citealt{Huitson2017}), significantly in the atmosphere of the hot Saturn WASP-96b (\citealt{Nikolov2018}) consistent with a cloud-free atmosphere, and the detection of Na, Li, and K absorption along with the signatures of scattering due to haze in the atmosphere of hot Neptune WASP-127b (\citealt{Chen2019}) are some examples that demonstrate that ground-based MOS observations are capable of estimating absolute abundances in some of the gas giants with clear atmospheres. 

Notably, transit observations using large ground-based telescopes like Gemini and VLT have yielded transit depth precision comparable to space-based observations from HST in their white light curves (e.g. \cite{Bean2010}, \cite{Todorov2019}). Observations for the same planet repeated over multiple epochs and from different instruments have helped in ascertaining the robustness of results (e.g. WASP-4b \cite{May2018}, \cite{Bixel2019}; WASP-19b \cite{Sedaghati2017}, \cite{Espinoza2019}) and in interpreting and mitigating the transit light source effect due to stellar photospheric heterogeneity which has the strongest observable effect on a transmission spectrum in the optical wavelength range (\cite{Rackham2018}). Furthermore, ground-based MOS observations have pushed the limits of atmospheric characterisation down to terrestrial planets (\citealt{Diamond-Lowe2018}, \citeyear{Diamond-Lowe2020a}, \citeyear{Diamond-Lowe2020b}) for which the optical transmission spectra have been able to rule out the presence of clear and low mean molecular weight atmospheres.

Spectrophotometric observations obtained using ground-based multi-object spectrographs are affected by telluric and instrumental systematics at levels comparable or even more than the amplitude of variations due to the planetary atmosphere in the transmission spectrum that we aim to measure. The conventional technique to compensate for systematics in ground-based low resolution spectra has been to simultaneously observe one or more reference or comparison stars in the instrument's field of view (\citealt{Bean2010}) and use that to correct for the systematics similarly affecting the target star light curve through differential photometry. The Rossiter-McLaughlin effect based observations measure changes in line shape for detecting transits (\citealt{Sluijs2019}) and deriving low-resolution transmission spectra (\citealt{Oshagh2020}, \citealt{DiGloria2015}). Such observations follow a parallel approach to measure transmission spectrum that does not need a comparison star but typically yield low resolution transmission spectra at sub-optimal precision. 

All the aforementioned ground-based MOS studies, however, have always used comparison stars to deal with systematics in the transit light curves and extract the signals of planetary atmosphere. The leftover systematics after Target/Comparison star light curve normalisation, arising due to brightness or differences in spectral types between the target and comparison stars, are then conventionally modelled by parametric models constructed using polynomials based on a set of decorrelation parameters (e.g. \cite{Gibson2014}, \cite{Stevenson2014}, \cite{Todorov2019}), or a non-parametric approach using Gaussian Processes (GP) regression (e.g. \citeauthor{Gibson2012} (\citeyear{Gibson2012}, \citeyear{Gibson2013})). 

Some previous works, instead of dividing the target star light curve by the comparison star light curve, fit the target star light curves directly by using comparison star light curves or a PCA components of multiple comparison stars as linear regressors (see e.g. \citealt{Jordan2013}, \citealt{Espinoza2019}). The \texttt{Divide-White} method introduced by \cite{Stevenson2014} extracts the transmission spectrum from target star light curves by using non-analytic models of wavelength dependent systematics derived from comparison star light curves. Note that all the aforementioned approaches: doing differential spectrophotometry, using comparison star light curves as linear regressors or non-analytic models of systematics, assume a linear relationship between the systematics in target and comparison star flux variations. Differential photometric corrections in particular perform best when the comparison stars are similar to the target stars in brightness and spectral type (\citealt{Broeg2005}, \citealt{Croll2015}). 

In most cases, it is likely that light from the target and comparison stars may not have travelled through the same column of atmosphere, especially in scenarios where the separation between the target and comparison stars in the sky is comparable or more than the typical spatial scale of variations in atmospheric turbulence. Systematics at the instrumental level and stellar variability in the comparison star can further cause complex non-linear variations between the target and comparison star fluxes. This implies that the linear functional forms of mapping between the two assumed by conventional methods are sub-optimal and may even be a source of additional systematics.   

The conventional strategy of MOS observations has relied on the availability of suitable close-by comparison stars which presents some issues. In situations when comparison stars are fainter than the target star or of a different spectral type, the Target/Comparison normalisation is photon-limited by the brightness of the comparison stars (in the whole bandpass or a range of wavelengths where the spectral shape and relative brightness of the comparison and target star differs the most, see e.g. \citeauthor{Diamond-Lowe2020a} (\citeyear{Diamond-Lowe2020a}, \citeyear{Diamond-Lowe2020b})). On the other hand, if the comparison stars are brighter than the target stars (as it happens to be the case for comparison star for the GMOS observations of HAT-P-26b presented in this paper), the duty cycle of the observations gets limited. Moreover, if the target star is in a sparse field, which is often the case for bright host stars, then there is less choice of an optimal comparison star given the limited instrument's field of view which has been a limiting factor in ground-based high precision spectrophotometric follow-up of exoplanets orbiting bright stars. 

In view of these several limitations, there is a need for a more generalised and robust approach to marginalise systematics in ground-based spectrophotometric light curves which accounts for non-linear relationship between target and comparison star fluxes, and does not explicitly rely on the availability of comparison stars.
% \textbf{For the latter, we note that methods using comparison stars or their PCA components as linear regressors to fit target star light curves (e.g. \citealt{Espinoza2019}) have explored the parameter space by allowing the weights corresponding to the comparison star flux in their regression models to go to zero.}

We present a novel alternative method in this paper which takes a more generalised approach when using a set of auxiliary time series (e.g., comparison star light curves, target star PSF width, airmass, etc.) to model systematics in the target star transit light curves. Our new method in essence lets a Gaussian Process model explore the underlying unknown and likely non-linear functional form between the regressors used to model the systematics in the target star transit light curves. This can be achieved for both the integrated white light curve and spectroscopic light curves. Through our method, we also demonstrate that remarkably precise wavelength-dependent transit depth measurements of exoplanet spectra can be reached when not using the comparison star light curves at all. We describe the method and its application in detail to our observations of the warm Neptune HAT-P-26b observed by Gemini/GMOS in Section \ref{sec_analysis}.  

The paper is distributed as follows: in Section \ref{sec_obs} we describe in detail our observational setup for the 6 transits of HAT-P-26b observed by GMOS to which we apply the new method we introduce in this paper. In Section \ref{sec_data_reduc}, we describe the data reduction steps to extract stellar spectra from raw data, and in Section \ref{sec_analysis} we discuss the analysis to model the GMOS transit light curves. Specifically, in Section \ref{noise_model} we introduce our new method to model the telluric and instrumental systematics directly in the target star light curves. In Section \ref{sec:method_comparison} we compare our new analysis method with the conventional approach, and discuss its caveats and implications for future ground-based observations of exoplanet atmospheres. In Section \ref{sec:interpretation} we interpret the optical to infrared transmission spectrum for HAT-P-26b from combined GMOS, HST, and Spitzer measurements using atmospheric models. We discuss the indications of transit timing variations for the planet in Section \ref{sec:ttv}, and in Section \ref{sec:conclusions} we present our conclusions.

%__________________________________________________________________
%__________________________________________________________________
%__________________________________________________________________
%__________________________________________________________________

% ---------------------
% SECTION
% Observations
% ---------------------

\section{Observations}
\label{sec_obs}

%%%%%%%%%%%%%%%%%%
% subsection
% GMOS Transmission spectroscopy
%%%%%%%%%%%%%%%%%%
\subsection{The warm Neptune HAT-P-26b}

HAT-P-26b is a low density warm (T$_{\rm eq}$ = 990 K) Neptune discovered by \cite{Hartman2011} orbiting its chromospherically quiet K1 host star in a close orbit of period $\sim$ 4.23 days. Given its large scale height the planet has been the subject of multiple atmospheric characterisation studies, including those constraining its atmospheric metallicity. Constraining the atmospheric metallicity of exo-Neptunes is crucial for tracing the dominant planet formation scenarios governing the formation of these planets, and distinguishing between scenarios of core accretion (\citealt{Pollack1996}) and in situ formation (\citealt{Bodenheimer2000}). Both of these scenarios can lead to significantly different metal enrichment of the atmosphere of a Neptune mass planet like HAT-P-26b. Initial studies of HAT-P-26b using Magellan/LDSS-3C and Spitzer by \cite{Stevenson2016} indicated tentative evidence of water vapour features in the red optical. \cite{Wakeford2017} reported a strong detection of the 1.4 $\mu$m water vapour feature muted by a grey opacity cloud as evident from the near-infrared and visible observations from HST/WFC3 and STIS respectively. From these observations \cite{Wakeford2017} retrieve  a near solar metallicity atmosphere with a high altitude cloud deck suppressing the transit spectral features. \cite{MacDonald2019} further combined the observations from \cite{Stevenson2016} and \cite{Wakeford2017} to perform a comprehensive retrieval analysis reporting the presence of several species of metal hydrides with absorption features ranging from optical to near infrared and a tentative hint of Rayleigh scattering.
% Moreover HAT-P-26b will be observed by the JWST Cycle 1 GTO using NIRISS-SOSS and NIRSpec-G395H modes (GTO 13121 , PI Lewis), NIRCam (GTO 11852, PI Greene), and MIRI-LRS (GTO 11773, PI Greene), which will capture the transmission spectrum in the wide range of 0.6 to 11 $\mu$m and investigate the presence and abundances of a host of molecular and atomic absorbers in the planet's atmosphere. 

In this paper, we present 6 Gemini/GMOS transit observations to measure the transmission spectrum of HAT-P-26b in the visible from 490 to 900 nm, extending the wavelength coverage of the transmission spectrum published by \cite{Wakeford2017} further towards blue optical. The primary motivations of our study are to investigate the exoplanet spectrum in the optical, expanding the wavelength coverage blueward, and independently test the presence of clouds and Rayleigh scattering. Additionally, from the precise mid-transit times obtained from our high SNR GMOS transit light curves we also investigate the transit timing variations (TTVs) for the planet previously indicated by \cite{Stevenson2016} and \cite{vonEssen2019}.

\subsection{GMOS Transmission spectroscopy}
\label{gmos_obs}

We observed a total of six transits of HAT-P-26b using the Gemini North telescope located at Mauna Kea, Hawaii, and the Gemini South telescope located at Cerro Pachon, Chile. Three transits were observed using Gemini North and three transits were observed using Gemini South. The observations used the same technique and setup as described in \cite{Huitson2017} (hereafter referred to as \citetalias{Huitson2017}), which is similar to that of previous observations using GMOS (e.g. \citeauthor{Bean2010} (\citeyear{Bean2010}, \citeyear{Bean2011}, \citeyear{Bean2013}), \cite{Gibson2013}, \cite{Stevenson2014}). All transits were observed as part of two survey programs of hot Jupiter atmospheres from GMOS North and South (P.I. J-.M. D\'esert) described in \citetalias{Huitson2017} (see Table \ref{obsstats} for program numbers). 

For each observation, we used the MOS mode of GMOS to observe the time series spectrophotometry of HAT-P-26b and a comparison star TYC 320-426-1, simultaneously. HAT-P-26 and the comparison star are separated by $\sim 3.8$ arcmin. HAT-P-26 has a V magnitude of 11.76 and TYC 320-426-1 has a V magnitude of 11.08, and similar spectral type from visual inspection of prominent stellar spectral features. Each observation lasted approximately 5 to 5.5 hours. To avoid slit losses, our MOS mask had wide slits of 10 arcsec width for each star. The slits were 30 arcsec long to ensure adequate background sampling for each star. 

In order to provide similar wavelength coverage between HAT-P-26 and the comparison star, the PA of the MOS mask needs to be as close as possible to the PA between the two stars. The PA for HAT-P-26 and the comparison star is 23 deg. E of N. However, at this PA, no suitable guide stars fell into the patrol field of Gemini's guider, the On Instrument Wave Front Sensor (OIWFS). We therefore used the Peripheral Wavefront Sensor (PWFS) instead for three of our observations from Gemini South (see Table \ref{obsstats} for details). The PWFS has a larger patrol field, but a lower guiding precision and so is used as a backup option if there are no suitable guide stars available for OIWFS. This setup enabled us to orient the instrument so that the instrument PA matched the PA between the two stars.

However, from the initial analysis, we found that the photometric precision was lower when using the PWFS than for our previous survey observations obtained using the OIWFS due to higher dispersion direction drift in case of PWFS as compared to OIWFS. The dispersion direction drift over a night is $\sim$15 pixels for PWFS as compared to $\sim$1 pixel for OIWFS. We therefore modified the setup for three of our observations at Gemini North (see Table \ref{obsstats}) to be able to continue using OIWFS. In this new setup, we selected the PA of the MOS mask to be 7 deg. E of N. While this meant that the wavelength coverage was different for both stars, it meant that we could orient the GMOS field of view such that a suitable guide star fell within the range of the OIWFS. We therefore achieved improved guiding in exchange for the loss of approximately 1/3rd of the wavelength coverage. 

Three transits were observed in the red optical with the R150 grating, covering a wavelength range of 530-900 nm with ideal resolving power $R=631$. Two transits were observed in the blue optical with the B600 grating, covering a wavelength range of 490-680 nm with ideal resolving power $R=1688$. The ideal resolving powers assume a slit width of 0.5 arcsec. In our case, due to using a wide slit, our resolution was seeing limited. Given the range of seeing measured in Table \ref{obsstats}, our resolution is up to $4\times$ lower than the ideal value depending on observation. For each observation, we used the gratings in first order. For the R150 observation, the requested central wavelength was 620 nm and we used the OG515\_G0330 filter to block light below 515 nm. The blocking filter was used to avoid contamination from light from higher orders. For B600 observation, the requested central wavelength was 520 nm and no blocking filter was needed. 

For all observations, we windowed regions of interest (ROI) on the detector to reduce readout time. We used one ROI for each slit, with each ROI covering the whole detector in the dispersion direction and approximately 40 arcsec in the cross-dispersion direction. We binned the output $1\times2$, binning in the cross-dispersion direction, to further reduce readout time. For the observations at Gemini South, the detector was read out with 3 amplifiers. For the observations at Gemini North, the detector was read out with 6 amplifiers. All amplifiers had gains of approximately 2 $e^-$/ADU. Exposure times were chosen to keep count levels between 10,000 and 30,000 peak ADU and well within the linear regime of the CCDs. Table \ref{obsstats} shows the observation log for each transit, as well as which observations were obtained at Gemini South and which at Gemini North. The numbers given under `No' in the table are the numbers by which we will refer to each transit observation in this paper. 

\begin{table*}
\centering
\caption{Observing Conditions for GMOS Runs. The numbers in the first column are the numbers by which we will refer to each transit observation throughout the rest of the paper. Observation IDs starting with ``GS" were observed at Gemini South using the ideal PA and the PWFS, while those starting with ``GN" were observed at Gemini North using the non-ideal PA and OIWFS (see Section \ref{sec_obs} for more details).}
\begin{tabular}{cccccccccc}
\hline
\hline
No. & Program ID & UT Date & Grating & Guider and PA & Exposure & No. of & Duty & Seeing & Airmass \\
& & & & &Time (s) & Exposures & Cycle (\%) &  (arcsec) & Range \\
\hline
1 &  GS-2013A-Q-27 & 2013 Mar 20  & R150 & PWFS, ideal PA& 50 & 226 & 63  & 0.6 - 1.8 & 1.21 - 2.06  \\
2 &  GN-2013A-Q-38 & 2013 Apr 10  & R150 & OIWFS, non-ideal PA & 15 & 574 & 45 & 0.3 - 0.9 & 1.04 - 1.97  \\
3 &  GS-2014A-Q-59 & 2014 May 09 & R150  & PWFS, ideal PA & 20 & 318 & 40 & 0.3 - 1.0 & 1.21 - 2.00 \\
6 &  GS-2014A-Q-59 & 2014 Jun 29 & R150  & PWFS, ideal PA & 25-40 & 299 &  47 & 0.6 - 1.0 & 1.21 - 1.96   \\    
4 &  GN-2016A-LP-6 & 2016 Mar 12 & B600  & OIWFS, non-ideal PA & 90-150 & 161 & 85 & 0.4 - 1.6 & 1.04 - 1.74 \\     
5 &  GN-2016A-LP-6 & 2016 Apr 15 & B600  & OIWFS, non-ideal PA & 110-150 & 131 & 88 & 0.6 - 1.3 & 1.04 - 1.97 \\

\hline
\end{tabular}
 %Note that the ``duty cycle" includes observation set up time, acquisition etc., which varies for each observation. There is therefore not an exact correspondence between exposure time and duty cycle.}
\label{obsstats}
\end{table*} 

% ---------------------
% SECTION
% Data Reduction
% ---------------------

\section{Data Reduction}
\label{sec_data_reduc}

%%%%%%%%%%%%%%%%%%
% subsection
% GMOS DATA Reduction
%%%%%%%%%%%%%%%%%%

\subsection{GMOS data}
\label{gmos_data_reduc}

We used our custom pipeline designed for reducing the GMOS data, the steps for which are described in more detail in \citetalias{Huitson2017}. We extract the 1D spectra and apply corrections for additional time- and wavelength-dependent shifts in the spectral trace of target and comparison stars on the detector due to atmospheric dispersion and airmass. In this section, we describe the main points of the pipeline and the additional corrections we apply to the data before extracting and analysing the transit light curves.

For the R150 grating, we only use 2/3 of the detector in the dispersion direction. For all observations we use a moving boxcar median of 20 frames in time for each pixel to compare its value in the frames immediately before and after. We flag pixels deviating more than 5 times the boxcar median value as cosmic rays and replace it with the median boxcar value. The cosmic-ray removal flagged a few percent of pixels per observation. Our pipeline flagged 1.8-3.8 \% of columns as bad depending on observation. The majority (80 \%) of flagged columns are consistent between the transits for each detector. For observations 1 and 3, these include columns of shifted charge occurring mostly in the transition regions between amplifiers, as discussed in \citetalias{Huitson2017}. These columns are not present on the GMOS-North detector (observations 2, 4, and 5) and are also not present in observation 6, which was taken after a detector upgrade at GMOS-South. 

We tested our extraction with and without flat-fielding and find that flat-fielding does not significantly affect the scatter of the resulting transit light curves. For this reason, and since flat-fielding did not improve the scatter blueward of 700 nm, we chose not to perform flat-fielding for all transit observations. We notice no slit tilt in the spectra of HAT-P-26 and the comparison star unlike as seen in \citetalias{Huitson2017} and \cite{Todorov2019}. The sky lines in the frames for all transits are parallel to the pixel columns. Thus, we choose to not perform any tilt correction.

We subtracted the background while performing optimal extraction (\citealt{Horne1986}), and found that taking the median background value in each cross-dispersion column provided the best fit to the background fluxes compared to using fits to the flux profile in each cross-dispersion column. The background fluxes were 1-10 \% of the stellar flux for the R150 observations and 2-20 \% of the stellar flux for the B600 observations, depending on the wavelength range and exposure number. 

After spectral extraction, we performed wavelength calibration using CuAr lamp spectra taken on the same day as each science observation. To obtain the CuAr spectra at high resolution, we used a separate MOS mask to that used for science, which had the same slit position and slit length as the science mask of only 1 arcsec width. We used the same grating and filter setup as the corresponding science observation.

We used the \textsc{identify} task in the Gemini \textsc{iraf} package to identify spectral features in the CuAr spectra. A wavelength solution was then constructed by a linear fit to the pairs of wavelength vs. pixel number in each ROI and then refined by comparison with known stellar and telluric feature locations. The final uncertainties in the wavelength solution are approximately 1 nm for all observations, which is $\sim$5 \% of the bin widths used to construct the final transmission spectrum.

Before generating the transit light curves, we performed further reduction of the extracted 1D spectra. This is because, as in \citetalias{Huitson2017}, we found that there is a dispersion-direction shift of the spectra on the detector during each observation that is a function of time and of wavelength, such that the spectra `stretch' over time. The result is that wavelength bins identified in fixed pixel space will not sample the same wavelength in each exposure. The effect therefore needs to be accounted for in order to build transit light curves that sample a constant wavelength region over time. Failure to account for this effect can introduce spurious slopes in the transmission spectra, as discussed in \citetalias{Huitson2017}. In \citetalias{Huitson2017}, we found that a model for differential atmospheric refraction explained the shifts well for our previous observations. This is consistent with the fact that GMOS has no atmospheric dispersion compensator, and so we expect an effect from differential atmospheric refraction. However, the differential atmospheric refraction model does not adequately fit the shifts observed in the HAT-P-26 data studied here. We therefore use the alternative method developed in \citetalias{Huitson2017}, in which we use multiple spectral features for cross-correlation as a function of time to account for the wavelength-dependent shifting empirically. 

However, instead of simply using the shifted spectra corresponding to the measured shift value from cross correlation with respect to the nearest feature for constructing light curves for each spectral bin (as done in \citetalias{Huitson2017}), we proceed further to use the information from cross-correlation with the spectral features to apply corrective shifts to each pixel in the 1D spectrum. From the measured shifts of the spectral features for an exposure, we estimate the shifts in the spectra for pixels in between and away from the features by a linear interpolation between the shift values for three features used for the cross-correlation. The interpolated shift values thus obtained for each pixel for an exposure are then applied to the whole 1D spectrum. We then repeat this step for every exposure so that in the end we have the same wavelength solution for the spectra across all exposures. We repeat this step for the comparison star as well, using the same set of spectral features as those used for the target star spectrum for cross-correlation. Finally, we interpolate the comparison star's spectrum from all exposures onto the wavelength solution of the target star, omitting detector gaps and bad columns, which ensures that both the target and comparison star spectra for every exposure have the same wavelength solution within the uncertainty of our estimates on the shifts derived from cross-correlation.

As final step in our reduction process, we also correct for the dispersion direction offset between the target and comparison star spectra. This occurs because the PA of the instrument is not exactly the same as the PA between the target and comparison star for observations taken using OIWFS guider. We used cross-correlation to measure the offset, which was between -18.2 and -16.0 pixels for the southern observations. For the northern observations, the offset was between -830 and -600 pixels due to the non-deal PA. We then interpolated the comparison star's spectrum onto the target star's wavelength grid, while omitting bad columns for both spectra (which are the same columns on the detector but are at different wavelengths for each star). 

\begin{figure*}

  \centering
  \includegraphics[scale=0.45]{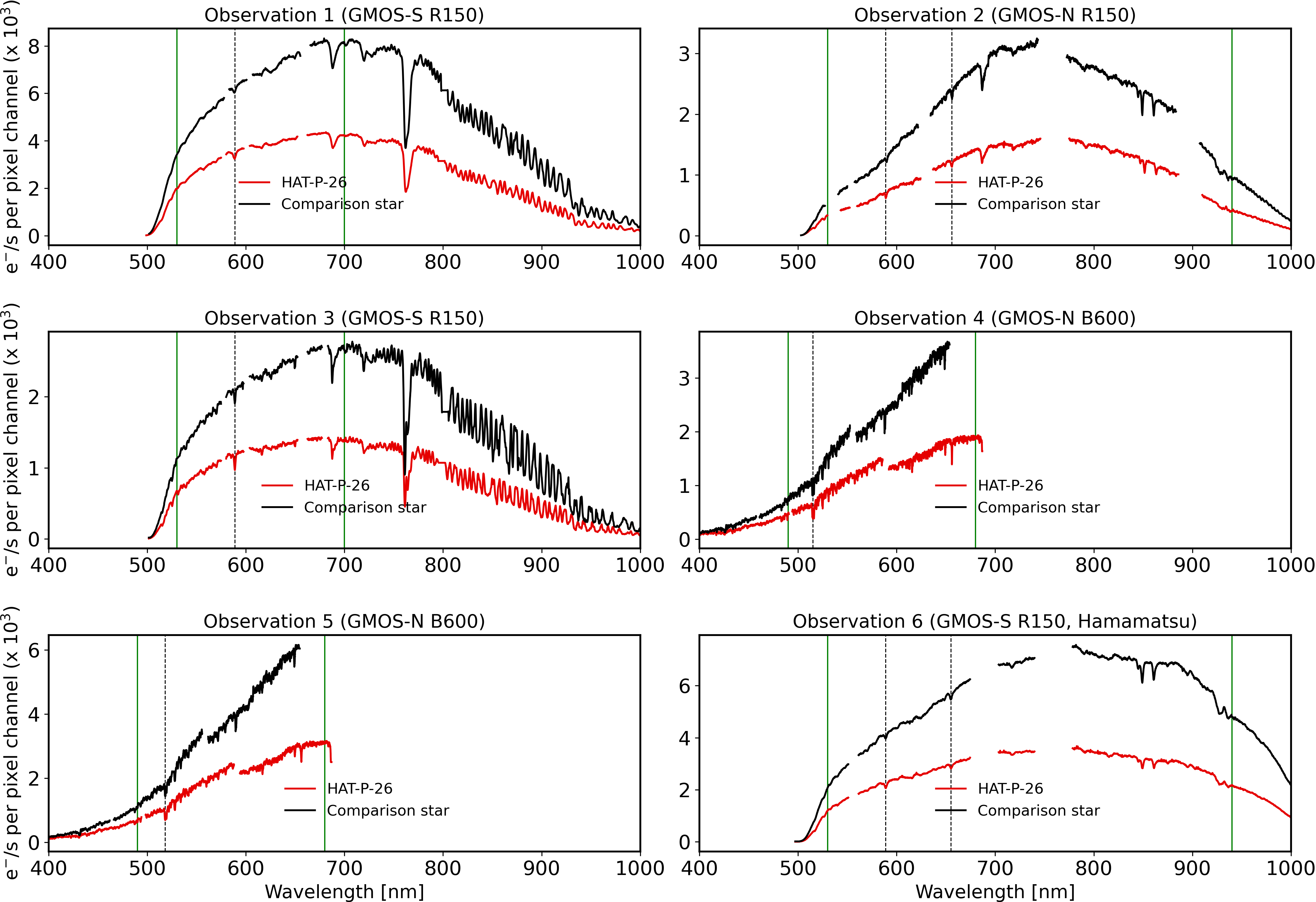}
  \caption{Optimally extracted spectra for HAT-P-26 and the comparison star from an arbitrarily chosen exposure, corrected for dispersion direction shifts and normalised by their exposure time for the 6 GMOS observations of HAT-P-26. Each panel shows one exposure for each observation, and the observation numbers correspond to the programs described in Table \ref{obsstats}. For all observations, especially the GMOS-N observations taken using non-ideal PA, the comparison star spectrum has been shifted to the same wavelength grid as the target star spectrum using prominent common stellar features in the spectra which have been marked by the black dashed vertical line. The green vertical lines show the wavelength range considered for obtaining the transmission spectrum for each observation. The gaps in the spectra correspond to physical gaps in the CCD and bad columns.    
  }
  \label{fig:1Dspec}
\end{figure*}

We show the final wavelength-calibrated 1D spectra for an arbitrarily chosen exposure for the target and comparison star in Figure \ref{fig:1Dspec} for all the observations. Note that some residual shifts (at the order of a few pixels) still remain between the target and comparison star spectra, especially towards the redder end for R150 observations (beyond 710 nm where fringing also becomes strong) as seen in Figure \ref{fig:1Dspec}. This is because the shifts between target and comparison star spectra vary both in time and wavelength, and constant offsets followed by interpolation to a common wavelength grid does not entirely correct for it. We refrain from further empirical corrections at this stage and to minimize the effects of any residual shifts we choose to use broad 20 nm wide bins for our spectroscopic light curves. This wavelength width is significantly larger than spatial scale of the shifts between the target and comparison star spectra. We nevertheless do not include the spectra beyond 730 nm for Observation 1 and 3 for computing the transmission spectrum due to excessive fringing in that region. We also emphasize that the residual shifts between the comparison star and HAT-P-26 spectra are not an issue for the new method we introduce in the paper of using only the target star to extract the transmission spectrum (see Section \ref{noise_model}).

% ---------------------
% SECTION
% ANALYSIS
% ---------------------

\section{Transit Light Curve Analysis}
\label{sec_analysis}
We now describe our light curve analysis methods that we apply to the 6 transit observations of HAT-P-26b. We first briefly discuss the noise models that we use to correct for the systematics in the light curves in Section \ref{noise_model}. In this section, we also introduce and motivate a new method to directly model the systematics in the target star light curves. 

We have summarised the the conventional method used to date and the new method introduced in this paper and their various types of applications to white and spectroscopic light curves in Table \ref{tab:method_summary}.  

The novel aspect of the new method in the context of both the white light curves and the spectroscopic light curves is that instead of assuming a linear functional form, we explore a distribution of functions (described by a GP) to explore the likely non-linear functional form of the mapping between the target transit light curves and one or more decorrelation time series (e.g. comparison star light curves). The new method in the context of all its applications is an alternative to the applications of the conventional linear method to fit for systematics in MOS transit light curves as described in Table \ref{tab:method_summary}. We describe the shortcomings of the conventional method and motivate the need for the new method in Sections \ref{sec:conv_method} and \ref{sec:new_method} respectively.

The conventional method to fit the white light curves specifically has two different types of applications: 1) \texttt{Conv1:WLC} : two step method of first performing differential spectrophotometry (normalising the target star light curve by the comparison star light curve) and then fitting the resultant light curve with a GP, and 2) \texttt{Conv2:WLC} : one step method of using a linear model with one or more comparison star light curves or their PCA components as one of the regressors to fit the target transit light curves. \texttt{Conv2:WLC} is especially suited for when there are more than one comparison stars available, which is not the case in this paper. In Section \ref{sec:WLC} we apply the conventional method \texttt{Conv1:WLC}, and the new methods \texttt{New:WLC} and \texttt{New:WLC;No\_Comp} to fit white-light curves for each observation.

In the context of fitting spectroscopic light curves, a frequently used method to correct for wavelength independent systematics in particular, in addition to using the comparison star light curves, is to perform a `common-mode correction'. This is the approach of the \texttt{Conv1:$\lambda$LC} method which subtracts a white light curve derived common-mode trend from each wavelength binned light curve. However, this approach also assumes a linear relationship between the common-mode trend and the spectroscopic light curves. Our new method \texttt{New:$\lambda$LC} explores the likely non linear relationship in this context (e.g. arising from wavelength dependent effects with changing airmass) by using the common-mode trend as a GP regressor. We fit the spectroscopic light curves using \texttt{Conv1:$\lambda$LC} and \texttt{New:$\lambda$LC} in Section \ref{sec:binned_LC_old_method} and \ref{sec:binned_LC_new_method} respectively.

\begin{table*}
\caption{Summary of the conventional and new methods used to model the systematics in white light curves (WLC) and spectroscopic light curves ($\lambda$LC) in this paper in Section \ref{noise_model}. The `Application' column specifies the different ways of applying the methods with a more detailed description in the column `Description'. `Abbreviation' specifies how we refer to each of these applications in this paper.}
\label{tab:method_summary}
\begin{tabular}{lllll}
\hline
\hline
Method              &   Application                                            & Description                             & Abbreviation                     & Example  \\
                    &                                                          &                                         &                                  & References \\
                                          \hline
\multirow{12}{*}{Conventional} &\multirow{2}{*}{Differential spectrophotometry}& Target/Comparison WLC: fit with GP     & \texttt{Conv1:WLC}               & \multirow{4}{*}{\cite{Gibson2013}} \\
\multirow{12}{*}{method} & \multirow{2}{*}{using comparison star LCs}          &                                         &                                  &                                   \\
                    &                                                          & Target/Comparison $\lambda$LC:          & \texttt{Conv1:$\lambda$LC}       &                                   \\
                    &                                                          & common-mode subtracted, fit with GP     &                                  &                                   \\
                                                              \\ \cline{2-5} \\
                    & \multirow{6}{*}{Comparison star(s) LC}                   & Target WLC:                            &  \texttt{Conv2:WLC}              & \multirow{6}{*}{\cite{Espinoza2019}}          \\
                    & \multirow{6}{*}{as linear regressor}                     & fit with linear model                   &                                  &                                   \\
                    &                                                          & including comparison star(s) white LC   &                                  &                                   \\
                    &                                                          & or their PCA as regressors              &                                  &                                   \\ \\
                    &                                                          & Target $\lambda$LC:                     &  \texttt{Conv2:$\lambda$LC}      &                                   \\
                    &                                                          & fit with linear model                   &                                  &                                   \\
                    &                                                          & including comparison star(s) $\lambda$LC&                                  &                                   \\
                    &                                                          & or their PCA  as regressors             &                                  &                                   \\
                    &                                                          &                                         &                                  &                                   \\
                                          \hline
\multirow{10}{*}{New method}          & Comparison star LC                     & Target WLC:                            & \texttt{New:WLC}                 & \multirow{10}{*}{This work}         \\
                    & as GP regressor                                          & fit with comparison star                &                                  &                                       \\
                    &                                                          & as a GP regressor                       &                                  &                                       \\
                                                              \\ \cline{2-4} \\
                    & No comparison stars                                       & Target WLC:                           & \texttt{New:WLC;No\_Comp}         &                                   \\
                    &                                                          &  fit with GP regressors                 &                                  &                                   \\
                    &                                                          &  excluding comparison star LC           &                                  &                                   \\
                                                              \\ \cline{2-4} \\
                    & common-mode trend as                                     & Target $\lambda$LC:                     & \texttt{New:$\lambda$LC}         &                                   \\
                    & GP regressor                                             & fit with common-mode trend              &                                  &                                   \\
                    &                                                          & as a GP regressor                       &                                  &                                   \\
                    \hline
\end{tabular}
\end{table*}

%%%%%%%%%%%%%%%%%%
% subsection
% Modelling the Transit Light Curves : Model description 
%%%%%%%%%%%%%%%%%%

\subsection{Modelling systematics in transit light curves}
\label{noise_model}

In the following sections, we model the instrumental and telluric time-dependent systematics in the HAT-P-26 transit light curves by following both the conventional method and the new method we introduce in this paper. We describe them both in the next two subsections to compare them and motivate the need for the new method. 

\subsubsection{Conventional method: using comparison star or common-mode trend as linear regressor}
\label{sec:conv_method}

The conventional method involves first dividing the target star light curve by comparison star light curve and then fitting the transit signal and systematics in the Target/Comparison light curve simultaneously using a transit model and a GP respectively. In the case of spectroscopic light curves, there is an additional step of removing the common-mode trend before fitting with a GP. The GP model takes as regressors, or inputs, a set of decorrelation time series which include e.g., time (time stamps of individual exposures), width (FWHM) and spatial shifts of the traces of target and comparison stars on the detector (e.g. \cite{Nikolov2018}, \cite{Diamond-Lowe2020b}). However, the step of doing differential spectrophotometry itself in this approach raises concerns on the relevance of decorrelation parameters derived from the individual target and comparison star spectral traces in the context of modelling the differential Target/Comparison light curve. In general, the step of doing differential spectrophotometry, assumes that the target and comparison star fluxes are affected by the same or linearly related time and wavelength-dependent systematics. Subtracting common-mode trend also assumes a linear relationship between the white light curve and the spectroscopic light curves. Given the complex nature of both instrumental and telluric systematics this is likely not the case. Considering the transit depth precisions ($\sim$ 100-500 ppm per $\sim$ 20 nm bins) we are aiming for, dividing the target star light curve by the comparison star light curve or subtracting common-mode trend can potentially propagate unwanted systematics and deteriorate the light curve SNRs which can be difficult to correct for when fitting the Target/Comparison light curve. An example of this are the B600 observations of HAT-P-26b we present in this paper where the target and comparison star light curves have systematics significantly different from each other. In this context, simply normalising the target star by the comparison star contaminates the transit signal originally present in the Target star light curve (see white light curves for observation 4 and 5 in Figure \ref{fig:WLC_all}). 

In cases when the instrument's field of view is large, in the order of $\sim$ 10 arcminutes, recent works (\citealt{Jordan2013}, \citealt{Espinoza2019}) have used Principal Component Analysis (PCA) to optimally use information from multiple comparison stars in the field of view. This approach (\texttt{Conv2:WLC} and \texttt{Conv2:$\lambda$LC}) relies on the availability of multiple comparison stars, and involves using the PCA components of more than one comparison star in log-space as regressors in a linear regression model to fit the systematics in the target star light curve. Specifically \cite{Espinoza2019}, and other studies analysing IMACS/Magellan observations e.g., \cite{Weaver2020}, \cite{McGruder2020}, \cite{Kirk2021} use the model averaging scheme for linear regression models outlined by \cite{Gibson2014} to incorporate the number of relevant PCA components as an additional uncertainty in their model. Since we only observed one comparison star, we do not test the PCA based approach in this work.

\subsubsection{New method: using comparison star or common-mode trend as a GP regressor}
\label{sec:new_method}

The intrinsic limitations of differential spectrophotometry using one or more comparison stars to correct for systematics in the light curves (also described in more detail in Section \ref{sec_intro}) narrows down the set of exoplanets around bright host stars that can be followed up for atmospheric characterisation from ground-based multi-object spectrographs. Hence, there is a need for a new method that doesn't explicitly rely on the comparison stars and can model the transit light curve systematics and extract the transmission spectra solely from the target star. 

With the new method we introduce in this paper, we present a way to directly fit the target star light curves using a set of time series recorded at the same time as target light curve as GP regressors. This includes the comparison star light curve (when fitting the white light curves, described in more detail in Section \ref{sec:WLC}), and common-mode trend when fitting the spectroscopic light curves (see Section \ref{sec:binned_LC}). This is essentially the novel aspect of our method: for both white and spectroscopic light curves, we use the set of time series, which have traditionally been used linearly to correct them either as simple normalising factors or as linear regressors, directly as regressors in a GP model. This allows for letting the GP itself explore an exhaustive set of non-linear mappings between the target transit light curve and regressors like comparison stars or the common-mode trend. This approach is more capable of incorporating the complex differences in which the target and comparison stars are affected by systematics during any observations. Our method also provides more accurate uncertainties by propagating them through the Bayesian framework of GPs. 

The underlying GP framework we use for our new method is the same as that introduced by \cite{Gibson2012} to model the transit signal and systematics simultaneously in the wavelength integrated white light curves and the wavelength binned light curves for each transit. In our new method, instead of fitting the Target/Comparison light curves with a GP model, we model the target star light curves directly as a numerical transit model combined with an additive GP model to account for systematics affecting the light curve. This means that we skip the step of dividing the target star light curve by one or multiple comparison star light curves, and instead use the comparison star light curve as one of the GP regressor. In the case of fitting spectroscopic target light curves, we use the common-mode trend derived from the white light curve as a GP regressor(see Section \ref{sec:binned_LC}). We describe the GP formalism used for both the conventional and new methods in more detail in the next section.

\subsubsection{Gaussian Process regression model}

A Gaussian Process model to account for the systematics in a transit light curve means that we model the observed transit light curve (which for the conventional method is Target/Comparison and for the new method just the target star light curve) as a multivariate Gaussian Process distribution with the mean function as the numerical transit model, and a covariance matrix $\mathbf{\Sigma}$:

\begin{equation}
f  = \mathcal{GP}(\mathbf T(t,\phi), \mathbf{\Sigma} (\mathbf X,\theta))
\label{gp_model}
\end{equation}

where $f$ is the flux time series representing a transit light curve, $t$ is time, $\phi$ is the set of planet transit parameters, $T(\bmath t,\phi)$ is the astrophysical transit light curve model, and $\mathbf{\Sigma} (\bmath X,\theta))$ is the covariance matrix described by a kernel function for a set of regressors or input parameter vectors $\mathbf X$ and hyperparameters $\theta$ : 

\begin{equation}
\mathbf\Sigma_{ij} = k({x}_i,{x}_j | \theta)
\end{equation}

Note that here we assume that the systematics we are attempting to model using the GP are additive, and we could just modify Equation \ref{gp_model} to instead have the GPs model multiplicative systematics, which subsequently gives identical results within the precision of our data as we tested and was also reported by \cite{Gibson2013}. The kernel function takes a set of input parameter vectors $\mathbf X$ ($\mathbf x_{1}$, $\mathbf x_{2}$, $\mathbf x_{3}$, ... $\mathbf x_{P}$) (each vector $\mathbf x_{p}$ of the same length as the number of points ($N$) in the light curve) which could be time, Cassegrain Rotator Position Angle (CRPA), the airmass, FWHM of the PSF of the spectra trace of the target star, and the measured position of the spectral trace corresponding to each exposure (averaged across the dispersion direction). This is analogous to using these time series quantities as decorrelation parameters to construct parametric models. In particular, in our new method we additionally also test the use of the comparison star light curve as one of the regressors to the GP. 

We choose to use the Mat\'ern 3/2 kernel function as it is known to provide a good prescription for time correlated noise at the time scales typically observed in GMOS transit light curves (\citealt{Gibson2012}): 

\begin{equation}
% \begin{multlined}
k({x}_i, {x}_j | \theta) = A \left( 1+{\sqrt{3}R_{ij}} \right) \exp \left( -{\sqrt{3}R_{ij}}\right) 
+ \delta_{ij}\sigma_\text{w}^2
\label{kernel}
% \end{multlined}
\end{equation}

where $A$ is the hyperparameter specifying the amplitude of covariance, $\sigma_\text{w}$ is the white noise term (which we fit for) and $\delta$ is the Kronecker delta. We emphasize that keeping the white noise term $\sigma_\text{w}$ free when fitting the light curves is an important aspect of our proposed method in this paper. The best fit value of $\sigma_\text{w}$ represents the combined noise variances in the target star light curve and in the individual decorrelation parameters used as GP regressors, assuming no heteroscedasticity in our observed light curves (see Equation 6 in \cite{Mchutchon2011}). When we use comparison star light curve as one of the GP regressors, the best fit value of $\sigma_\text{w}$ represents the combined noise variance from the comparison star light curve and the target star light curves. We highlight that this is a way to propagate the relevant uncertainties from the comparison stars within the Bayesian framework of GPs (which we use to fit for $\sigma_\text{w}$ as described below) in contrast to just adding them in quadrature as done in the case of differential photometry. This is analogous to fitting for a jitter term in the methods that use comparison stars as an input to linear regression models (e.g. \citealt{Espinoza2019}). 

The term $R_{ij}$ in the Equation \ref{kernel} is a quadrature summation of pairwise difference between regressor points ($\eta$ being the inverse length scale hyperparameter corresponding to each input vector). The $R_{ij}$ term for P number of input vectors can be described as:  

\begin{equation}
\mathit{{R}_{ij}} = \sqrt{ \sum_{p=1}^{P} \left( \frac{{x}_{p,i}-{x}_{p,j} }{\eta_{p}} \right)^{\!\!2}  }
\label{eq:Rij}
\end{equation}

This is one of the few ways in which information from multiple input parameters or regressors can be combined to describe the covariance matrix of the GP, and involves a single amplitude hyperparameter ($A$) and length scale hyperparameters ($\eta$) for each of the input parameters respectively. We also considered and tested another type of combination where we take the kernel in Equation \ref{kernel} for each regressor, and construct the final kernel as the sum of kernels for each regressor (similar to the approach followed by \cite{Aigrain2016}, \texttt{k2sc}). This combination leads to using more number of hyperparameters as each GP regressor now also has a respective amplitude hyperparameter in addition to a length scale hyperparameter. For all the observations we analyse in this paper, we find that the first type of kernel combination (described in Equation \ref{eq:Rij}) performs consistently better in terms of root mean square (RMS) of the residuals and consistency of best fit transit parameters with the literature values as compared to the other combinations.

The joint GP posterior probability distribution we marginalise over to estimate the transit parameters and hyperparameters corresponding to the best fit to the observed light curves is: 

\begin{equation}
p(f | \bmath t,\phi,\theta) = \pi(\phi,\theta) \times \mathcal{L}  [ \mathcal{GP} \left (\mathbf T(t,\phi) , \mathbf{\Sigma} (\mathbf X,\theta) \right) ]
\end{equation}

where $\pi(\phi,\theta)$ encodes the prior probability on the transit model parameters ($\theta$) and hyperparameters ($\phi$), and $\mathcal{L}  [ \mathcal{GP} \left (\mathbf T(\bmath t,\phi) , \mathbf{\Sigma} (\mathbf X,\theta) \right) ]$ is the GP likelihood, written in form of log-likelihood as:  

\begin{equation}
% \begin{multlined}
\label{eq:likelihood}
\log \mathcal{L}(r | \mathbf X,\phi,\theta) = -\frac{1}{2} r^T\, \mathbf{\Sigma}^{-1} \, r -\frac{1}{2}\log | \mathbf{\Sigma}|  -\frac{N}{2} \log\left(2\pi\right)
% \end{multlined}
\end{equation}
\\
where $r$ is the vector of residuals of the observed light curve from the mean function ($f - \mathbf T(t,\phi)$) and $N$ is the number of data points in the light curve. 

We used the transit modelling package {\texttt{batman}} (\citealt{Kreidberg2015} which is an implementation of the formalism of \cite{Mandel2002}) to calculate the numerical transit model $\mathbf T(t,\phi)$, and the package {\texttt{george}} (\citealt{Ambikasaran2015}) for constructing and computing the GP kernels and likelihoods.

%%%%%%%%%%%%%%%%%%
% subsection
% WLC Analysis  
%%%%%%%%%%%%%%%%%%

\subsection{Analysis of White Transit Light Curves}
\label{sec:WLC}

\subsubsection{Constructing white light curves}
\begin{figure*}

  \centering
  \includegraphics[scale=0.45]{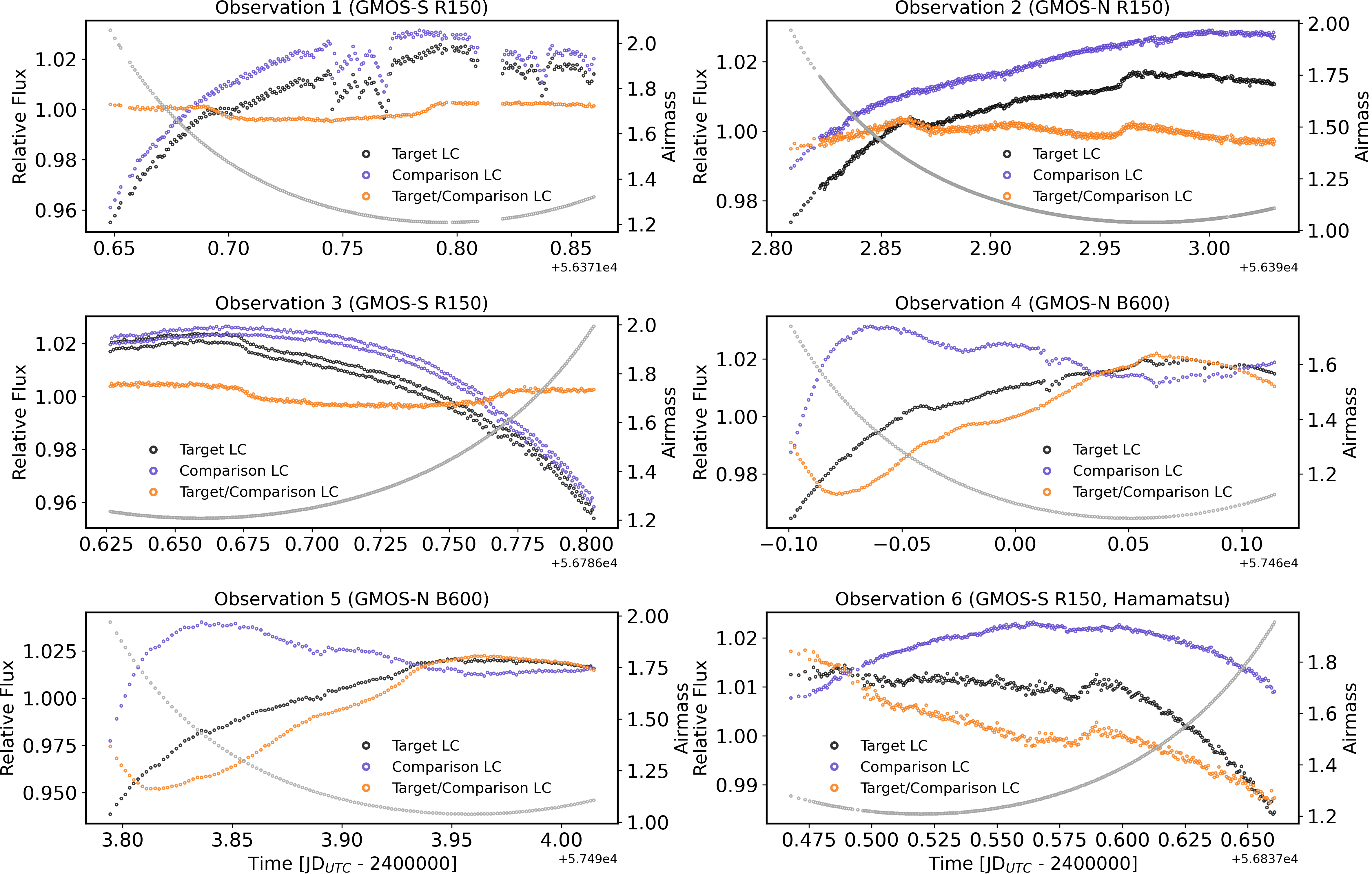}
  \caption{Raw wavelength integrated white target and comparison star light curves of GMOS observations of HAT-P-26b first normalised by the exposure times of the individual exposures and then by their out of transit median flux. Note the low frequency trend present in both the target and comparison star light curves due to the changing airmass (shown in grey) through the night. Observations 2, 4, and 5 were taken using non-ideal PA which is reflected here in the deviating trends between the target and comparison star light curves of the corresponding observations which subsequently contaminates the transit signal in the Target/Comparison light curve and are examples of sub-optimal results from Target/Comparison star light curve normalisation. Observation 6, which was taken with the newly installed Hamamatsu detector on Gemini South, also shows a similar deviating trend between the the target and comparison star light curves.    
  }
  \label{fig:WLC_all}
\end{figure*}

For each observation, we constructed the target and comparison star white light curves by summing the measured flux over 530 to 700 nm for observations 1 and 3, 530 to 900 nm for observations 2 and 6 (as these R150 observations do not show fringing redward of 700 nm), and 490 to 680 nm for observations 4 and 5. We then normalise the total flux in each exposure by the corresponding exposure time for both the target and comparison star. The white transit light curves thus obtained are shown in Figure \ref{fig:WLC_all}. The white light curves for each observation contain information on the dominant time-dependent systematics affecting all the wavelength channels, and analysing them prior to fitting the wavelength dependent light curves is an important step to constraining transit parameters and understanding the sources of systematics that can affect the final transmission spectrum.   

\subsubsection{Fitting the white transit light curves}

We obtain the best fits for each transit observation independently using both the conventional method \texttt{Conv1:WLC} and the two applications of the new method \texttt{New:WLC} and \texttt{New:WLC;No\_Comp} as described in \ref{noise_model}. For both methods and for each transit white light curve, we fix the orbital period ($P$) and eccentricity ($e$) to literature values, and fit for the orbital inclination ($i$), orbital separation ($a/R_\star$), mid transit time ($T_0$), and planet to star radius ratio ($R_{\rm P}/R_\star$). For $i$ and $a/R_\star$, we put a Gaussian prior with the mean and standard deviation as the mean and 3 times the 1 $\sigma$ uncertainty values measured by \cite{Stevenson2016} respectively. We use truncated wide uniform priors for $R_{\rm P}/R_\star$ and for the mid transit time ($T_0$) around the values predicted by a linear ephemeris. We adopt a linear stellar limb darkening law and calculate the limb darkening coefficients and uncertainties on them (stemming from uncertainties in stellar parameters) for the wavelength range integrated to obtain the white light curve, and the wavelength bins we adopt for spectroscopic light curves (see Section \ref{sec:binned_LC}) using \texttt{PyLDTk} (\citealt{Parviainen2015}), which uses the spectral library in (\citealt{Husser2013}), based on the PHOENIX stellar models. We put a Gaussian prior on the linear limb darkening coefficient with the mean value and the standard deviation as the mean and 3 times the 1 $\sigma$ uncertainty calculated from \texttt{PyLDTk} respectively. We summarise all the priors we use in this paper in Table \ref{tab:priors}.

\begin{table}
\centering
\caption{Summary of priors and fixed values for the parameters (transit model and GP hyperparameters) used to fit the transit light curves of HAT-P-26b. We fixed the planet orbital period (P) and eccentricity (e) for all fits. $\mathcal{U}$ represents a uniform prior applied within the specified range, and $\mathcal{N}$ represents a Gaussian prior with the specified mean and standard deviation. T$_{c}$ is the predicted mid transit time for each epoch using the ephemeris from \citealt{Hartman2011}. For the linear limb darkening coefficient mean of the Gaussian prior is taken as the theoretically calculated value from \texttt{PyLDTk} (\citealt{Parviainen2015}) for the B600 and R150 wavelength ranges.}
\begin{tabular}{ccc}
\hline
\texttt{batman} model \\
\hline
\hline
Parameter & Prior/Fixed value & Reference \\
\hline
P [d] &  4.2345 & \cite{Hartman2011} \\
$e$ & 0.124  &  \cite{Hartman2011} \\
$i$ [$^\circ$] & $\mathcal{N}$ (88.09, 1.5)  & \cite{Wakeford2017}\\
$R_{\rm P}/R_\star$ & $\mathcal{U}$ (0, 1)   & -- \\
$a/R_\star$ & $\mathcal{N}$ (11.89, 1.2) & \cite{Wakeford2017} \\
$T_{0}$[d] & $\mathcal{U}$ (T$_{c}$-0.001, T$_{c}$+0.001) & \cite{Hartman2011} \\
u$_{1}$[B600] & $\mathcal{U}$ (0.603, 0.03)  & \texttt{PyLDTk} \\  
u$_{1}$[R150] & $\mathcal{U}$ (0.73, 0.03)  & \texttt{PyLDTk}  \\  \\ \\
\hline
GP model \\
\hline
\hline
ln (A) & $\mathcal{U}$ (-100, 100)  & -- \\
ln ($\eta_{p}$) & $\mathcal{U}$ (-100, 100) & -- \\
$\sigma_{w}$ &  $\mathcal{U}$ (0.00001, 0.005) & -- \\
\hline
\end{tabular}

\label{tab:priors}
\end{table} 

We also fit for the white noise hyperparameter $\sigma_{w}$ as described in Section \ref{noise_model} which lets the GP model fit for the white noise variance in the target star light curves along with contributions from the variance in the GP regressors (e.g. the comparison star light curve). This is one of the key advantages and an important feature of our method as instead of propagating the variance from comparison star light curve by simply adding in quadrature (as is the case when the target star light curve is normalised by the comparison star light curve), our method provides a way to propagate uncertainties from the comparison star light curve to our fit of the target star light curve within the Bayesian framework of GPs described in Section \ref{noise_model}. We further emphasize that fitting for $\sigma_{w}$ is crucial for allowing the GP model to capture the white noise in the target star light curves.

\begin{figure*}

  \centering
  \includegraphics[scale=0.4]{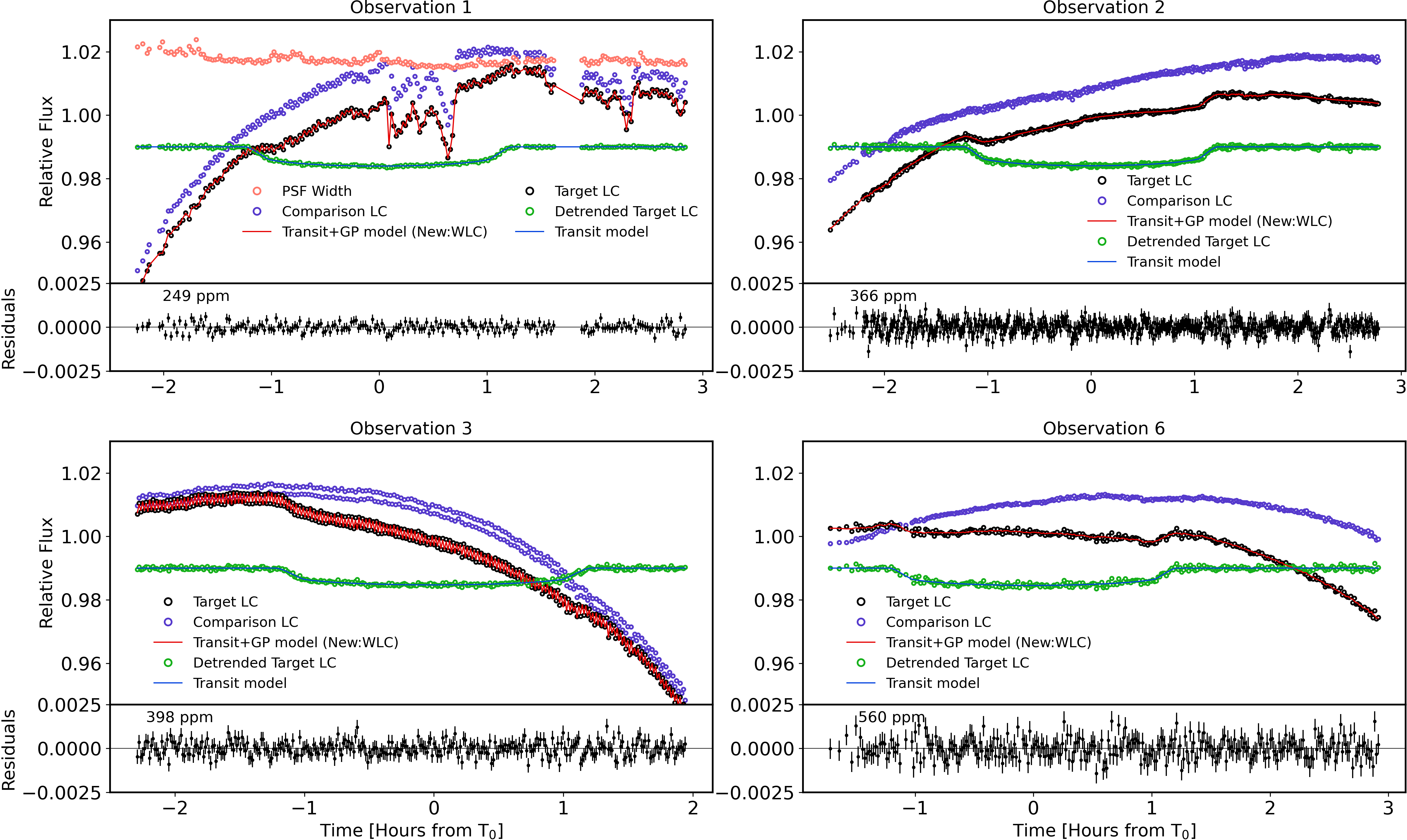}
  \caption{White transit light curves for HAT-P-26b obtained using the GMOS-R150 grism integrated in the range of 530 to 700 nm for observations 1 and 3, and in the range of 530 to 900 nm for the observations 2 and 6. Purple points show the comparison star light curve, black points show the target star (HAT-P-26) light curve overplotted with the best fit \texttt{New:WLC} model in red and the corresponding residuals plotted in the bottom panel of each observation, and green points show the detrended target star light curve overplotted with the {\texttt{batman}} transit model corresponding to the best fit transit parameters in blue. For observation 4, in pink is overplotted the PSF width time series for the spectral trace of target star. Note that the target and comparison star light curve are affected by the known odd-even pattern in GMOS datasets due to unequal travel times of the GMOS shutter blades which are known to differ slightly with the direction of motion (\citealt{Stevenson2014}, \citealt{Jorgensen2009}) as seen significantly in observations 1 and 3.   
  }
  \label{fig:r150_WLC}
\end{figure*}

\begin{figure*}

  \centering
  \includegraphics[scale=0.4]{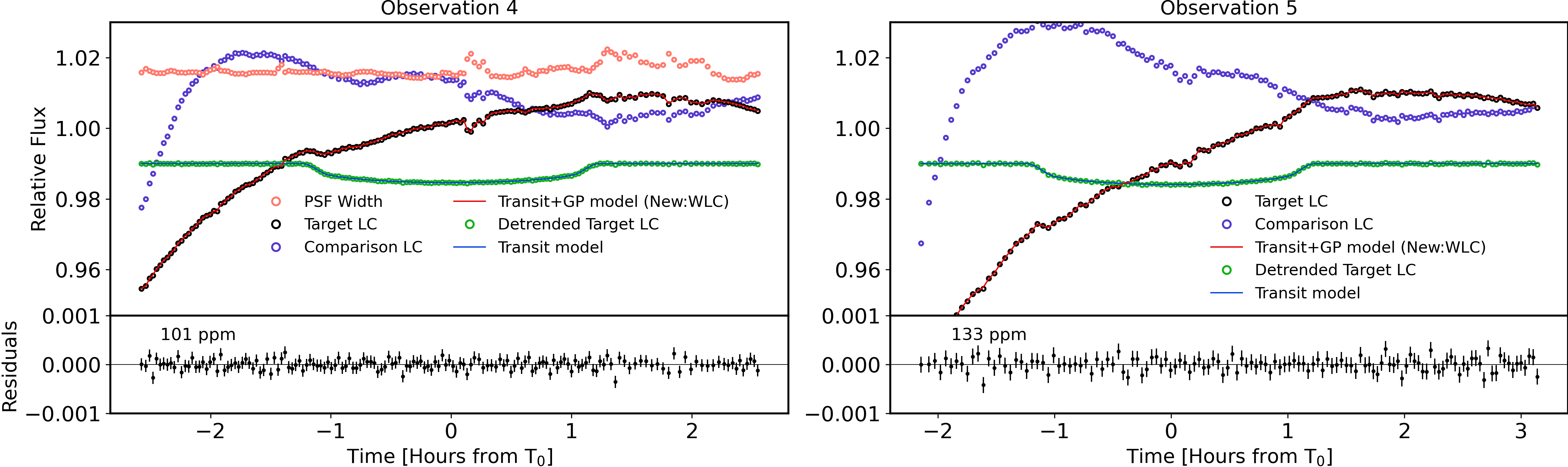}
  \caption{White transit light curves for HAT-P-26b obtained using the GMOS-B600 grism integrated in the range of 490 to 680 nm for observations 4 and 5. Purple points show the comparison star light curve, black points show the target star (HAT-P-26) light curve overplotted with the best fit \texttt{New:WLC} model in red and the corresponding residuals plotted in the bottom panel of each observation, green points show the detrended target star light curve overplotted with the {\texttt{batman}} transit model corresponding to the best fit transit parameters in blue. For observation 4, in pink is overplotted the PSF width time series for the spectral trace of target star.      
  }
  \label{fig:b600_WLC}
\end{figure*}

For the conventional method application \texttt{Conv1:WLC}, we perform fits for both R150 and B600 observations using all combinations of following GP regressors common to both target and comparison star light curves: time, CRPA, and airmass. For the new method application \texttt{New:WLC} we use all combinations of the following GP regressors: comparison star light curve, time, CRPA, airmass, PSF full width at half maxima (FWHM) of the spectral trace for every exposure (averaged across the dispersion direction) for the target star. For \texttt{New:WLC;No\_Comp} we use all GP regressors as for \texttt{New:WLC} except comparison star light curve to demonstrate the performance of fits without using the comparison star at all. We determine the GP regressor combination that best describes the systematics for all the methods in Sections \ref{model_selection} and \ref{app:model_selection}. 

For all applications of the conventional and new methods, we first find the Maximum a-Posteriori (MAP) solution by optimising the GP posterior using the Powell optimiser in the {\texttt{SciPy}} python package. We put wide uniform priors on the GP hyperparameters and sample them logarithmically. The logarithmic sampling of hyperparameters effectively puts a shrinkage prior on the hyperparameters which pushes them to smaller values if the corresponding GP regressor truly does not represent the correlated systematics in the data (\citealt{Gibson2012}, \citealt{Gibson2014}). Using the MAP solution as the starting point we marginalise the GP posterior over all the hyperparameters and transit model parameters through an MCMC using the package {\texttt{emcee}}, a pure-Python implementation of the affine-invariant Markov chain Monte Carlo (MCMC) ensemble sampler (\citealt{Goodman2010}, \citealt{ForemanMackey2013}). We use 50 walkers for 10000 steps and check for the convergence of chains by estimating the integrated auto-correlation times for each walker following the method described in (\citealt{Goodman2010}). We ensure that the total length of our chains are at least 50 times the auto-correlation times to make sure our samples are effectively independent and have converged. 

We discard the first 1000 samples as burn-in. We judge the final goodness of fit based on the consistency of best fit transit parameters with the literature values, and the model selection criteria described in Section \ref{model_selection} for each combination of GP regressors and the various forms of kernel combinations (described in Section \ref{noise_model} ). We also tested the robustness of our fits using a nested sampler {\texttt{dynesty}} (\citealt{Speagle2020}) and obtain posteriors consistent with those from {\texttt{emcee}}. We measure the best fit transit parameters as the median of the corresponding posteriors and $\pm 34$ percentile from the median as their 1 $\sigma$ uncertainties. 

\subsubsection{Selecting the best GP regressor combinations for white light curve fits}
\label{model_selection}

We select the best combination of GP regressors for both new and conventional methods of fitting the white light curves separately by comparing the Bayesian Information Criterion (BIC, \citealt{Schwarz1978}) and Bayesian evidence estimated using \texttt{dynesty}. For each GP regressor combination we calculate the BIC corresponding to the GP likelihood computed for the best fit parameters. BIC computed using the GP likelihood takes into account the covariance structure in the data through the covariance matrix (see Equation \ref{eq:likelihood}). We discuss the model selection threshold in more detail in Appendix \ref{app:model_selection}.

We have highlighted the best GP regressor combinations in the Tables \ref{tab:obs1_bicevi} to \ref{tab:obs5_bicevi} for the following applications of new and conventional methods that we compare further in Section \ref{sec:method_comparison}:

1) \texttt{New:WLC} - Target LC fit with Time and comparison LC, and additional regressors if that is favoured by higher log$_{e}$Z in some cases,

2) \texttt{New:WLC;No\_Comp} - Target LC fit without comparison LC as a regressor (Time and/or an additional regressor),

3) \texttt{Conv1:WLC} - Target/Combination LC fit using the best regressor combination.

For each of the three cases above, we perform the model selection by separately comparing the log$_{e}$Z for the set of GP regressor combinations applicable to each case. Also note that since the new and conventional methods are not fitting the same light curves exactly, we do not use log$_{e}$Z or BIC to perform comparison between the methods themselves but only to choose the best GP regressor combinations for each of them.  

% \textbf{When selecting the best GP regressor combination for these three cases, we look at high log$_{e}$Z, and also ensure that the length scale hyperparameter ($\eta_{p}$) for the each GP regressor in that combination is well constrained. We show the best fit parameters for each of these cases in Table \ref{tab:WLC_bestfitparams}. We further compare these cases in more detail in Section \ref{sec:method_comparison}.}

\subsubsection{Odd-even effect in GMOS light curves}
\label{odd_even}
The consecutive exposures in the GMOS light curves have been known to suffer from an odd-even effect due to the unequal travel times of the GMOS shutters (\citealt{Jorgensen2009}) with respect to the direction of motion. This has also been previously observed by \cite{Stevenson2014}. We observe the level of this effect for our HAT-P-26b observations to be as high as 700 ppm just for the target star light curves, and as high as 200 ppm for the Target/Comparison light curves, varying with the observation and the corresponding exposure time, and observed most significantly in the R150 observations 1, 2, and 3 (Figures \ref{fig:WLC_all} and \ref{fig:r150_WLC}). The comparison star light curve also suffers from the same odd-even effect at similar time scales as the target star as confirmed from the Lomb Scargle periodograms of both the light curves. Normalising the target light curve by the comparison light curve does not correct for this effect entirely as can be seen for observation 2 (which has the shortest exposure time among all observations) in Figure \ref{fig:WLC_all} where the odd even effect is still visible in the Target/Comparison light curve. This shows that the odd-even effect prevalent in GMOS observations doesn't affect the target and comparison light curves in the same manner from one exposure to the next and hence cannot be corrected for completely through a linear method like differential spectrophotometry. This is especially true for observations with shorter exposure times. Instead, the odd-even effect is superimposed on existing high frequency noise in the Target/Comparison light curves due to other variations between systematics affecting the target and comparison light curves individually. This further motivates the need for methods alternative to performing differential spectrophotometry to correct for the effect in the target star light curves directly, which is what our new method does. In particular when considering the residual RMS for observation 2, which has the shortest exposure time of all observations (and hence the largest amplitude of odd-even effect difference between the target and comparison stars), the new method \texttt{New:WLC} performs much better at modelling the odd-even effect in the target star light curves compared to the conventional method \texttt{Conv1:WLC}. In \texttt{New:WLC} the odd-even effect is accounted for by using the comparison star as one of the GP regressors.  

\cite{Stevenson2014} use different flux offsets on odd and even frames respectively to correct for this effect in another target HAT-P-7 in their survey, but the method they ultimately use for WASP-12 in \cite{Stevenson2014} is \texttt{Divide-White} which corrects for this effect automatically. Essentially \cite{Stevenson2014} use a linear mapping between the Target/Comparison light curves and an analytical functional form (different offsets for alternative exposures) or a non-analytical form (the \texttt{Divide-White} method) to correct for this effect. In this paper we correct for the effect in the target star light curve directly by letting the GP model do the non-linear mapping between the target star light curves and the odd-even effect information in the comparison star light curve. For spectroscopic light curves the white light curve derived common-mode trend when used as a GP regressor accounts for this effect for \texttt{New:$\lambda$LC} as described in Section \ref{sec:binned_LC_new_method}.

It should be noted that it is not just because of the presence of differential odd-even effect in the data that makes our new method more effective than the conventional method. We performed a simple transit injection and retrieval test by applying both methods to a pair of synthetic target and comparison star light curves both sharing the same correlated systematics but different levels of white noise. We find that when the comparison star light curve has higher level of white noise than the target star light curve, our new method performs much better than the conventional method in terms of both accuracy and precision of retrieving injected transit parameters.

We highlight that besides the odd-even effect, there are additional possible sources of instrumental and atmospheric systematics that can affect the comparison and target star fluxes differently, which would be potentially be present in data from other multi-object spectrographs as well. These effects can range from low-frequency trends e.g., due to changing CRPA through the night, or high and low frequency telluric absorption variations. The latter effect could be even more significant in near-infrared bands due to second-order colour-dependent extinction effect (e.g. \citealt{Blake2011}, \citealt{Young1991}).

After performing fits to the white transit light curves and gleaning information about the dominant time-dependent systematics affecting each of our observations, we fit the spectroscopic light curves to obtain the transmission spectrum, as described in more detail in the following section. 

%%%%% Table of the best fit parameters for the White Light Curve fits 
\begin{landscape}
\begin{table}
% [htb!]
\centering
\caption{Best fit transit parameters obtained from the fits to white transit light curves of 6 GMOS observations of HAT-P-26b presented in this work. Three sub-rows for each observation number (specified in the first column) show the best fit transit parameters and residual RMS from the applications of our new method (two sub-rows marked \texttt{New:WLC} and \texttt{New:WLC;No\_Comp}) and from the conventional method (third sub-row \texttt{Conv1:WLC}). The third column shows the light curve to which the method is applied to (`Targ' referring to Target and `Targ/Comp' referring to Target/Comparison light curve). The fourth column shows the the combination of regressors for the GP noise model in Section \ref{noise_model}, where `Time' is the time of each exposure in the light curves, `Comp' is the comparison star light curve, and `PSF' is the full width half maxima of the spectral trace PSF. The bottom section of the table shows the weighted average of transit parameters measured for R150 and B600 observations from the applications of both new and conventional methods, and the transit parameters (weighted average of \texttt{New:WLC} from B600 and R150) eventually used to derive the common-mode trend used to fit the spectroscopic light curves ($\lambda$LC) in Section \ref{sec:binned_LC}.}  
\label{tab:WLC_bestfitparams}
\begin{tabular}{ccccccccccc}
\hline
\hline 
\\
No.& Method & Light Curve & GP  & $R_{\rm P}/R_\star$ & $T_{0}$[BJD$_{\rm TDB}]$ & $a/R_\star$ & $i$ [$^\circ$] & u$_{1}$ &$\sigma_{w}$ & RMS   \\
   &        & type        & regressors &               &                      &           &                &         &             & [ppm]   \\ \\
\hline 
\\
1 (R150) & \texttt{New:WLC} & Targ & Time, Comp, PSF & 0.0725 $\pm$ 0.0023 & 2456371.74717 $\pm$ 0.00024 & 11.35 $\pm$ 0.48 & 87.18 $\pm$ 0.49 & 0.611 $\pm$ 0.025 & 0.000268 & 249 \\
        %  & Targ & Time, Comp & 0.0721 $\pm$ 0.0028 & 2456371.74731 $\pm$ 0.00026 & 11.26 $\pm$ 0.48 & 87.04 $\pm$ 0.48 & 0.607 $\pm$ 0.024 & 0.000291 & 270 \\
         & \texttt{New:WLC;No\_Comp} & Targ & Time & 0.0685 $\pm$ 0.053 & 2456371.74731 $\pm$ 0.006 & 11.23 $\pm$ 0.44 & 87.03 $\pm$ 0.43 & 0.605 $\pm$ 0.029 & 0.0015 & 1254 \\
         & \texttt{Conv1:WLC} & Targ/Comp & Time & 0.0730 $\pm$ 0.0034 & 2456371.74728 $\pm$ 0.00029 & 11.21 $\pm$ 0.49 & 86.99 $\pm$ 0.47 & 0.605 $\pm$ 0.023 & 0.000299 & 281 \\ \\

2 (R150) & \texttt{New:WLC} & Targ & Time, Comp & 0.0703 $\pm$ 0.0034 & 2456392.92016 $\pm$ 0.00024 & 12.18 $\pm$ 0.45 & 88.09 $\pm$ 0.58 & 0.602 $\pm$ 0.026 & 0.00038 & 366 \\
         & \texttt{New:WLC;No\_Comp} & Targ & Time, PSF & 0.0713 $\pm$ 0.0042 & 2456392.91988 $\pm$ 0.00027 & 11.98 $\pm$ 0.47 & 87.82 $\pm$ 0.54 & 0.602 $\pm$ 0.026 & 0.000347 & 332 \\
         & \texttt{Conv1:WLC} & Targ/Comp & Time & 0.0703 $\pm$ 0.0045 & 2456392.91946 $\pm$ 0.00038 & 11.28 $\pm$ 0.61 & 87.18 $\pm$ 0.66 & 0.603 $\pm$ 0.026 & 0.000569 & 557 \\ \\

3 (R150) & \texttt{New:WLC} & Targ & Time, Comp & 0.0694 $\pm$ 0.0032 & 2456786.72782 $\pm$ 0.00032 & 10.88 $\pm$ 0.56 & 86.59 $\pm$ 0.52 & 0.595 $\pm$ 0.024 & 0.000412 & 398 \\
         & \texttt{New:WLC;No\_Comp} & Targ & Airmass & 0.0625 $\pm$ 0.0031 & 2456786.72849 $\pm$ 0.00038 & 11.67 $\pm$ 0.75 & 87.67 $\pm$ 0.86 & 0.603 $\pm$ 0.026 & 0.001567 & 1536 \\
         & \texttt{Conv1:WLC} & Targ/Comp & Time & 0.0696 $\pm$ 0.0029 & 2456786.72790 $\pm$ 0.0003 & 11.07 $\pm$ 0.72 & 86.81 $\pm$ 0.68 & 0.611 $\pm$ 0.029 & 0.000431 & 420 \\ \\

6 (R150) & \texttt{New:WLC} & Targ & Time, Comp & 0.0683 $\pm$ 0.0048 & 2456837.54218 $\pm$ 0.00038 & 12.12 $\pm$ 0.58 & 87.85 $\pm$ 0.70 & 0.599 $\pm$ 0.025 & 0.000580 & 560 \\
         & \texttt{New:WLC;No\_Comp} & Targ & Time & 0.0706 $\pm$ 0.0055 & 2456837.54214 $\pm$ 0.00046 & 12.05 $\pm$ 0.59 & 87.77 $\pm$ 0.67 & 0.598 $\pm$ 0.026 & 0.000589 & 562 \\
         & \texttt{Conv1:WLC} & Targ/Comp & Time & 0.0671 $\pm$ 0.0044 & 2456837.54211 $\pm$ 0.00043 & 11.92 $\pm$ 0.62 & 87.54 $\pm$ 0.69 & 0.602 $\pm$ 0.026 & 0.000609 & 590 \\ \\
         
4 (B600) & \texttt{New:WLC} & Targ & Time, Comp, PSF & 0.0654 $\pm$ 0.0045 & 2457460.01335 $\pm$ 0.00036 & 11.82 $\pm$ 0.55 & 87.71 $\pm$ 0.58 & 0.73 $\pm$ 0.026 & 0.000125 & 101 \\
        %  & Targ & Time, Comp & 0.0668 $\pm$ 0.0049 & 2457460.01309 $\pm$ 0.00035 & 12.08 $\pm$ 0.55 & 87.96 $\pm$ 0.65 & 0.731 $\pm$ 0.026 & 0.000192 & 109 \\
         & \texttt{New:WLC;No\_Comp} & Targ & Time, PSF & 0.0673 $\pm$ 0.0067 & 2457460.01315 $\pm$ 0.00062 & 11.50 $\pm$ 0.72 & 87.44 $\pm$ 0.74 & 0.733 $\pm$ 0.027 & 0.000193 & 166 \\
         & \texttt{Conv1:WLC} & Targ/Comp & Time, Airmass & 0.0735 $\pm$ 0.0057 & 2457460.01402 $\pm$ 0.00046 & 11.29 $\pm$ 0.54 & 87.26 $\pm$ 0.54 & 0.732 $\pm$ 0.025 & 0.000194 & 165 \\ \\

5 (B600) & \texttt{New:WLC} & Targ & Time, Comp & 0.0685 $\pm$ 0.0053 & 2457493.89006 $\pm$ 0.00031 & 11.97 $\pm$ 0.55 & 87.81 $\pm$ 0.64 & 0.733 $\pm$ 0.026 & 0.000164 & 133 \\
         & \texttt{New:WLC;No\_Comp} & Targ & Time, Airmass, PSF & 0.0726 $\pm$ 0.0021 & 2457493.89000 $\pm$ 0.00024 & 12.50 $\pm$ 0.44 & 88.46 $\pm$ 0.69 & 0.721 $\pm$ 0.024 & 0.000283 & 264 \\
         & \texttt{Conv1:WLC} & Targ/Comp & Time, Airmass & 0.0764 $\pm$ 0.0089 & 2457493.88987 $\pm$ 0.00048 & 11.64 $\pm$ 0.68 & 87.57 $\pm$ 0.73 & 0.730 $\pm$ 0.029 & 0.000218 & 181 \\ \\

\hline
\\

R150 & \texttt{New:WLC} & Targ & & 0.0703 $\pm$ 0.0014 &  & 11.67 $\pm$ 0.25 & 87.33 $\pm$ 0.28 & 0.602 $\pm$ 0.012 &  &  \\
R150 & \texttt{New:WLC;No\_Comp} & Targ & & 0.067 $\pm$ 0.0022 &  & 11.68 $\pm$ 0.25 & 87.45 $\pm$ 0.28 & 0.602 $\pm$ 0.013 &  &  \\
R150 & \texttt{Conv1:WLC} & Targ/Comp & & 0.0702 $\pm$ 0.0018 &  & 11.36 $\pm$ 0.29 & 87.09 $\pm$ 0.3 & 0.605 $\pm$ 0.013 &  &  \\ \\

B600 & \texttt{New:WLC} & Targ & & 0.067 $\pm$ 0.0034 &  & 11.89 $\pm$ 0.39 & 87.75 $\pm$ 0.42 & 0.732 $\pm$ 0.018 &  &  \\
B600 & \texttt{New:WLC;No\_Comp} & Targ & & 0.0721 $\pm$ 0.002 &  & 12.22 $\pm$ 0.38 & 87.98 $\pm$ 0.5 & 0.726 $\pm$ 0.017 &  &  \\
B600 & \texttt{Conv1:WLC} & Targ/Comp & & 0.0743 $\pm$ 0.0048 &  & 11.42 $\pm$ 0.42 & 87.37 $\pm$ 0.43 & 0.731 $\pm$ 0.019 &  &  \\
\\
For $\lambda$LC fits &  &  & & 0.0701 $\pm$ 0.0013 &  & 11.73 $\pm$ 0.21 & 87.45 $\pm$ 0.23 &  &  &  \\

\hline
\hline

\end{tabular}
% \end{rotatetable*}

\end{table}
\end{landscape}

%%%%%%%%%%%%%%%%%%
% subsection
% SPTLC Analysis  
%%%%%%%%%%%%%%%%%%

\subsection{Analysis of Spectroscopic Light Curves}
\label{sec:binned_LC}

\subsubsection{Construction of spectroscopic light curves}
We constructed the spectroscopic transit light curves ($\rm \lambda$LC) for both target and comparison stars by summing the flux in $\sim$ 20 nm wide bins within the same wavelength range as the respective white light curves. We normalise each exposure in the individual target and comparison $\rm \lambda$LC by the corresponding exposure times. Similar to our white light curve analyses, we fit the $\rm \lambda$LC for each observation using the conventional method \texttt{Conv1;$\lambda$LC} and the new method \texttt{New:$\lambda$LC} as described in Section \ref{noise_model}. 

\subsubsection{Fitting the spectroscopic light curves using conventional method}
\label{sec:binned_LC_old_method}

We first describe the application \texttt{Conv1;$\lambda$LC} of the conventional method of fitting $\rm \lambda$LCs. We divide the target $\rm \lambda$LCs by the corresponding comparison star $\rm \lambda$LCs. GMOS observations, like many other ground-based MOS observations, have been conventionally corrected for wavelength-independent systematics through common-mode correction (e.g. \cite{Stevenson2014}, \citetalias{Huitson2017}, \cite{Todorov2019}, \cite{Wilson2021}), which leverages the information about the time dependent systematics contained in the white light curve to correct the individual wavelength bins. However, while performing common-mode correction using the white light curve provides an effective way to remove dominant time-dependent systematics, it also implies that we effectively lose information on the absolute value of transit depths and obtain relative transit depths, which is nevertheless useful to search for dominant features in the transmission spectrum. 

In \texttt{Conv1;$\lambda$LC}, we follow the conventional common-mode correction approach and derive the common-mode trend as the residuals obtained by subtracting the transit model computed using the weighted averaged transit parameters for the white light curves (last row of Table \ref{tab:WLC_bestfitparams}) from the white light curves for all observations. Except the limb darkening coefficient we use the same transit parameters for both B600 and R150 observations to construct the transit model. We perform the common-mode correction by subtracting the white light curve derived common-mode trend from the corresponding Target/Comparison $\rm \lambda$LC. Note that using the weighted averaged transit parameters to derive the common-mode trend across all observations is valid here as HAT-P-26 is known to be inactive and any potential contamination from stellar activity for the individual epochs that could affect the transit depth is below the precision of our measurements (discussed in more detail in Section \ref{sec:interpretation}.

We then fit the common-mode corrected $\rm \lambda$LCs with the model described in \ref{noise_model}, using only time as a GP regressor. This is mainly to account for wavelength-dependent trends not removed by the common-mode correction and arising likely due to wavelength-dependent differential atmospheric extinction between the target and comparison stars with changing airmass through the night (discussed in more detail in Section \ref{sec:method_comparison}). Since our main goal is to measure the wavelength-dependent transit depths, we fix the orbital inclination ($i$), orbital separation ($a/R_\star$), and mid transit time ($T_0$) to the best fit values for the corresponding white light curve in Section \ref{sec:WLC} (see Table \ref{tab:WLC_bestfitparams}), and orbital period and eccentricity to literature values. We use a linear limb darkening law for each wavelength bin and fix the limb darkening coefficients to pre-calculated values by \texttt{PyLDTk} (approximating a top hat transmission function for each wavelength bin). 

We find that doing common-mode correction prior to fitting the Target/Comparison $\rm \lambda$LCs improved the precision of measured transit depths in R150 observations by $\sim15$\% on average per wavelength bin compared to when not performing common-mode correction. The Target/Comparison R150 $\rm \lambda$LCs along with their best fit models, detrended light curves, and the residuals are shown in the top three panels of Figures \ref{fig:sptlcs_1}, \ref{fig:sptlcs_2}, \ref{fig:sptlcs_3}, and \ref{fig:sptlcs_6}. 

For the B600 observations, the target and comparison star light curves suffer from significantly different trends through the night as already discussed in Section \ref{sec:WLC}. Hence, doing Target/Comparison normalisation contaminates the transit signal. This can be noticed by visually inspecting the B600 white light curves in Figure \ref{fig:WLC_all} and also in the B600 $\rm \lambda$LCs. Nevertheless, same as done for R150 $\rm \lambda$LCs, we apply \texttt{Conv1;$\lambda$LC} to B600 $\rm \lambda$LCs by doing common-mode correction followed by fitting common-mode corrected Target/Comparison light curves. The resultant fits, detrended light curves, and the residuals are shown in Figures \ref{fig:sptlcs_4} and \ref{fig:sptlcs_5}. 

%%%%%% 
\subsubsection{Fitting the spectroscopic light curves using the new method}
\label{sec:binned_LC_new_method}

We now describe the application \texttt{New:$\lambda$LC} of our new method to fitting the $\rm \lambda$LCs. One of the motivations behind our new method is that as we observe the planetary transit through a range of airmass during a night, the differential atmospheric extinction between the target and the comparison star across the optical wavelength range due to the difference in brightness and/or spectral type between the stars implies that simple normalisation of the target $\rm \lambda$LCs by the comparison $\rm \lambda$LCs introduces wavelength dependent systematics in the light curves. This is also evident as residual trends in the $\rm \lambda$LCs after conventional common-mode correction as done Section \ref{sec:binned_LC_old_method}. When inspecting the individual target and comparison star spectra, we find that for our GMOS observations the bluest end of the stellar spectrum suffers from $\sim$ 5 to 10 \% more extinction at high airmasses compared to the reddest end. For the B600 observations specifically, the difference between the atmospheric extinction between the target (fainter than the comparison star) and the comparison star is 5 \% more at the bluest end than the reddest end of the spectrum.  

The conventional method to mitigate this residual wavelength-dependent noise remaining after common-mode correction as mentioned in Section \ref{sec:binned_LC_old_method} is to fit the common-mode corrected Target/Comparison $\rm \lambda$LCs using a linear or quadratic function of airmass or time as the baseline, or a GP model with time as a regressor. In this conventional method, however, there is no straightforward way to ascertain additional systematics propagated to the light curves during the division by comparison $\rm \lambda$LCs and then common-mode correction. The linear approach of conventional method is also sub-optimal due to non-linear wavelength dependent difference between target and comparison $\rm \lambda$LCs, lack of wavelength dependent information present in the common-mode trend, and other potential non-linear differences between the target $\rm \lambda$LCs and common-mode trend. 

With our new method to fit the $\rm \lambda$LCs, we propose to neither perform normalisation by the comparison star $\rm \lambda$LCs nor perform common-mode correction to the $\rm \lambda$LCs. We instead use the information of the time dependent systematics contained in the white light curves as one of the regressors in the GP noise model described in Section \ref{noise_model} for fitting the corresponding $\rm \lambda$LCs. This is possible through two different combinations of GP regressors : 

1) Time and GP noise model of the white light curve (from Section \ref{sec:WLC}), 

2) Time and normalised residuals between the white light curve and its best fit transit model (these residuals are same as the conventional common-mode trend). 

The first combination effectively still uses information from the comparison star (which was used to fit the white target light curve and obtain the GP noise model in Section \ref{sec:WLC}). The second combination however for both R150 and B600 observations doesn't rely on the comparison star directly and simply leverages the information contained in the common-mode trend to inform the GP systematics model for each $\rm \lambda$LC. This combination is in part analogous to the combination employed by \texttt{Divide-White} method (\citealt{Stevenson2014}) which uses the white target light curve residuals as a common-mode correction factor in combination with non-analytic models of wavelength dependent systematics derived from the comparison star $\rm \lambda$LCs. In contrast to the conventional \texttt{Divide-White} method, we do not use any information from the comparison star $\rm \lambda$LCs and simply subtract the transit model from the white target light curve and use the residuals or the common-mode trend hence obtained as a regressor in the GP model for the individual $\rm \lambda$LCs. We eventually use the second combination (time and common-mode trend) to fit the target $\rm \lambda$LCs.

It should be noted that the white light curve transit parameters we obtain from not using comparison stars at all (\texttt{New:WLC;No\_Comp} sub-row in Table \ref{tab:WLC_bestfitparams}) are consistent with those obtained from using comparison star as one of the GP regressor (\texttt{New:WLC} sub-row in Table \ref{tab:WLC_bestfitparams}, see detailed comparison in Section \ref{sec:method_comparison_WLC}). Hence, the derived common-mode trend is consistent between whether we use the comparison stars or not to fit the white light curves, and hence the common-mode trend is not a function of the comparison star light curve in our new method.

Similar to the conventional method described in Section \ref{sec:binned_LC_old_method}, we use the same weighted averaged transit parameters (except the limb darkening coefficient) for both the B600 and R150 observations and the respective transit models used to obtain the common-mode trend from the white light curves. The common-mode trend is then used as a GP regressor to fit the target $\rm \lambda$LCs. Similar to the conventional method, when fitting $\rm \lambda$LCs we keep all the transit model parameters except the transit depth fixed to the best weighted average values derived from the white light curve, and also fix the linear limb darkening coefficients to the pre-computed values from \texttt{PyLDTk} for each spectral bin.

Using the common-mode trend as a GP regressor to fit target $\rm \lambda$LCs as an alternative to subtracting it from $\rm \lambda$LCs is a novel approach and we test its robustness using a transit injection and recovery test described in detail in Appendix \ref{cmode_gp_test}. We find from this test that using the common-mode trend as a GP regressor yields transmission spectra consistent with and on average 25 \% better precision than that obtained from the conventional common-mode correction.

Through our transit injection test in Appendix \ref{cmode_gp_test} (see right panel of Figure \ref{fig:cmode_gp_test}) we also demonstrate the choice of using time as a GP regressor in addition to the common-mode trend to fit the target $\rm \lambda$LCs. The common-mode trend by itself models the high frequency systematics in the target $\rm \lambda$LCs which also includes the odd-even effect described in Section \ref{odd_even}. Time as an additional GP regressor models the wavelength dependent low frequency trend across the $\rm \lambda$LCs. It is possible to use additional GP regressors to fit $\rm \lambda$LCs, but since we independently fit each $\rm \lambda$LCs in this paper, it is not possible to perform model selection for all the $\rm \lambda$LCs together as done for the white light curves in Section \ref{model_selection}. Hence, we stick to the simplest choice of using only time as the additional GP regressor to model the wavelength-dependent trend. It would be possible for a future study into joint modelling of systematics for all $\rm \lambda$LCs in both time and wavelength dimension to comprehensively explore the use of additional regressors.

\begin{figure*}

  \centering
  \includegraphics[scale=0.4]{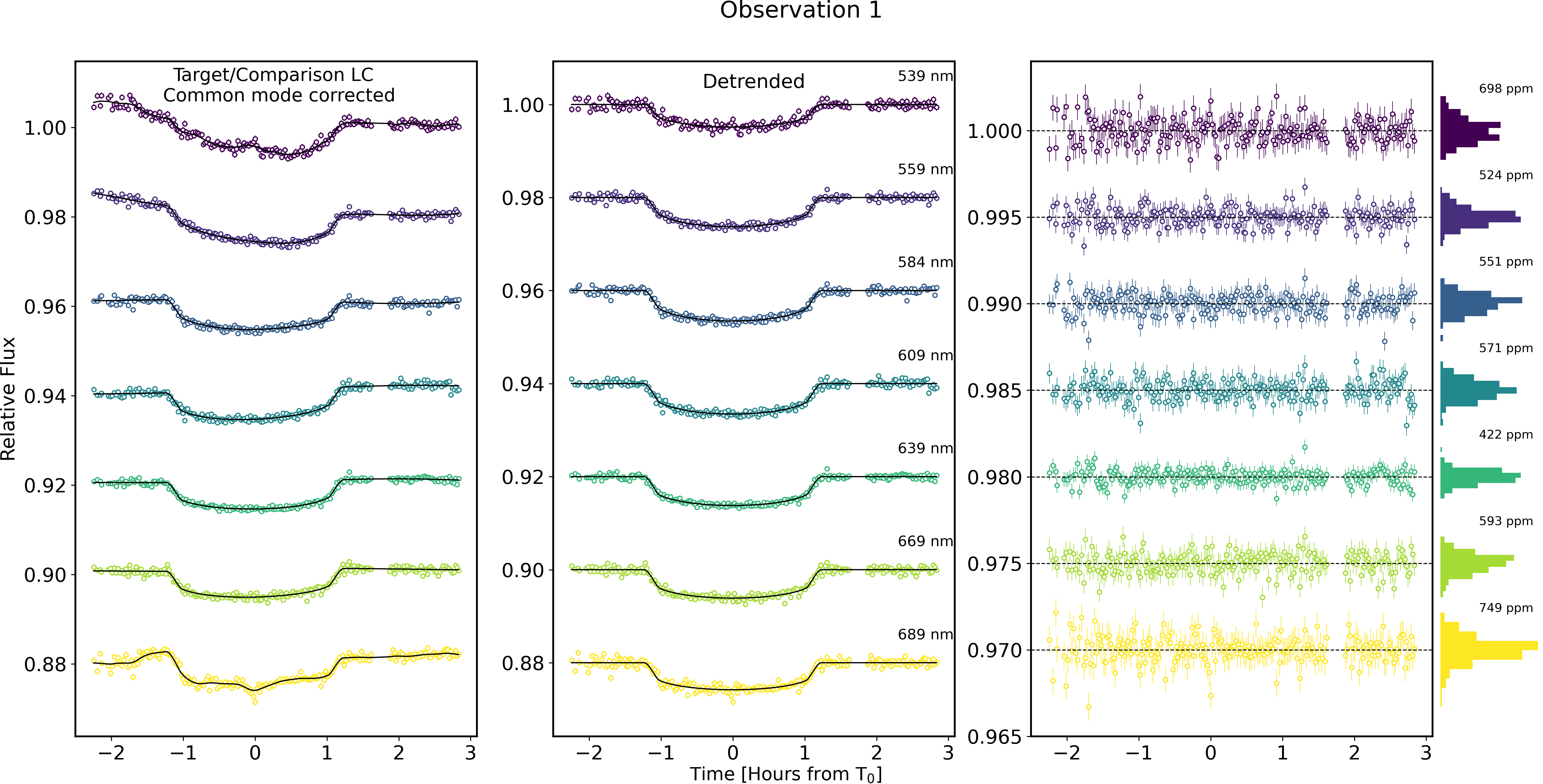}
  \includegraphics[scale=0.4]{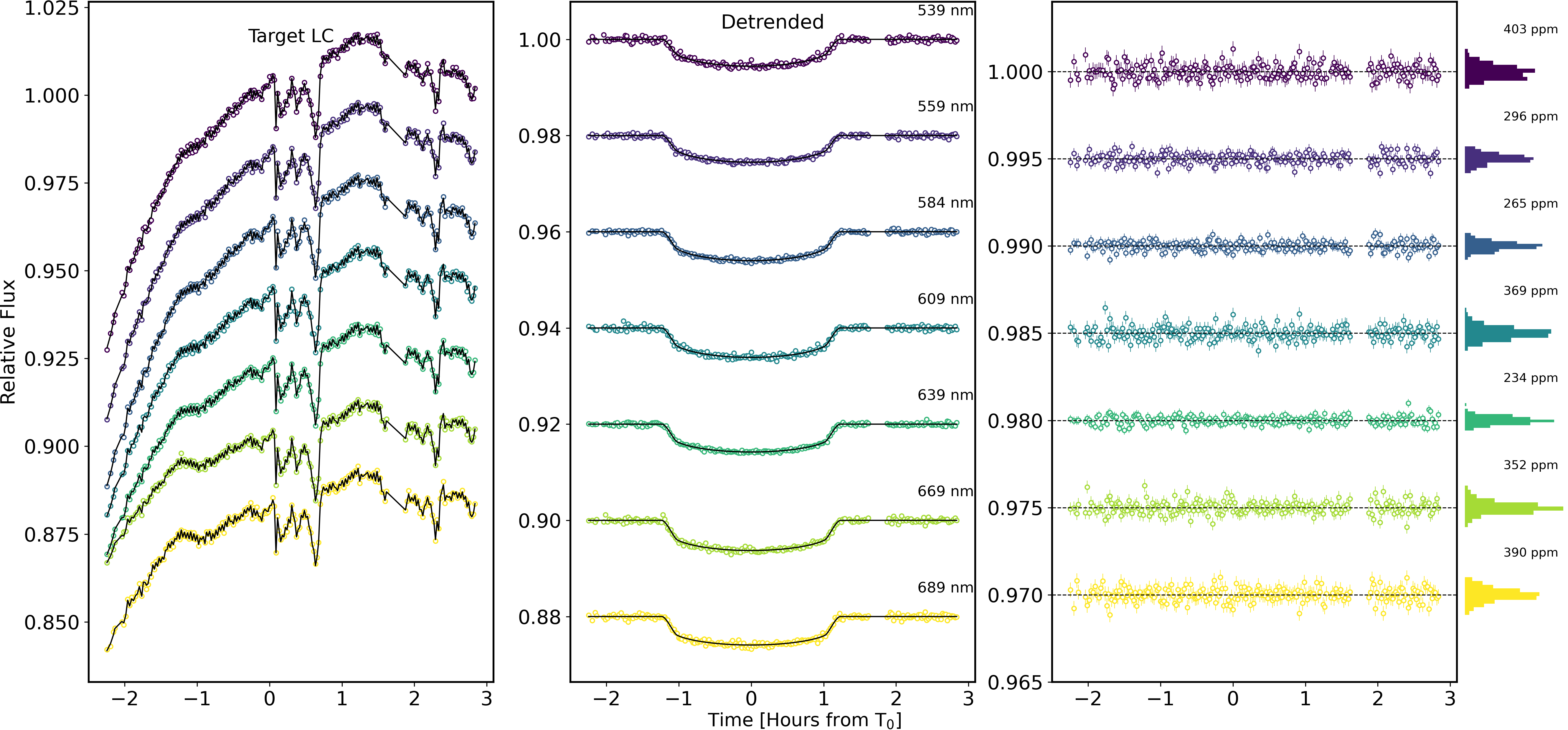}

  \caption{Spectroscopic light curves for observation 1 (R150) fit using the conventional method (\texttt{Conv1:$\lambda$LC}, top three panels) and the new method introduced in this paper (\texttt{New:$\lambda$LC}, bottom three panels). The leftmost panel for each method shows the best fit to the light curves for each wavelength bin, the middle panel shows the detrended light curves, and the rightmost panel shows the corresponding residuals, their histograms, and the RMS of their scatter. The target $\lambda$LCs show a wavelength dependent low frequency trend due to changing airmass through the night.       
  }
  \label{fig:sptlcs_1}
\end{figure*}

\begin{figure*}

  \centering
  \includegraphics[scale=0.4]{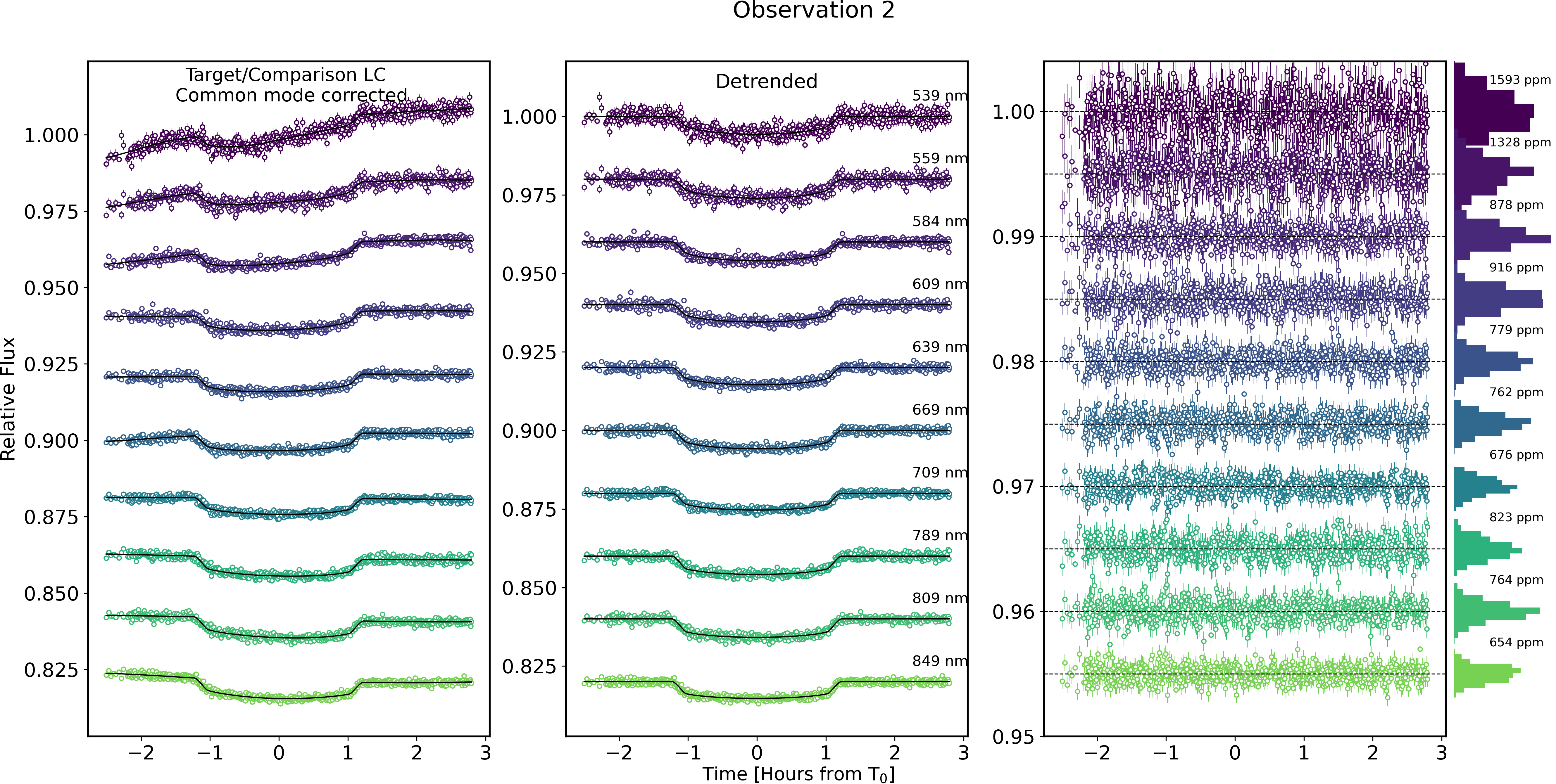}
 \includegraphics[scale=0.4]{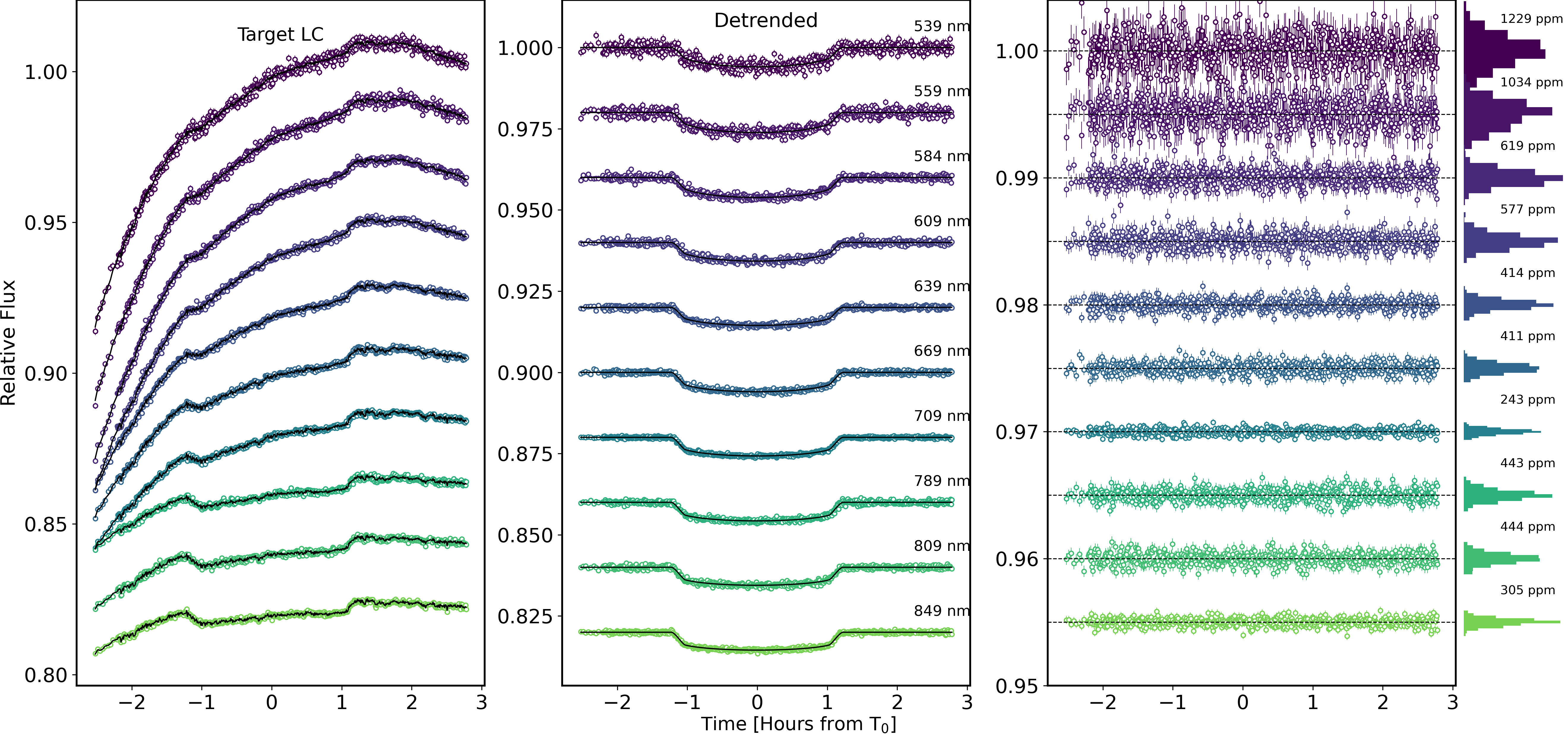}
  \caption{Same as Figure \ref{fig:sptlcs_1} for observation 2 (R150). 
  }
  \label{fig:sptlcs_2}
\end{figure*}

\begin{figure*}

  \centering
  \includegraphics[scale=0.4]{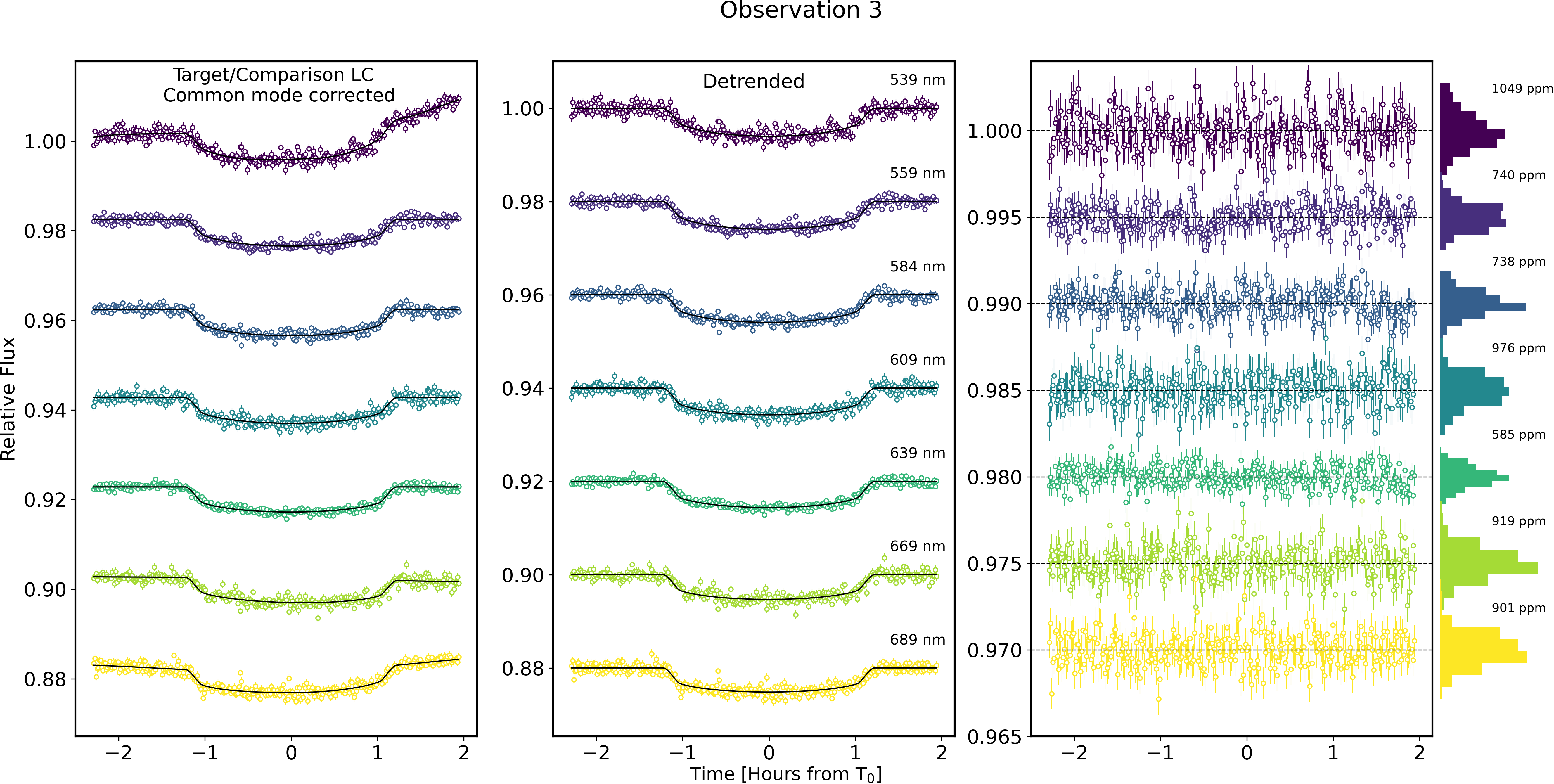}
    \includegraphics[scale=0.4]{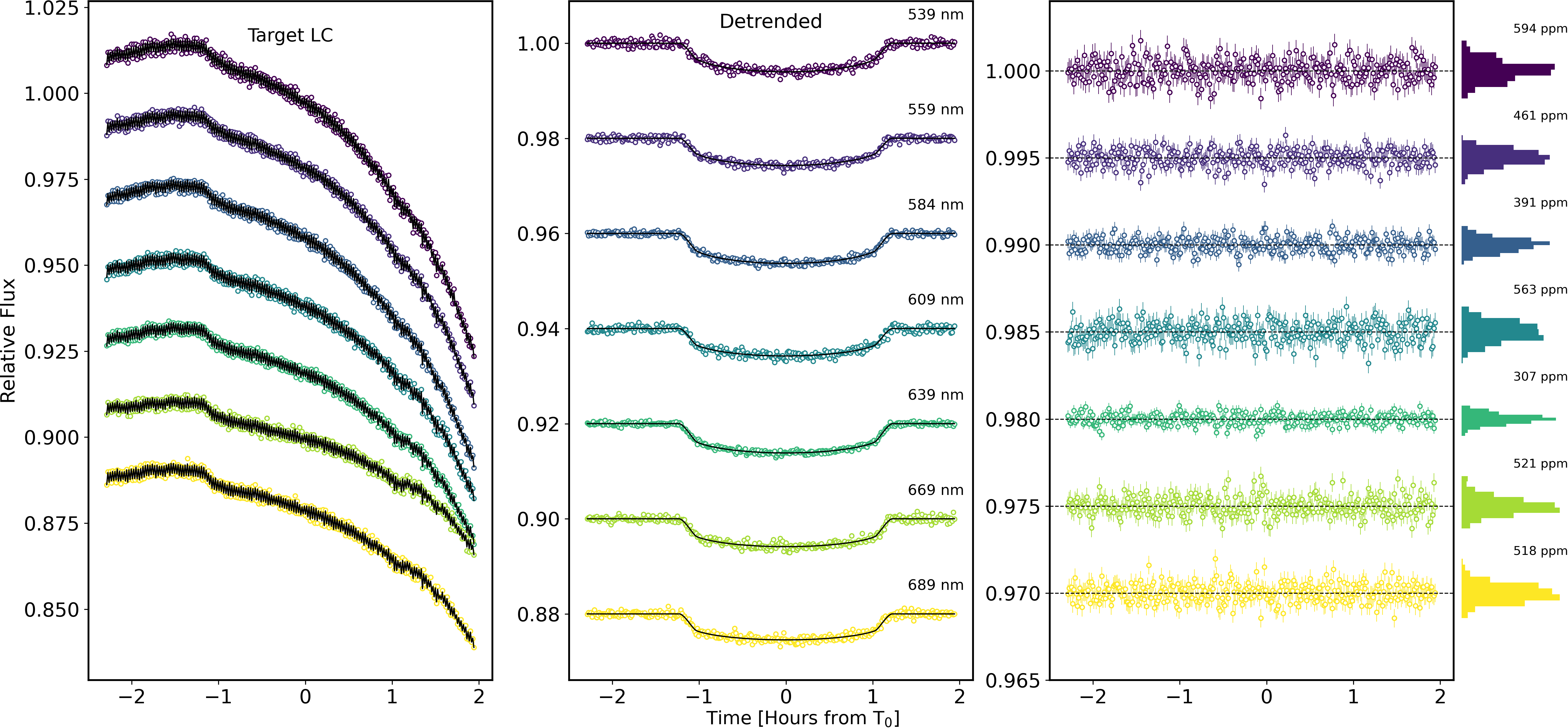}
  \caption{Same as Figure \ref{fig:sptlcs_1} for observation 3 (R150).
  }
  \label{fig:sptlcs_3}
\end{figure*}

\begin{figure*}

  \centering
  \includegraphics[scale=0.4]{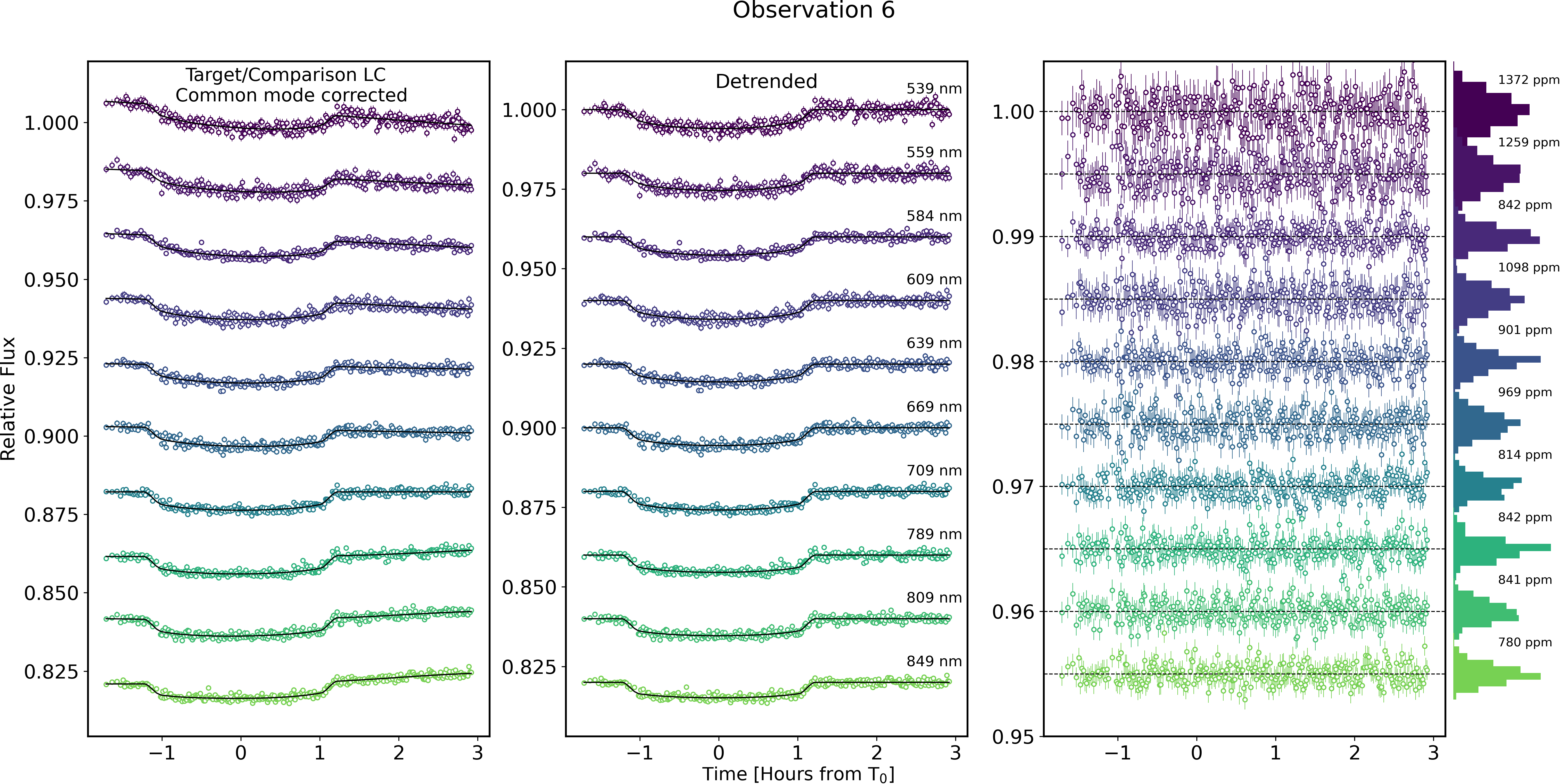}
    \includegraphics[scale=0.4]{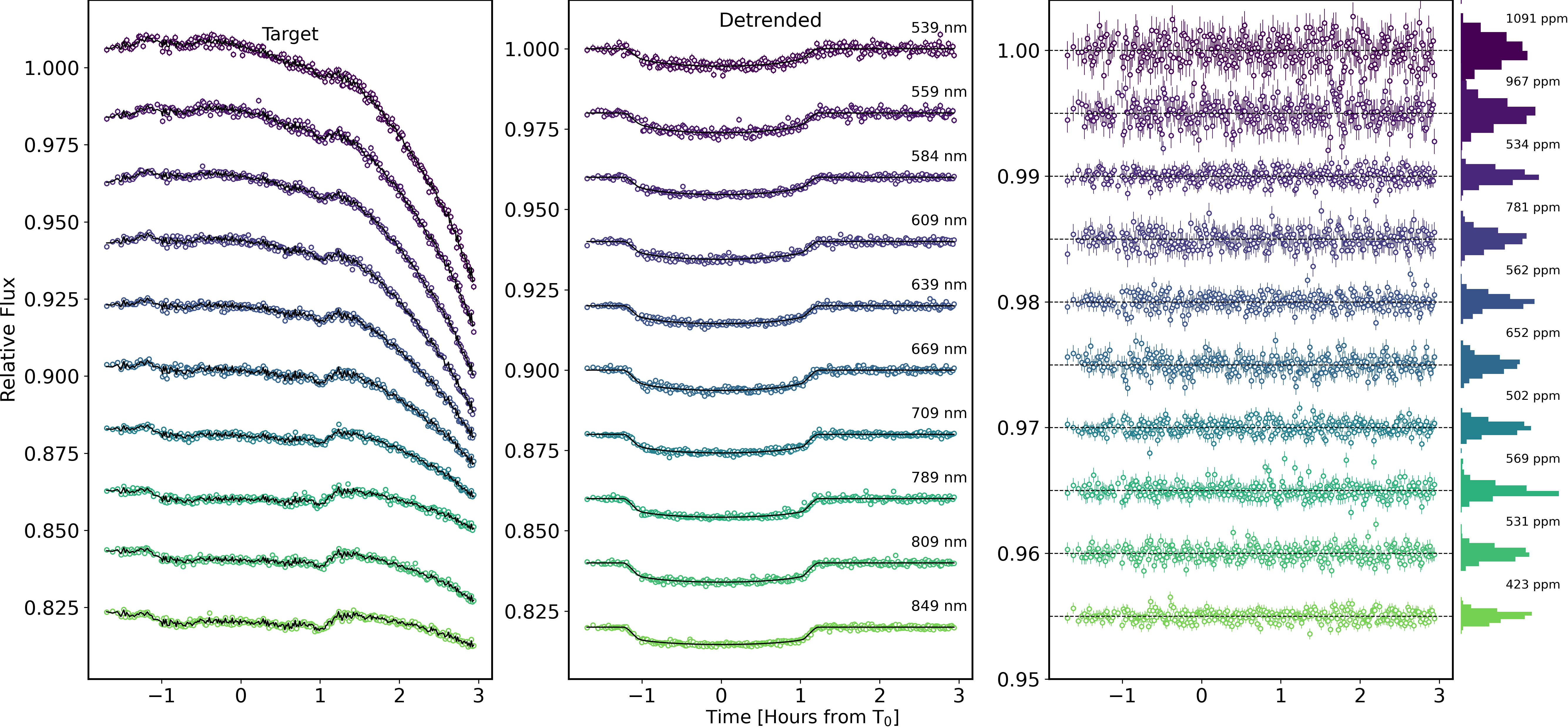}
  \caption{Same as Figure \ref{fig:sptlcs_1} for observation 6 (R150).
  }
  \label{fig:sptlcs_6}
\end{figure*}

\begin{figure*}

  \centering
  \includegraphics[scale=0.4]{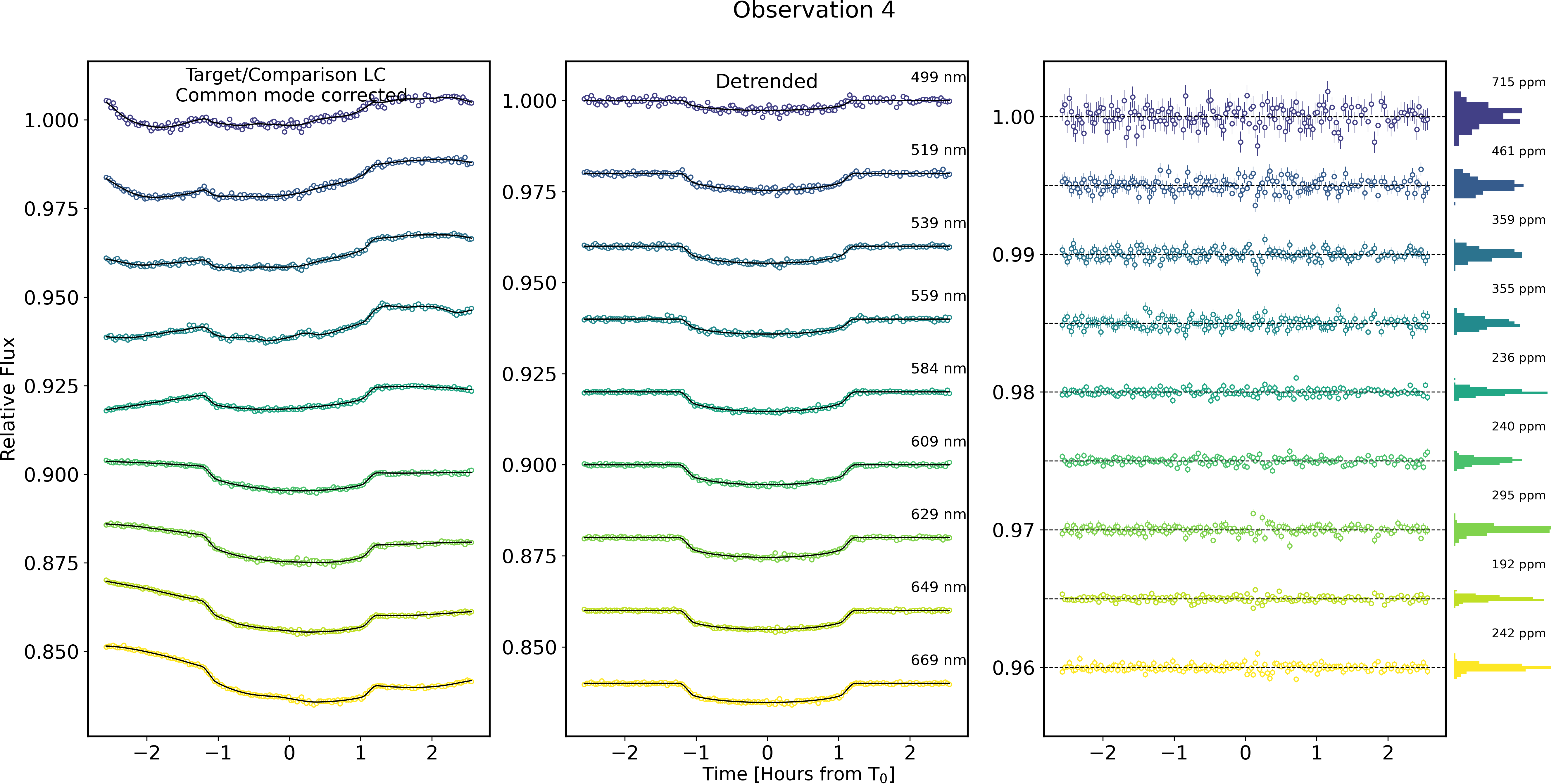}
  \includegraphics[scale=0.4]{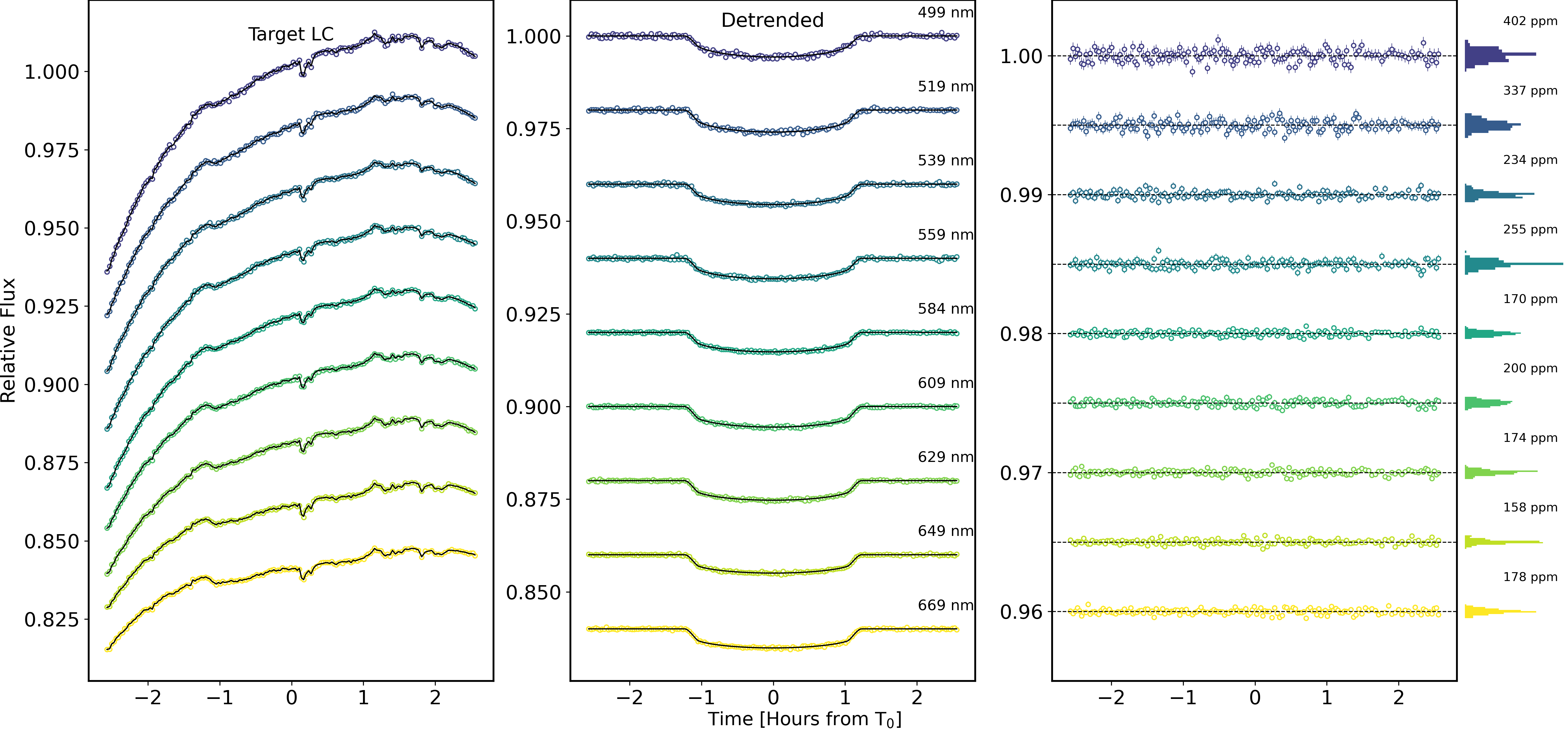}
  \caption{Spectroscopic light curves for observation 4 (B600) fit using the conventional method (\texttt{conv1:$\lambda$LC}, top three panels) and the new method introduced in this paper (\texttt{New:$\lambda$LC}, bottom three panels). The leftmost panel for each method shows the best fit to the light curves for each wavelength bin, the middle panel shows the detrended light curves, and the rightmost panel shows the corresponding residuals, their histograms, and the RMS of their scatter. The target $\lambda$LCs show a wavelength dependent low frequency trend due to changing airmass through the night.    
  }
  \label{fig:sptlcs_4}
\end{figure*}

\begin{figure*}

  \centering
  \includegraphics[scale=0.4]{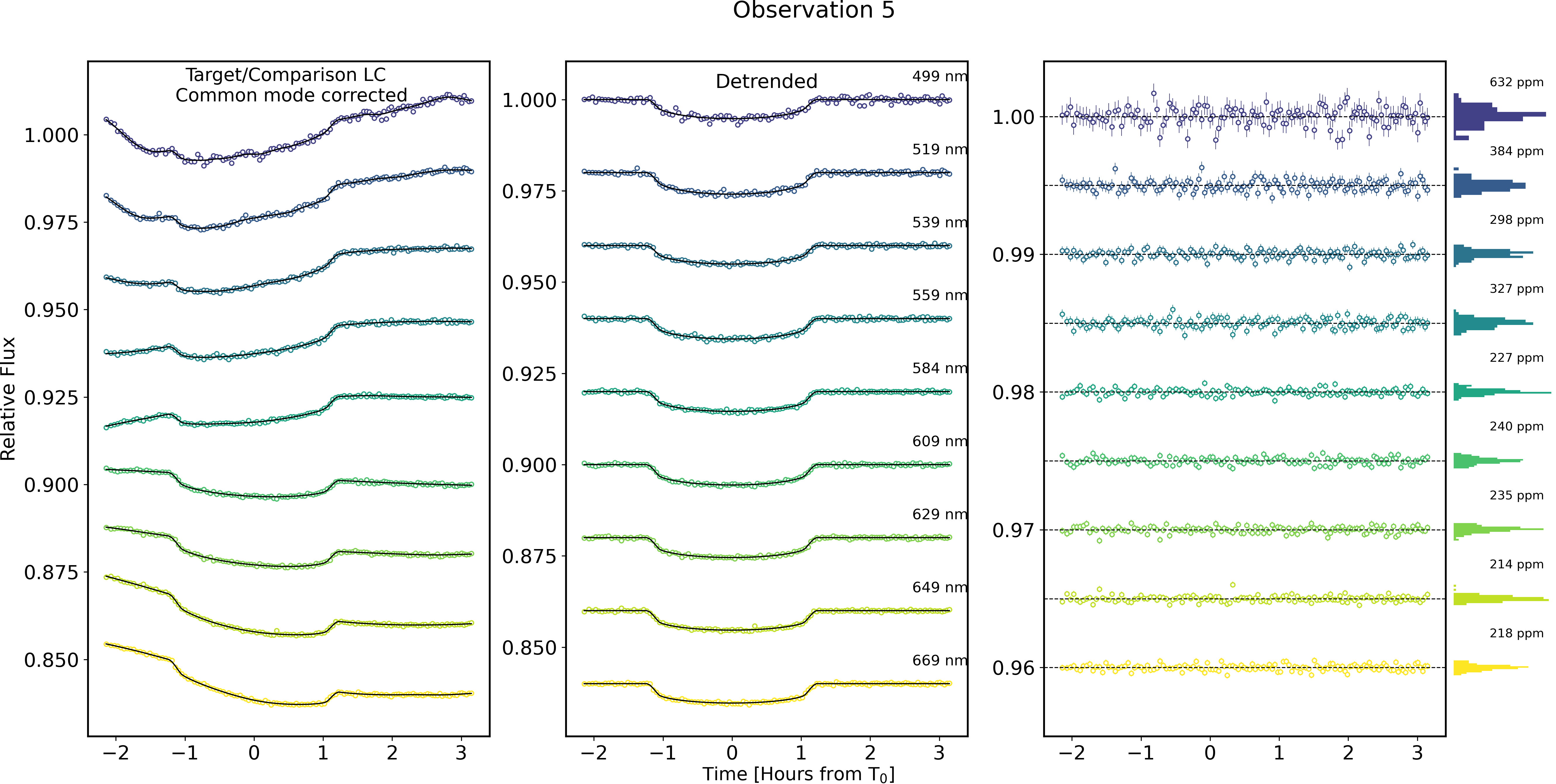}
  \includegraphics[scale=0.4]{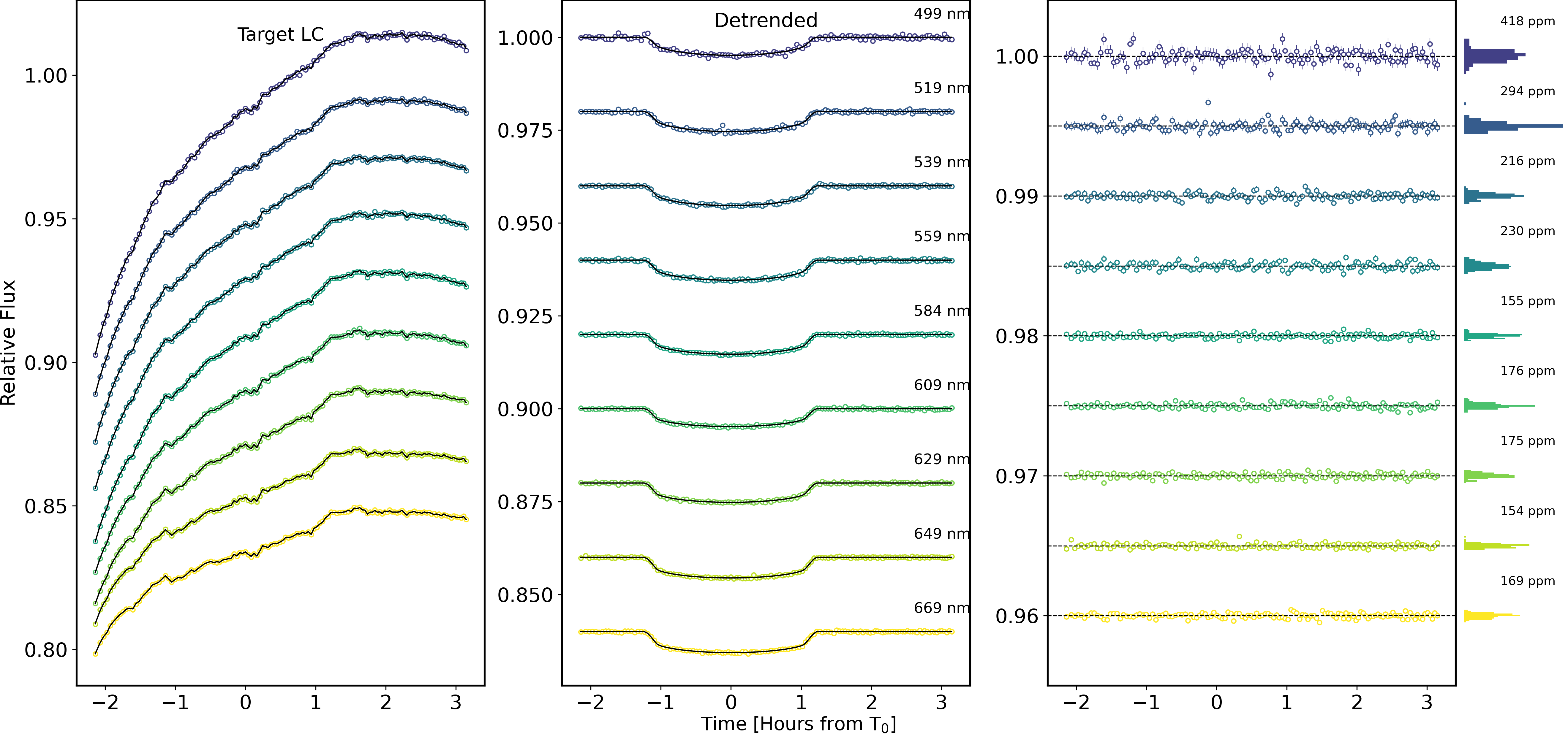}
  \caption{Same as Figure \ref{fig:sptlcs_4} for observation 5 (B600).
  }
  \label{fig:sptlcs_5}
\end{figure*}

The Target $\rm \lambda$LCs for both B600 and R150 observations along with their best fit models from the new method, detrended light curves, and the residuals are shown in the bottom three panels of Figures \ref{fig:sptlcs_1}, \ref{fig:sptlcs_2}, \ref{fig:sptlcs_3}, \ref{fig:sptlcs_6}, \ref{fig:sptlcs_4}, and \ref{fig:sptlcs_5}. The resulting transmission spectra are tabulated in Tables \ref{tab:r150_ts_targ} to \ref{tab:b600_ts_targ}, and shown in Figure \ref{fig:r150_ts} and \ref{fig:b600_ts}. 

We compare the transmission spectrum of HAT-P-26b constructed from the best fit wavelength-dependent transit depths for each observation obtained from the conventional and the new method introduced in this paper, and interpret and discuss them in the context of previous transmission spectroscopy measurements of HAT-P-26b.

%% Wavelength dependent transit depths 
%% R150 Target LC 

\begin{table*}
\begin{center}

\caption{\texttt{New:$\lambda$LC}, R150 : Wavelength dependent transit depths (in ppm) for the individual GMOS-R150 observations (marked by the columns) and combined from all observations obtained using the new method described in Section \ref{sec:binned_LC}. }  
\label{tab:r150_ts_targ}
\begin{tabular}{cccccc}
\hline
Wavelength [\AA] & & & Transit Depth [ppm] \\ 
 & 1 & 2 & 3 & 6 & Combined \\ 
\hline

5301 - 5501 & 4530 $\pm$ 473 & 4813 $\pm$ 850 & 4897 $\pm$ 526 & 4404 $\pm$ 995 & 4681 $\pm$ 301 \\
5501 - 5701 & 4530 $\pm$ 406 & 5038 $\pm$ 809 & 4684 $\pm$ 385 & 5034 $\pm$ 922 & 4682 $\pm$ 252 \\
5701 - 5999 & 4958 $\pm$ 339 & 5073 $\pm$ 578 & 5199 $\pm$ 397 & 4440 $\pm$ 593 & 4971 $\pm$ 216 \\
5999 - 6199 & 5099 $\pm$ 277 & 4773 $\pm$ 464 & 4797 $\pm$ 99 & 4608 $\pm$ 576 & 4830 $\pm$ 90 \\
6199 - 6600 & 4867 $\pm$ 183 & 4655 $\pm$ 332 & 5177 $\pm$ 176 & 4663 $\pm$ 414 & 4982 $\pm$ 106 \\
6600 - 6800 & 5308 $\pm$ 364 & 5047 $\pm$ 191 & 4962 $\pm$ 503 & 5356 $\pm$ 316 & 5148 $\pm$ 138 \\
6800 - 7000 & 5037 $\pm$ 609 & - & 4642 $\pm$ 369 & - & 4752 $\pm$ 310 \\
6799 - 7399 & - & 4896 $\pm$ 91 & - & 4937 $\pm$ 183 & 4903 $\pm$ 87 \\
7799 - 7999 & - & 4968 $\pm$ 211 & - & 5018 $\pm$ 411 & 4978 $\pm$ 184 \\
7999 - 8201 & - & 4831 $\pm$ 273 & - & 5205 $\pm$ 364 & 4954 $\pm$ 211 \\
8201 - 8801 & - & 4836 $\pm$ 240 & - & 4740 $\pm$ 391 & 4809 $\pm$ 204 \\

\hline

\end{tabular}
\end{center}
\end{table*}

%% R150 Target/Comparison LC 

\begin{table*}
\begin{center}

\caption{\texttt{Conv1:$\lambda$LC}, R150 : Wavelength dependent transit depths (in ppm) for the individual GMOS-R150 observations (marked by the columns) and combined from all observations obtained using the conventional method described in Section \ref{sec:binned_LC}. }  
\label{tab:r150_ts_targ_by_comp}
\begin{tabular}{cccccc}

\hline
Wavelength [\AA] & & & Transit Depth [ppm] \\ 
 & 1 & 2 & 3 & 6 & Combined \\ 
\hline

5301 - 5501 & 3947 $\pm$ 1112 & 4637 $\pm$ 673 & 4886 $\pm$ 766 & 4853 $\pm$ 331 & 4762 $\pm$ 272 \\
5501 - 5701 & 5145 $\pm$ 413 & 4951 $\pm$ 474 & 4850 $\pm$ 81 & 4420 $\pm$ 287 & 4837 $\pm$ 77 \\
5701 - 5999 & 5396 $\pm$ 311 & 4921 $\pm$ 290 & 4875 $\pm$ 109 & 4729 $\pm$ 228 & 4892 $\pm$ 100 \\
5999 - 6199 & 5457 $\pm$ 220 & 4546 $\pm$ 202 & 4776 $\pm$ 110 & 4907 $\pm$ 334 & 4867 $\pm$ 82 \\
6199 - 6600 & 5222 $\pm$ 183 & 4601 $\pm$ 159 & 4716 $\pm$ 65 & 4678 $\pm$ 203 & 4760 $\pm$ 54 \\
6600 - 6800 & 5178 $\pm$ 171 & 4936 $\pm$ 243 & 4491 $\pm$ 177 & 4722 $\pm$ 199 & 4841 $\pm$ 98 \\
6800 - 7000 & 4943 $\pm$ 1096 & - & 4400 $\pm$ 353 & - & 4441 $\pm$ 321 \\
6799 - 7399 & - & 4569 $\pm$ 111 & - & 4989 $\pm$ 93 & 4844 $\pm$ 74 \\
7799 - 7999 & - & 5074 $\pm$ 193 & - & 4709 $\pm$ 275 & 4976 $\pm$ 140 \\
7999 - 8201 & - & 5115 $\pm$ 223 & - & 4667 $\pm$ 293 & 4975 $\pm$ 161 \\
8201 - 8801 & - & 4654 $\pm$ 214 & - & 4271 $\pm$ 289 & 4525 $\pm$ 164 \\

\hline

\end{tabular}
\end{center}
\end{table*}

%% B600 Target

\begin{table*}
\begin{center}
\caption{\texttt{New:$\lambda$LC}, B600 : Wavelength dependent transit depths (in ppm) for the individual GMOS-B600 observations (marked by the columns) and combined from all observations obtained using the new method described in Section \ref{sec:binned_LC}. }   
\label{tab:b600_ts_targ}
\begin{tabular}{cccc}

\hline
Wavelength [\AA] & & Transit Depth [ppm] \\ 
 & 4 & 5 & Combined \\ 
\hline

4900 - 5100 & 4925 $\pm$ 75 & 4266 $\pm$ 550 & 4913 $\pm$ 75 \\
5100 - 5300 & 5197 $\pm$ 418 & 4755 $\pm$ 448 & 4992 $\pm$ 306 \\
5300 - 5500 & 4879 $\pm$ 324 & 4713 $\pm$ 289 & 4787 $\pm$ 216 \\
5500 - 5700 & 4925 $\pm$ 255 & 4819 $\pm$ 393 & 4894 $\pm$ 214 \\
5700 - 6000 & 4735 $\pm$ 197 & 4784 $\pm$ 212 & 4757 $\pm$ 145 \\
6000 - 6200 & 5026 $\pm$ 41 & 4401 $\pm$ 223 & 5006 $\pm$ 40 \\
6200 - 6400 & 4830 $\pm$ 187 & 4759 $\pm$ 381 & 4816 $\pm$ 168 \\
6400 - 6600 & 4583 $\pm$ 226 & 5091 $\pm$ 307 & 4761 $\pm$ 182 \\
6600 - 6800 & 4760 $\pm$ 410 & 5196 $\pm$ 398 & 4984 $\pm$ 286 \\

\hline

\end{tabular}
\end{center}
\end{table*}

%% B600 Target/Comparison
\begin{table*}
\begin{center}
\caption{\texttt{Conv1:$\lambda$LC}, B600 : Wavelength dependent transit depths (in ppm) for the individual GMOS-B600 observations (marked by the columns) and combined from all observations obtained using the new method described in Section \ref{sec:binned_LC}. }   
\label{tab:b600_ts_corr}
\begin{tabular}{cccc}

\hline
Wavelength [\AA] & & Transit Depth [ppm] \\ 
 & 4 & 5 & Combined \\ 
\hline

4900 - 5100 & 2577 $\pm$ 1280 & 4572 $\pm$ 1373 & 3504 $\pm$ 937 \\
5100 - 5300 & 4151 $\pm$ 791 & 5174 $\pm$ 774 & 4674 $\pm$ 553 \\
5300 - 5500 & 4199 $\pm$ 553 & 4510 $\pm$ 429 & 4393 $\pm$ 339 \\
5500 - 5700 & 3776 $\pm$ 1130 & 4974 $\pm$ 294 & 4898 $\pm$ 284 \\
5700 - 6000 & 4825 $\pm$ 229 & 4868 $\pm$ 246 & 4845 $\pm$ 168 \\
6000 - 6200 & 4976 $\pm$ 166 & 5041 $\pm$ 127 & 5017 $\pm$ 101 \\
6200 - 6400 & 4945 $\pm$ 283 & 4968 $\pm$ 212 & 4960 $\pm$ 170 \\
6400 - 6600 & 4819 $\pm$ 238 & 4906 $\pm$ 237 & 4862 $\pm$ 168 \\
6600 - 6800 & 4860 $\pm$ 383 & 4903 $\pm$ 260 & 4889 $\pm$ 215 \\

\hline

\end{tabular}
\end{center}
\end{table*}

%%%% Wavelength dependent light curves: plot of fits to light curves, detrended light curves, and residuals
%% Observation 1

%% Observation 2

%% Observation 3

%% Observation 6

%% Observation 4

%% Observation 5

\section{Results and Discussion} 
\label{sec:discuss}

\subsection{Comparison of the two Methods and Implications}
\label{sec:method_comparison}

\subsubsection{Comparing the white light curve fits}
\label{sec:method_comparison_WLC}

We first compare the performance of the conventional and new methods applied to fitting the white light curves. We compare the three cases \texttt{Conv1:WLC}, \texttt{New:WLC}, and \texttt{New:WLC;No\_Comp} used to fit the white transit light curves for each observation highlighted in Table \ref{tab:WLC_bestfitparams}.

From Table \ref{tab:WLC_bestfitparams} we find that the new method (\texttt{New:WLC} and \texttt{New:WLC;No\_Comp}) provides similar results compared to the conventional method \texttt{Conv1:WLC} at a precision better than 2$\sigma$ level. 

\texttt{New:WLC} yields on average lower residual RMS compared to \texttt{Conv1:WLC} for all observations. For observations 1, 2, 4, and 5 \texttt{New:WLC} also yields marginally smaller (by $\sim$20\% on average) uncertainties on $R_{\rm P}/R_\star$ as compared to the \texttt{Conv1:WLC}. With \texttt{New:WLC;No\_Comp} when not using the comparison star at all, we achieve marginally larger (by $\sim$10-20\% on average) $R_{\rm P}/R_\star$ uncertainties for all observations except observations 1,3, and 5. For observation 5, \texttt{New:WLC;No\_Comp} gives $\sim$80\% smaller uncertainty on $R_{\rm P}/R_\star$. For observations 1 and 3, \texttt{New:WLC;No\_Comp} leads to an order of magnitude larger uncertainty on $R_{\rm P}/R_\star$. This is because for these two R150 observations the odd-even effect is particularly high and using comparison star light curve, either as a GP regressor or linearly (as in \texttt{Conv1:WLC}), is crucial to account for the odd-even effect in the target light curve.    

For the B600 observations (4 and 5) specifically, the comparison star light curves have time dependent trends significantly different from the target star light curve due to the non-ideal PA of the observational setup, which significantly contaminates the transit signal in the resulting Target/Comparison light curves (as seen in Figure \ref{fig:WLC_all}). From a visual inspection of the B600 light curves in Figure \ref{fig:WLC_all}, Target/Comparison corrects for the odd-even effect in the transit light curve but adds additional low frequency trend not present in the original target transit light curve. This effect especially strong before and during the transit. Nevertheless, the conventional method \texttt{Conv1:WLC} for fitting the B600 Target/Comparison light curves using a GP with time and airmass as regressors retrieves transit parameters consistent with R150 observations. Notably, \texttt{New:WLC} achieves lower uncertainties on $R_{\rm P}/R_\star$ as compared to \texttt{Conv1:WLC}. Not using the comparison star with \texttt{New:WLC;No\_Comp} yields lower (observation 5) or marginally larger but comparable (observation 4) uncertainties on $R_{\rm P}/R_\star$.

We conclude from comparison across the aforementioned three cases of fitting the white light curves that both applications of our proposed new method perform consistently and even better in some instances compared to the conventional method. We also conclude that in most instances it is possible to detrend the target light curves and achieve decent precision on transit parameters even without using the comparison stars at all.

\subsubsection{Comparing the spectroscopic light curve fits}
\label{sec:method_comparison_sptlc}

We now compare the transmission spectra obtained from the conventional method \texttt{Conv1:$\lambda$LC} and the new method \texttt{New:$\lambda$LC} to fit $\lambda$LCs. We emphasize here that when not using comparison star at all to fit the white target light curves we obtain consistent transit parameters and hence the common-mode trend as when we use the comparison star (also see Section \ref{sec:binned_LC_old_method}). Hence all conclusions below about the new method are valid whether the common-mode trend is derived by using the comparison star indirectly as in \texttt{New:WLC} or not using it at all in \texttt{New:WLC;No\_Comp}. 

\begin{figure*}
  \centering
  \includegraphics[scale=0.45]{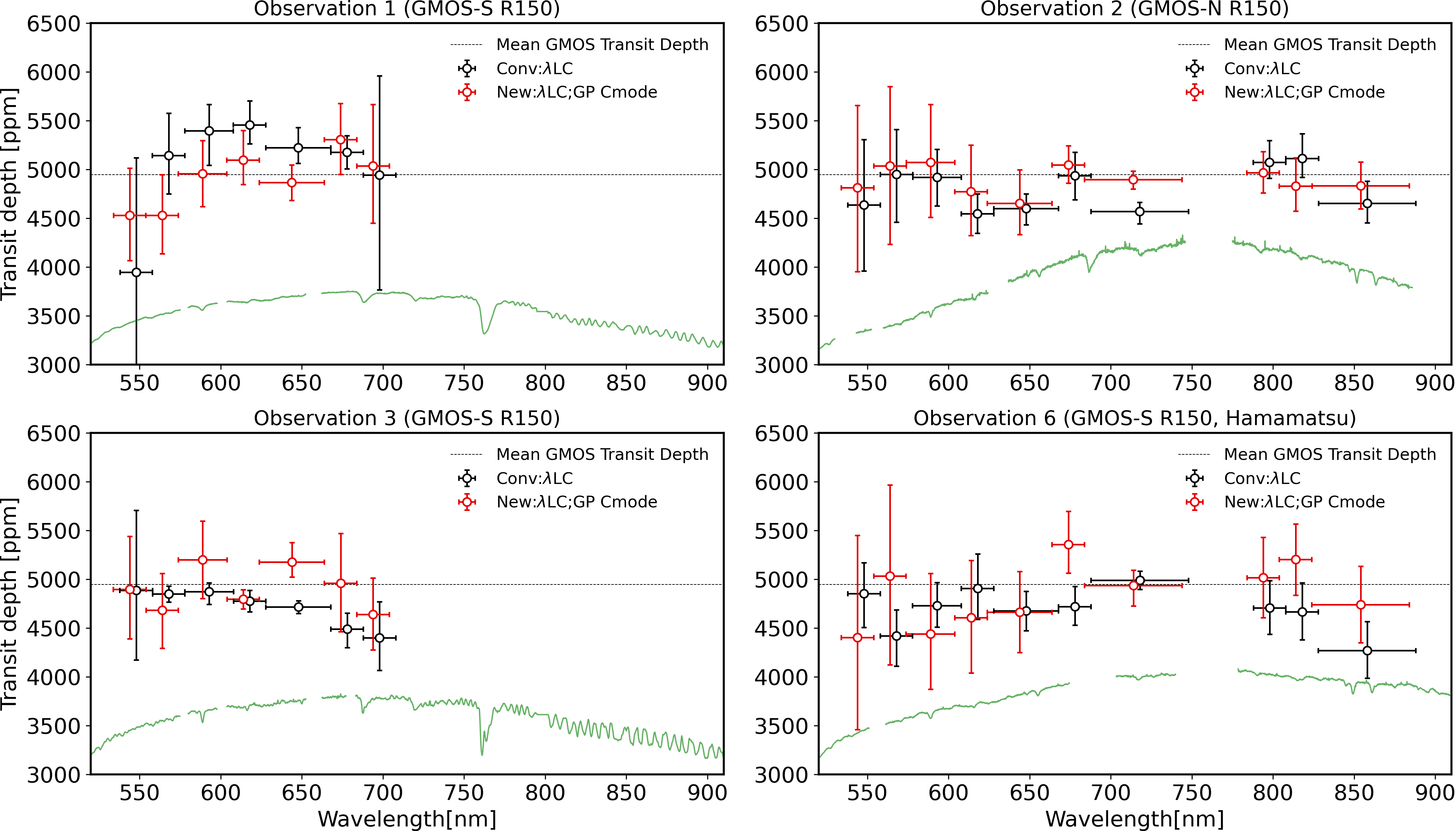}
  \caption[1]{Transmission spectra for the GMOS-R150 observations obtained using \texttt{Conv1:$\lambda$LC} and \texttt{New:$\lambda$LC} (slightly shifted in wavelength for clarity) described in Section \ref{sec:binned_LC}. The average GMOS optical transit depth (corresponding to the weighted average white light curve $R_{\rm P}/R_\star$, $0.0701^{2} = $4914 ppm), which is consistent with the median HST STIS/G750L transit depth from \cite{Wakeford2017}, is marked by the dashed line. For each observation, in black are shown the spectra obtained through conventional method \texttt{Conv1:$\lambda$LC} of fitting Target/Comparison $\rm \lambda$LCs using a GP model with time as a regressor. In red are the transmission spectra obtained by using the new method \texttt{New:$\lambda$LC} to extract transmission spectra only from Target $\rm \lambda$LCs: using a GP model with time and common-mode trend as regressors are shown in red. Overplotted is the observed stellar spectrum for the target star (HAT-P-26b) in green.}
%   Transmission spectra for the GMOS-R150 observations obtained using \texttt{Conv1:$\lambda$LC} and \texttt{New:$\lambda$LC} (slightly shifted in wavelength for clarity) described in Section \ref{sec:binned_LC}. The average GMOS optical transit depth (corresponding to the weighted average white light curve $R_{\rm P}/R_\star$, $0.0701^{2} = $4914 ppm), which is consistent with the median HST STIS/G750L transit depth from \cite{Wakeford2017}, is marked by the dashed line. For each observation, in black are shown the spectra obtained through conventional method \texttt{Conv1:$\lambda$LC} of fitting Target/Comparison $\rm \lambda$LCs using a GP model with time as a regressor. In red are the transmission spectra obtained by using the new method \texttt{New:$\lambda$LC} to extract transmission spectra only from Target $\rm \lambda$LCs: using a GP model with time and common-mode trend as regressors are shown in red. Overplotted is the observed stellar spectrum for the target star (HAT-P-26b) in green.
  \label{fig:r150_ts}
\end{figure*}

\begin{figure*}

  \centering
  \includegraphics[scale=0.45]{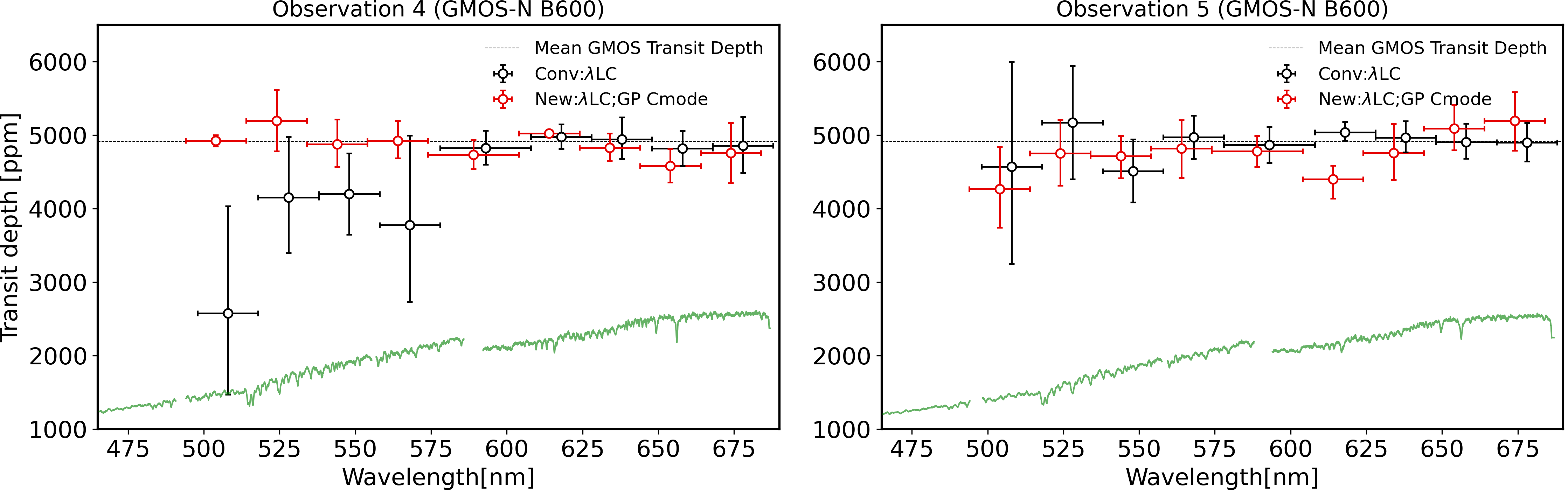}
  \caption[1]{Transmission spectra for the GMOS-B600 observations obtained using \texttt{Conv1:$\lambda$LC} and \texttt{New:$\lambda$LC} (slightly shifted in wavelength for clarity) as described in Section \ref{sec:binned_LC}. The average GMOS optical transit depth (corresponding to the weighted average white light curve $R_{\rm P}/R_\star$, $0.0701^{2} = $4914 ppm), which is consistent with the median HST STIS/G750L transit depth from \cite{Wakeford2017}, is marked by the dashed line. For each observation, in black are shown the spectra obtained through conventional method \texttt{Conv1:$\lambda$LC} of fitting Target/Comparison $\rm \lambda$LCs using a GP model with time as a regressor. In red are the transmission spectra obtained by using the new method \texttt{New:$\lambda$LC} to extract transmission spectra only from Target $\rm \lambda$LCs: using a GP model with time and common-mode trend as regressors are shown in red. Both the B600 observations were obtained using non-ideal PA which manifests as widely different time dependent trends in the target and comparison $\rm \lambda$LCs. This leads to contamination of the transit signal in \texttt{Conv1:$\lambda$LC} (black points), especially for the bluest wavelength bins as seen here for both B600 observations. Overplotted is the observed stellar spectrum for the target star (HAT-P-26b) in green.
  }
  \label{fig:b600_ts}
\end{figure*}

For both B600 and R150 observations, the transmission spectra shown in Figures \ref{fig:r150_ts} and \ref{fig:b600_ts} and corresponding wavelength-dependent transit depth values in Tables \ref{tab:r150_ts_targ}, \ref{tab:r150_ts_targ_by_comp} reveal that the individual and combined transmission spectra across observations from the \texttt{Conv1:$\lambda$LC} and \texttt{New:$\lambda$LC} are on average consistent within their 2$\sigma$ uncertainties. \texttt{New:$\lambda$LC} on average yields $\sim 40\%$ smaller RMS of the residuals per wavelength bin (see Figures \ref{fig:sptlcs_1} to \ref{fig:sptlcs_5}). 

The per wavelength bin uncertainties on the transmission spectra are on average $\sim 30\%$ larger from \texttt{New:$\lambda$LC} as compared to that from \texttt{Conv1:$\lambda$LC} for the R150 observations. For the B600 observation 4 \texttt{New:$\lambda$LC} yields $\sim 50\%$ smaller uncertainties, especially for the bluest wavelength bins. \texttt{New:$\lambda$LC} also performs similarly well in terms of precision for the three bluest bins for the B600 observation 5, but yields nearly $\sim 30\%$ larger uncertainties for the redder bins. This difference in uncertainties on the transmission spectra points towards fundamental differences between the two methods in their approach of dealing with the systematics which we elaborate on. One clear difference is the number of free hyperparameters used for the GP models in both methods. For \texttt{Conv1:$\lambda$LC}, the GP model uses only one regressor (time) and hence two hyperparameters (amplitude and length scale, see Equation \ref{kernel} and \ref{eq:Rij} in Section \ref{noise_model}). \texttt{New:$\lambda$LC} in comparison uses two regressors (time and common-mode trend) and hence three hyperparameters (common amplitude, and one length scale hyperparameters for each of the regressors). Using a more flexible model with more hyperparameters is one of the reasons behind larger uncertainties in the transmission spectra from \texttt{New:$\lambda$LC}. 

Note that \texttt{Conv1:$\lambda$LC} before fitting the GP model also involves two additional steps: dividing by comparison $\lambda$LCs and subtracting the common-mode trend. Both of these steps are linear corrections which do not explicitly propagate uncertainties arising from non-linear differences between the target $\lambda$LCs and the comparison $\lambda$LCs or common-mode trend. It can be seen from the target $\lambda$LCs in Figures \ref{fig:sptlcs_1} to \ref{fig:sptlcs_5} that target star light curves suffer from a low frequency trend in time that varies with wavelength due to wavelength-dependent extinction that is changing with the airmass. These low frequency trends still remain after division by comparison $\lambda$LCs and subtracting the common-mode trend, as seen in the Target/Comparison $\lambda$LCs in Figures \ref{fig:sptlcs_1} to \ref{fig:sptlcs_5}. There is also a high frequency trend (e.g. odd-even effect described in Section \ref{odd_even}) which affects every wavelength bin in a similar manner. 

Our new method \texttt{New:$\lambda$LC} for fitting target $\lambda$LCs doesn't use comparison $\lambda$LCs and accounts for trends at both frequencies in addition to accounting for their wavelength dependence in one step. The Bayesian framework of GPs propagates the uncertainties in the information from the common-mode trend as relevant to each target $\lambda$LCs. Specifically, the common-mode trend helps in accounting for the high frequency trends while time accounts for the low frequency trend varying with respect to wavelength (as demonstrated in Appendix \ref{cmode_gp_test}). Moreover, when not using comparison star $\lambda$LCs, using common-mode correction as a GP regressor can potentially provide better precision as compared to conventional common-mode correction as we demonstrate in Appendix \ref{cmode_gp_test}. By not using the comparison star $\lambda$LCs, we prevent possible introduction of additional systematics due to different instrumental systematics or differential atmospheric extinction between the target and comparison stars with changing airmass. This is supported by the superior performance of \texttt{New:$\lambda$LC} for $\lambda$LCs for the bluest bins of observation 4 as compared to \texttt{Conv1:$\lambda$LC} in terms of precision and accuracy of transit depths. 

\subsubsection{Implications of measuring transmission spectra without using comparison stars}

In the previous subsection, we show that decent precision on fits to both white transit light curves and $\lambda$LCs can be achieved even when not using the comparison star at all. We highlight the usefulness of this aspect of our method for cases when the comparison star is not a suitable reference for systematics in the target star light curve either due to large differences in brightness or spectral type, or issues with the observational setup, as is the case of our GMOS-B600 observations. In fact, our new method essentially removes the transit signal from the white target light curves and uses the information in the residuals (common-mode trend) to fit the target $\lambda$LCs. In this context, a further step could be that we may not need to fit the white target light curves and we can rely on using the previously measured planet transit parameters from other observatories e.g. TESS, HST/STIS in the bandpass significantly overlapping with GMOS and theoretical priors on the limb darkening for the star to compute the transit signal used to obtain the common-mode trend which can be used to fit $\lambda$LCs. Caveats of this approach of bypassing the fit of the white light curve are observations of planets with variable broadband transit depths due to e.g., stellar host variability over multiple epochs. In such cases, it would be essential to fit the white light curve transit depths for individual epoch first to be able to obtain common-mode trend with normalisation to the transit depth for that epoch leading to accurate absolute transmission spectra. In particular, for active host stars, instead of using the same transit depth to derive the common-mode trend across all epochs (as we do for HAT-P-26 in this paper), we advise using the individual best fit white light curve transit depths for each epoch.

Our new method of extracting transmission spectra from solely target star light curves has further implications for ground-based follow-up atmospheric observations of exoplanets orbiting bright host stars especially those discovered by TESS. In particular, the majority of TESS stellar host stars are bright in optical, with median V$_{mag}$ $\sim$ 11 as indicated by simulations from \cite{Barclay2018}, and may not have a choice of comparison stars with similar brightness and spectral type in the limited field of view of up to 10 arcminutes for most ground based multi object spectrographs. We recommend using \texttt{New:WLC;No\_Comp} followed by \texttt{New:$\lambda$LC} for obtaining ground-based transmission spectra of such exoplanets orbiting bright stars. Furthermore, another strength of our new method is that it can potentially mitigate significant second order colour dependent extinction effects arising due to differences in target and comparison star spectral types (\citealt{Young1991}, \citealt{Blake2011}).

%%% Figure: R150 Transmission spectra 

%%% Figure: B600 Transmission spectra  

\subsection{Interpretation of the Optical to NIR Transmission Spectrum}
\label{sec:interpretation}

We generate the combined transmission spectrum by weighted averaging the wavelength-dependent transit depths across common wavelength bins covered by individual R150 and B600 observations, taking the squared reciprocal of the transit depth uncertainties as weights for the respective observations. The combined transmission spectrum values from both methods for R150 observations are shown in Tables \ref{tab:r150_ts_targ} and \ref{tab:r150_ts_targ_by_comp}, and for B600 observations in Tables \ref{tab:b600_ts_targ} and \ref{tab:b600_ts_corr}. Since for the B600 observations, \texttt{New:$\lambda$LC} performs much better than \texttt{Conv1:$\lambda$LC}, we only consider the combined transmission spectra obtained from \texttt{New:$\lambda$LC} for further comparison with atmospheric models.  

We use the open source atmospheric modelling code \texttt{platon} (\citealt{Zhang2019}, \citeyear{Zhang2020} ) based on \texttt{ExoTransmit} (\citealt{Kempton2016}) to conduct a simple retrieval analysis for the atmosphere of HAT-P-26b to interpret our combined GMOS observations in conjunction with the near infrared transmission spectra measurements from HST and Spitzer reported by \cite{Wakeford2017}. For the self-consistent retrieval framework of \texttt{platon} we consider equilibrium chemistry models for three cases: 1) both metallicity and C/O fixed to solar values, and 2) both metallicity and C/O free to fit, 3) metallicity free and C/O fixed to solar value. For all three cases we also let free the pressure level of a grey opacity cloud deck. 

Since early measurements of chromospheric activity indicator S$_{HK}$ index (\citealt{Hartman2011}) and subsequent photometric follow-up observations by \cite{vonEssen2019} show no signs of activity or significant spot modulated variability of stellar photospheric brightness, we do not include contributions from transit light source effect (\citealt{Rackham2018}) in our retrieval analysis. From the stellar photometry reported by \cite{vonEssen2019} no signatures of spot modulations of stellar flux are observed and the upper limit on V band photometric variability for HAT-P-26 (a K1 dwarf) is 2.3 parts per thousand or 0.23 \% (which is the maximum scatter in the light curves). Referring to empirical relationship between the peak to peak optical variability amplitude vs spot covering fraction for K dwarfs from \cite{Rackham2019}, we note that 0.2 \% variability would correspond to less than 1 \% spot covering fraction. Considering the upper limit of 1 \% spot covering fraction, we use Equation 3 from \cite{Rackham2019} to estimate the upper limit on the amplitude of wavelength dependent stellar contamination factor on the transmission spectrum and find it to be 0.9901. Considering the average transit depth of HAT-P-26b to be around 5000 ppm, this would correspond to a maximum offset of ~50 ppm to the transmission spectrum, which is about a factor 5 to 10 less as compared to the average precision of the transmission spectrum in the individual epochs. We hence conclude that given the precision of our observations, we would not be able to detect the offsets due to stellar contamination corresponding to the available upper limits from stellar photometry.

Note that our GMOS-B600 observation 4 was taken at the same time (on 12/03/2016 UT) as one of the HST/WFC3 observations of \cite{Wakeford2017} and the consistency of the median wavelength-dependent transit depth between both observations taken simultaneously from two different further underscores the suitability of combining them. Hence, we do not introduce any vertical offset between the measurements from GMOS, HST, and Spitzer in further analysis.  

\begin{figure*}

  \centering
  \includegraphics[scale=0.45]{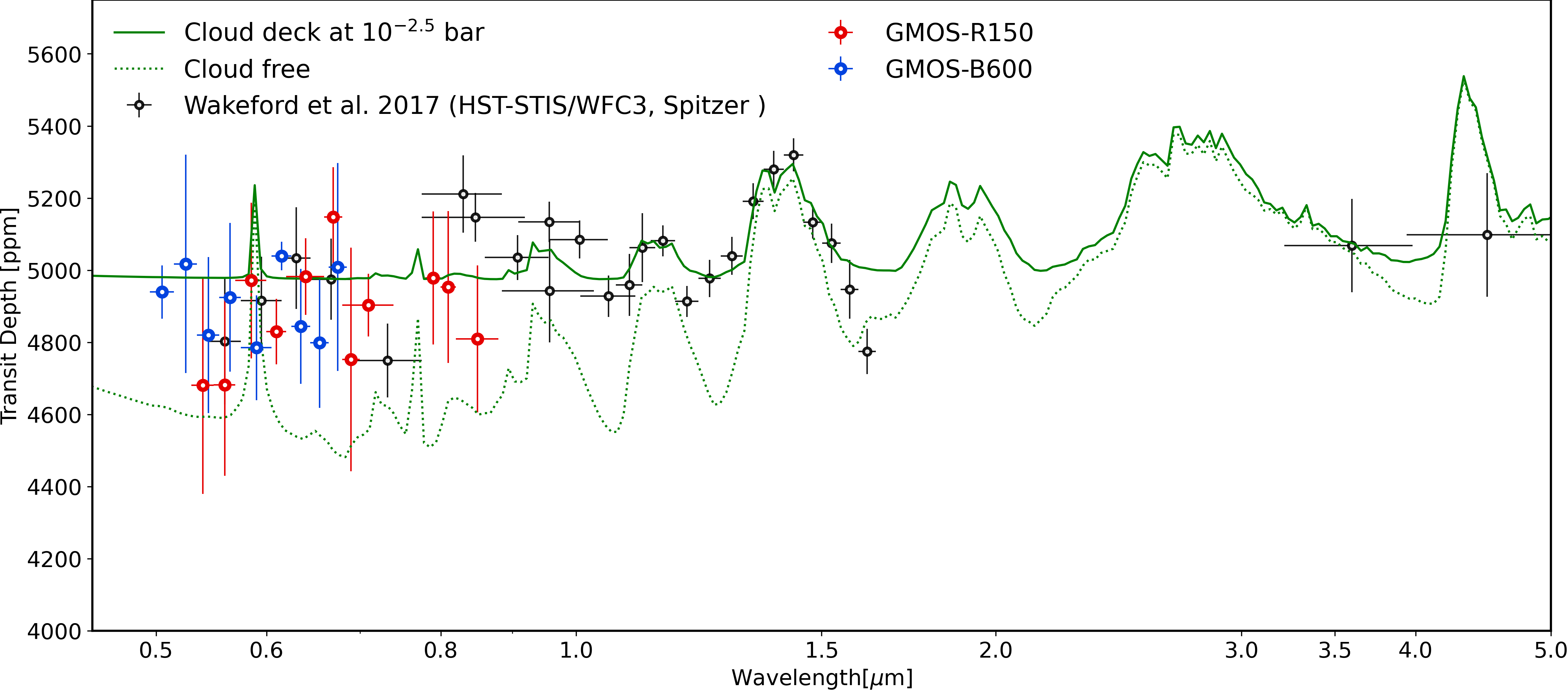}
  \caption[1]{Combined optical transmission spectrum from the 4 GMOS-R150 (red points) and 2 GMOS-B600 (blue points) observations obtained from \texttt{New:$\lambda$LC} presented in Section \ref{sec:binned_LC_new_method} along with the previous measurements in the optical and near infrared from HST/STIS-G750L and HST/WFC3-G102 and WFC3-G141, and in infrared from Spitzer as presented by \cite{Wakeford2017}. Overplotted is the best fit transmission spectroscopy model obtained using \texttt{platon} which has a cloud deck at 3.5 millibar (10$^{-2.5}$ bar), in solid green, and a cloud free model in dotted green for comparison. }
  \label{fig:platon_model_TS}
\end{figure*}

We find that the transmission spectrum of HAT-P-26b from the combined GMOS, HST, and Spitzer measurements are best explained by a model corresponding to a solar metallicity and solar C/O atmosphere with a grey opacity cloud deck at log$_{10}$P (bar) = -2.5$^{+0.53}_{-0.28}$ , which is consistent with the pressure level of the cloud deck constrained by \cite{Wakeford2017} (log$_{10}$P (bar) $\sim$ -2 ) using STIS/G750L observations. The $\chi^{2}_{red}$ for the best fit model with a grey opacity cloud is 1.68 compared to 17.4 for a cloud-free model. The resulting best fit model along with the cloud-free model for comparison is shown in Figure \ref{fig:platon_model_TS}.
Given the lack of coverage at the bluest optical end of the transmission spectrum due to the drop in throughput of GMOS observations blueward of 490 nm our observations cannot constrain the signatures of tentative Rayleigh scattering predicted by \cite{MacDonald2019}. We also do not confirm or rule out the $\sim$ 400 ppm TiH feature at 0.54 $\mu$m predicted by \cite{MacDonald2019} due to our precisions around this region (see Tables \ref{tab:r150_ts_targ} to \ref{tab:b600_ts_targ}) being comparable to the amplitude of the feature as well as our seeing limited resolution restricting us to 20 nm wide wavelength bins. 

\subsection{Transit Timing Variations}
\label{sec:ttv}
Ground based transit observations from multi object spectrographs like GMOS can provide high precision (of the order of 10s of seconds) on the mid-transit time as a result of the high signal to noise nature of observations and continuous sampling of the transit including the ingress and egress without gaps. An example is the mid transit times from the Gemini/GMOS observations of WASP-4b (\citetalias{Huitson2017}) which when combined with other timing measurements including those from TESS by \citeauthor{Bouma2019} (\citeyear{Bouma2019}, \citeyear{Bouma2020}) have been used to study the transit timing variations of the planet at high precision. 

\begin{figure*}

  \centering
  \includegraphics[scale=0.5]{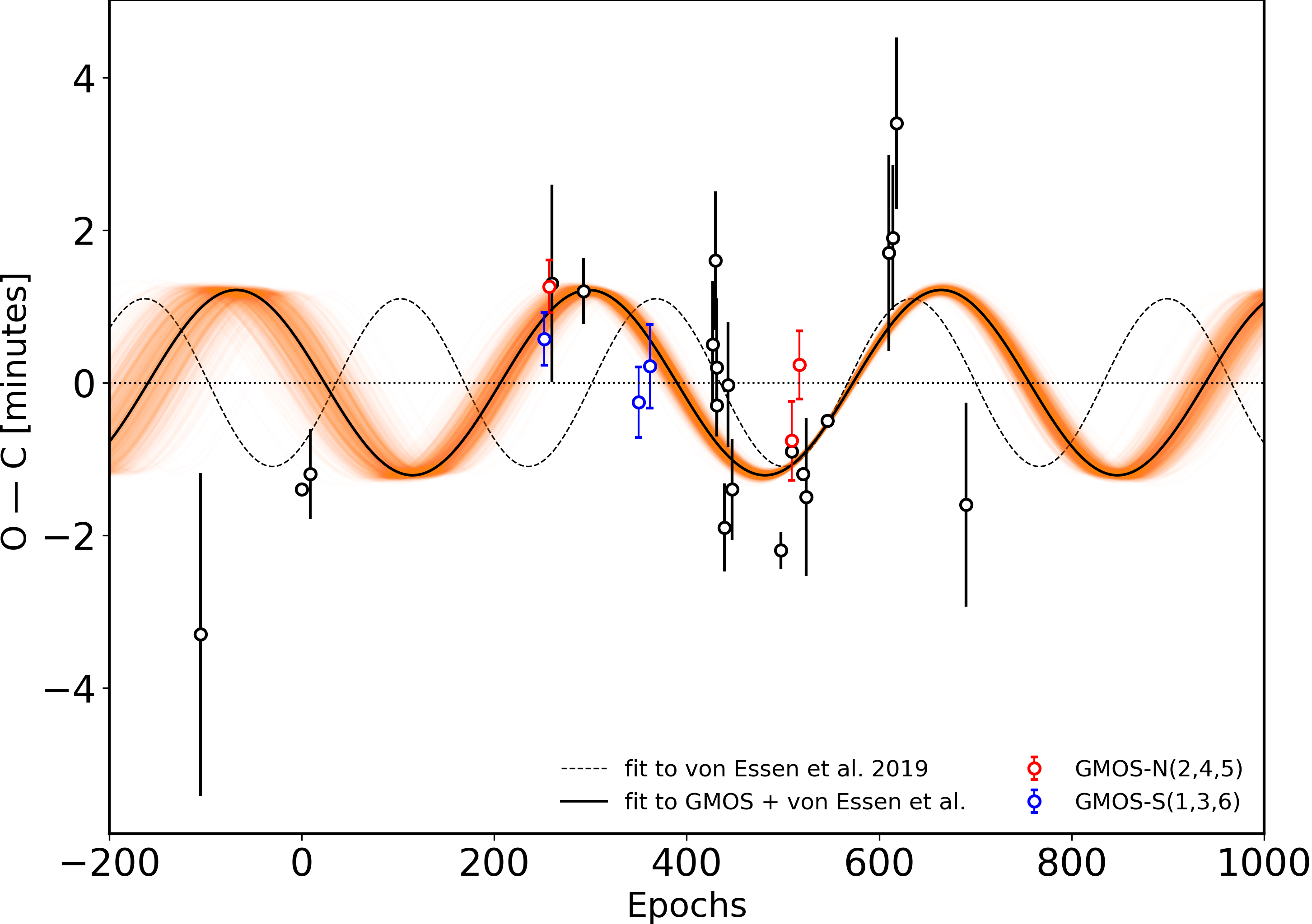}
  \caption[1]{Observed minus calculated mid transit times (O -- C, from a linear ephemeris) from the mid transit times presented by \cite{vonEssen2019} (black points, including a compilation of all the previously published mid transit times and those measured by them) and those presented in this paper (red and blue points, numbers corresponding to observation number in Table \ref{obsstats}). Overplotted in dashed black line is the best fit sinusoidal model to only the O -- C values from \cite{vonEssen2019}, in solid black is the best fit sinusoidal model fit to O -- C values from the \cite{vonEssen2019} and the GMOS observations, and in orange are the randomly sampled fits from the MCMC posteriors.}
  \label{fig:ttv}
\end{figure*}

For HAT-P-26b, we obtain an average precision of $\sim 25$ seconds on the mid-transit times across the 6 GMOS transit observations as shown in Table \ref{tab:WLC_bestfitparams} (mid-transit times from \texttt{New:WLC}). We combine our mid-transit times from \texttt{New:WLC} with those compiled by \cite{vonEssen2019}. The mid-transit time measured from GMOS for observation 4 is consistent with that measured from the simultaneous HST/WFC3 transit observation from \citealt{Wakeford2017} within the 1 $\sigma$ uncertainty. Taking the zeroth epoch same as that considered by \cite{vonEssen2019} we compute the observed minus calculated (O -- C)  for the GMOS mid-transit times assuming a linear ephemeris for the calculated or predicted mid-transit times. To these O -- C values combined with the measurements from \cite{vonEssen2019} we then fit a sinusoidal model with three free parameters : amplitude of the TTVs $A_{TTV}$, period (P, in number of epochs), and a phase value ($\phi_{TTV}$) using \texttt{emcee}. The resulting best fit and fits from random samples from the posteriors computed by \texttt{emcee} are shown in Figure \ref{fig:ttv}. 

Our best fit sinusoidal fit has an amplitude of $\rm A_{\rm TTV}$ $= 1.21^{+0.040}_{−0.039}$ minutes, with period P $ = 366.016^{+14.76}_{-14.19}$ epochs and $\phi_{TTV}$ $= −2.74^{+0.38}_{−0.37}$. The reduced chi-squared value (with a degree of freedom 22) for the sinusoidal fit to the  O $-$ C including the GMOS and \cite{vonEssen2019} measurements is $\sim$5 as compared to $\sim$288 for O $-$ C = 0 which is the case when the measured O $-$ C values would be consistent with a linear ephemeris. This is consistent with the indication of TTVs for HAT-P-26b previously reported by \cite{vonEssen2019} and also indicated by \cite{Stevenson2016}, and motivates future follow up using both transit and secondary eclipse measurements to determine the physical explanation behind the TTVs.

%%%%%%%%%%%% 

% Our best fit sinusoidal fit has an amplitude of $\rm A_{\rm TTV}$ $1.21^{+0.040}_{−0.039}$ minutes, with period P  
% $366.016^{+14.76}_{-14.19}$ epochs and $\phi_{TTV}$ $−2.74^{+0.38}_{−0.37}$.The reduced chi-squared value (with a degree of freedom 22) for the sinusoidal fit to the  O $-$ C including the GMOS and \cite{vonEssen2019} measurements is $\sim$ 5 as compared to $\sim$ 288 for O $-$ C  0 which is the case when the measured O $-$ C values would be consistent with a linear ephemeris. This is consistent with the indication of TTVs for HAT-P-26b previously reported by \cite{vonEssen2019} and also indicated by \cite{Stevenson2016}, and motivates future follow up using both transit and secondary eclipse measurements to determine the physical explanation behind the TTVs. 

% A thorough interpretation and modeling of the physical origin TTVs is beyond the scope of our study. 

\section{Conclusions}
\label{sec:conclusions}

We have introduced a new method to model systematics in ground based spectrophotometric observations that allows for a generalised non-linear mapping between the target star transit light curves and the time series used as regressors to detrend them. We test and demonstrate the performance of the new method in comparison to the conventional method by applying both methods to ground-based optical transmission spectra of the warm Neptune HAT-P-26b from 6 transits observed by Gemini/GMOS as part of our ground-based survey of exoplanet atmospheres in the optical.

We summarise the key aspects and conclusions for the new method we introduce in this paper: 

1) With the new method, we fit the systematics and transit signal in the target star white light curves directly by using a GP regression model conditioned with various combinations of regressors which include the simultaneously observed comparison star white light curve. This is a generalisation of conventional linear methods which have used comparison star white light curves as a linear regressor. The new method when using comparison star white light curves as a GP regressor lets the GP determine the underlying non-linear mapping between the comparison and target star light curves. This approach utilises the information about systematics from the comparison star light curves without introducing additional uncertainties as often is the case when doing differential photometry. It also propagates uncertainties appropriately within the Bayesian framework of GPs when using the comparison star light curve as a GP regressor. \\ 

2) The application of the new method \texttt{New:WLC;No\_Comp} to fit the target white light curves without using the comparison light curves emulates a scenario when suitable comparison stars may not be available. We show that even in the absence of suitable comparison stars, accurate transit parameters with comparable precisions can be obtained from the white target transit light curve fit using our new method.\\

3) The new method when applied to $\lambda$LCs lets the GP determine the non-linear mapping between the white target light curve derived common-mode trend and the individual target $\lambda$LCs. We show by application to observed and transit injected $\lambda$LCs that this approach without needing to perform normalization by comparison $\lambda$LCs is robust and achieves accurate transmission spectra. From the transit injection test, we conclude that using common-mode trend as a GP regressor achieves $\sim$ 20\% better precision on the transmission spectra compared to that from conventional common-mode correction. \\

4) Except for the bluest bins in B600 observations, the new method yields marginally higher uncertainties on the transmission spectra. We interpret this increase in uncertainties as an outcome of fitting for both low and high frequency systematics in $\lambda$LCs in one step and propagating the uncertainties in the process. In contrast, the conventional linear method with multiple steps of dividing by comparison $\lambda$LCs and subtracting the common-mode trend does not explicitly propagate uncertainties at each step.\\          

5) In the context of bluest bins in B600 observations, where in addition to effects due to non-ideal PA we also expect largest differential atmospheric extinction between the target and comparison star spectra due to changing airmass, we show that our new method is able to extract the transmission spectra for scenarios when the conventional Target/Comparison normalisation strongly contaminates the transit signal.  \\

6) We demonstrate that just the target white light curve itself can be used to model the time and wavelength-dependent systematics in the spectroscopic target light curves, albeit at the cost of $\sim$30 \% larger uncertainties on the transmission spectra. This approach can ultimately be used for future optical and near infrared ground-based atmospheric characterisation of exoplanets orbiting bright host stars with little or no available choice of comparison stars with similar brightness and spectral type in the instrument field of view. \\

7) The current prescription of the new method as applied to $\lambda$LCs in this paper fits each $\lambda$LC independently and hence does not explicitly model potential covariance in the wavelength dimension. A future possible extension to our method, especially when applied to medium resolution spectrophotometric observations, is to jointly model the $\lambda$LCs accounting for potential covariance due to systematics in wavelength dimension. \\  

Based on our analyses, we obtain the following conclusions about the atmosphere of HAT-P-26b : 

1) Through equilibrium chemistry retrieval analysis of combined GMOS optical observations with near infrared HST and Spitzer observations, we conclude that the terminator of HAT-P-26b is consistent with solar metallicity and C/O atmosphere with a grey opacity cloud layer at log$_{10}$P (bar) = -2.5$^{+0.53}_{-0.28}$  obscuring the alkali absorption features in optical and suppressing the water absorption features in the near infrared, consistent with the findings of \cite{Wakeford2017}. The low resolution nature of our observations and comparatively low precision on the transit depths preclude confirmation of presence of metal hydride features predicted by \cite{MacDonald2019}.\\

2) Based on the mid transit times constrained by the GMOS transits we find further indications of TTVs for HAT-P-26b in agreement with previous studies. This warrants future follow up primary and secondary eclipse observations of the planet to investigate the physical origin of TTVs. \\

Finally our results add to the growing library of optical transmission spectra of exoplanets obtained using ground-based low-resolution spectrographs. The precision and accuracy of our measurements combined with the repeatability of the observations over multiple epochs emphasize the importance of optical ground-based observations in complementing the upcoming observations of transiting exoplanets in the infrared using JWST.

\section*{Acknowledgements}

Based on observations obtained at the Gemini Observatory (acquired through the Gemini Observatory Archive and Gemini Science Archive), which is operated by the Association of Universities for Research in Astronomy, Inc. (AURA), under a cooperative agreement with the NSF on behalf of the Gemini partnership: the National Science Foundation (United States), the National Research Council (Canada), CONICYT (Chile), Ministerio de Ciencia, Tecnolog\'{i}a e Innovaci\'{o}n Productiva (Argentina), and Minist\'{e}rio da Ci\^{e}ncia, Tecnologia e Inova\c{c}\~{a}o (Brazil). Based in part on Gemini observations obtained from the National Optical Astronomy Observatory (NOAO) Prop. ID: 2012B-0398; PI: J-.M D\'{e}sert. We thank the anonymous reviewer for their thorough comments and feedback which helped in improving the paper. V.P. acknowledges stimulating discussions with Lorenzo Pino, Claire Baxter, Jacob Arcangeli, Niloofar Khorshid, Bob Jacobs, Saugata Barat, Ernst de Mooij, Neale Gibson, and Dan-Foreman Mackey which helped in shaping the paper. V.P. also acknowledges inspiring discussions on Gaussian Processes with Mehmet Deniz Aksulu. J.M.D acknowledges support from the Amsterdam Academic Alliance (AAA) Program, and the European Research Council (ERC) European Union’s Horizon 2020 research and innovation programme (grant agreement no. 679633; Exo-Atmos). This work is part of the research programme VIDI New Frontiers in Exoplanetary Climatology with project number 614.001.601, which is (partly) financed by the Dutch Research Council (NWO). This material is based upon work supported by the NWO TOP Grant Module 2 (Project Number 614.001.601).  This material is based upon work supported by the National Science Foundation (NSF) under Grant No. AST-1413663. This research has made use of NASA's Astrophysics Data System. The authors also acknowledge the significant cultural role and reverence the summit of Mauna Kea has within the indigenous Hawaiian community. This research has made use of \texttt{Astropy},\footnote{http://www.astropy.org} a community-developed core Python package for Astronomy \cite{astropy:2013, astropy:2018}, \texttt{NumPy} \cite{harris2020array}, \texttt{matplotlib}  \cite{Hunter2007}, \texttt{SciPy} \cite{Virtanen2020} and \texttt{IRAF} \cite{Tody1986} distributed by the NOAO, which is operated by AURA under a cooperative agreement with the NSF.

%%%%%%%%%%%%%%%%%%%%%%%%%%%%%%%%%%%%%%%%%%%%%%%%%%
% \section*{Data Availability}

% The inclusion of a Data Availability Statement is a requirement for articles published in MNRAS. Data Availability Statements provide a standardised format for readers to understand the availability of data underlying the research results described in the article. The statement may refer to original data generated in the course of the study or to third-party data analysed in the article. The statement should describe and provide means of access, where possible, by linking to the data or providing the required accession numbers for the relevant databases or DOIs.

\section*{Data Availability}
The data underlying this article and Python notebooks from which the results and figures of this paper can be obtained will be made available upon publication.



%\bibliographystyle{mnras}
%\bibliography{mnras_template} % if your bibtex file is called example.bib

%\begin{thebibliography}{99}
%\end{thebibliography}

%%%%%%%%%%%%%%%%%%%%%%%%%%%%%%%%%%%%%%%%%%%%%%%%%%

%%%%%%%%%%%%%%%%% APPENDICES %%%%%%%%%%%%%%%%%%%%%

\appendix

\section{Testing robustness of using common-mode trend as a GP regressor for $\rm \lambda$LC}
\label{cmode_gp_test}

Correcting for time only dependent systematics in $\rm \lambda$LCs has been conventionally done by dividing or subtracting each $\rm \lambda$LC by a common-mode trend derived from the white light curve. One of the novel aspects of the method we introduce in this paper is to use this common-mode trend as a GP regressor instead of subtracting it from each $\rm \lambda$LC. In this section, we perform a transit injection and recovery test to assess the robustness of using the common-mode trend as a GP regressor to fit $\rm \lambda$LCs and deriving the transmission spectrum. 

We take the observation 4 B600 comparison star white light curve and additively inject to it a transit signal with known transit parameters and linear limb darkening coefficient fixed to the \texttt{PyLDTk} for HAT-P-26b. We inject the same transit signal to each of the 20 nm spectroscopic light curves keeping the limb-darkening coefficient of the injected model same across wavelength bins.

\begin{figure*}

  \centering
  \includegraphics[scale=0.35]{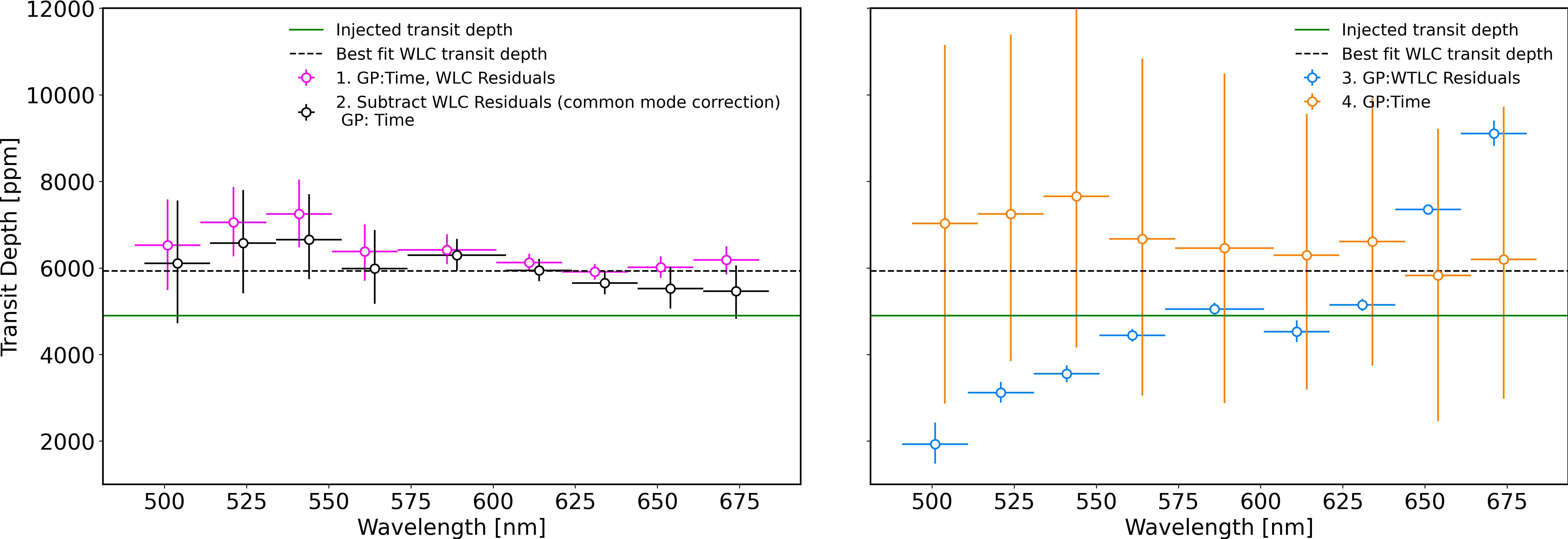}
  \caption{Transmission spectra from different GP regressor combination cases used to test the robustness of using common-mode trend as a GP regressor to fit the $\rm \lambda$LCs with the injected transit signal (horizontal green line in both panels) as described in Section \ref{cmode_gp_test}. The dashed black line in both panels shows the best fit transit depth for the WLC, obtained from fitting it using a GP model with only time as a regressor. Left panel shows the spectra resulting from fitting $\rm \lambda$LCs using 1) New method: Time and WLC residuals as a GP regressor (pink) and 2) Subtracting WLC residuals and fitting using Time as a GP regressor (black). The two cases are consistent with each other and with the best fit WLC transit depth within 1$\sigma$. The new method results in 25 \% smaller uncertainties on average. Right panel shows the spectra from fitting $\rm \lambda$LCs with a GP with regressor as 3) only WLC residuals (blue) and 4) only Time (orange). WLC residuals only case yields high precision and less accurate spectra, and vice versa for Time only case. This shows that WLC residuals are better at fitting high frequency wavelength independent systematics while Time helps in fitting the lower frequency wavelength dependent trend. Both are complementary to achieve better accuracy and precision.        
  }
  \label{fig:cmode_gp_test}
\end{figure*}

We first fit the synthetic white transit light curve (referred to as WLC for brevity) with the injected signal using a \texttt{batman} model for the transit plus a GP with time as a regressor for the systematics. We subtract the best fit \texttt{batman} transit model thus obtained from the WLC to obtain the residuals (i.e. common-mode trend) to be used for the next steps. For fitting the $\rm \lambda$LCs, we test 4 different cases : 1) Using time and WLC residuals as GP regressors to fit $\rm \lambda$LC, 2) Subtracting the WLC residuals from each $\rm \lambda$LC (conventional common-mode correction) and fitting the common-mode corrected $\rm \lambda$LC using time as GP regressor, 3) Using only WLC residuals as a GP regressor to fit $\rm \lambda$LC, and 4) Using only time as a GP regressor to fit $\rm \lambda$LC. The first two cases are the ones that we eventually use in the paper (in Section \ref{sec:binned_LC}). We discuss the latter two cases to demonstrate the individual contributions from time and WLC residuals as GP regressor, respectively. We show the resulting transmission spectra for each case in Figure \ref{fig:cmode_gp_test}. Note that the case 2) here involving conventional common-mode correction does not involve dividing $\rm \lambda$LCs by any corresponding comparison star $\rm \lambda$LCs, and hence is not exactly the same as the conventional method used in the paper (in Section \ref{sec:binned_LC_old_method} where we do divide the targets $\rm \lambda$LCs by comparison star $\rm \lambda$LCs.

We outline below the conclusions from our transit injection test below: 

1) The new method introduced by us in the paper of using the WLC residuals as a GP regressor along with time to fit the $\rm \lambda$LCs robustly retrieves the injected transit signal in the individual $\rm \lambda$LCs with respect to the corresponding best fit WLC transit depth (black dashed line in Figure \ref{fig:cmode_gp_test}). The mean values of the transit depths across the bins from this method (in pink in the Figure \ref{fig:cmode_gp_test}) are consistent within $\pm$1 $\sigma$ with those retrieved from the conventional common-mode correction (subtract the WLC residuals from each bin and fit using a time-dependent GP model, grey in the Figure \ref{fig:cmode_gp_test}).\\ 

2) The mean values from both the methods are centred around the best fit WLC transit depth within $\pm$1 $\sigma$ (black dashed line) and deviate by almost 2 to 3 $\sigma$ from the injected transit depth. This is a potential pitfall of both the new and conventional methods of using the common-mode trend, and shows that the accuracy of both the methods depends on the accuracy of WLC fit.\\

3) The uncertainties from the new method (GP regressors: time and WLC Residuals) is on average 25 \% lower as compared to that from the conventional common-mode correction followed by fitting using time as a GP regressor. This shows that the new method of using the common-mode trend as a GP regressor performs better than conventional common-mode correction in terms of retrieved precision on transmission spectra. \\

4) We also show the results from two additional cases : using only WLC residuals as GP regressor, and using only time as a GP regressor performs. We find that using only WLC residuals as a GP regressor performs poorly in terms of the accuracy of the retrieved transit depths (blue points in Figure \ref{fig:cmode_gp_test}). The $\rm \lambda$LCs suffer from a wavelength dependent low-frequency trend due to the changing airmass through the night. The shape of this trend varies across the wavelength bins due to wavelength dependent atmospheric extinction. This effect can also be seen in the target star spectroscopic light curves shown in the paper in Figures \ref{fig:sptlcs_1} to \ref{fig:sptlcs_5}. The WLC residuals by themselves when used as a GP regressor model the high frequency systematics but are unable to take this low frequency wavelength dependent effect into account. Using time as an additional GP regressor helps to take this account as shown in Figure \ref{fig:cmode_gp_test} (pink points). On the other hand using only time as the GP regressor (orange points), while performing well in terms of overall accuracy of transit depths, performs poorly in terms of precision. We interpret this as the inability of the time only GP regressor model to account for the high frequency systematics in $\rm \lambda$LCs, which is the reason the uncertainties on the corresponding transit depths are larger.

\section{Model selection criteria values for combinations of GP regressors used to fit white transit light curves}
\label{app:model_selection}

We summarise the BIC and log Bayesian evidence values for each GP regressor combination for all observations corresponding to the two methods used for fitting white light curves (Section \ref{sec:WLC}) in Tables \ref{tab:obs1_bicevi} to \ref{tab:obs5_bicevi}. The combinations are shown in decreasing order of log$_{e}$Z, which is broadly consistent with increasing order of BIC values.

We use the prescription of \citealt{KassRaftery1995} to define the threshold of $\Delta$BIC and difference of log Bayesian evidences to estimate the evidence in favour of a GP regressor combination against other combinations. According to this prescription, for two models M$_{1}$ and M$_{0}$, $\Delta$BIC = BIC$_{1}$ $-$ BIC$_{0}$ $\geq$ 10 implies a strong evidence in favour of the model M$_{0}$. In terms of Bayesian evidences, log$_{e}$Z$_{0}$ $-$ log$_{e}$Z$_{1}$ = log$_{e}$(Z$_{0}$/Z$_{1}$) $\geq$ 5 implies a strong evidence in favour of the model M$_{0}$ with log Bayesian evidence log$_{e}$Z$_{0}$. The BIC values and log$_{e}$Z for each GP regressor combination for both methods are shown in Tables \ref{tab:obs1_bicevi} to \ref{tab:obs5_bicevi}. The GP regressor combinations in these tables are shown in decreasing order of log$_{e}$Z which is broadly consistent with the increasing order of BIC. 

Note that for each of the three cases mentioned in Section \ref{model_selection}, during model comparison we neglect the regressor combinations for which one or more GP length scale hyperparameter ($\eta_{p}$) are unconstrained despite having higher Bayesian evidence. We check the posteriors for the corresponding combination sampled by \texttt{dynesty} and \texttt{emcee} for each $\eta_{p}$ to confirm if they are constrained. We also confirm that for models with $\Delta$log$_{e}$Z less than our threshold of 5, we obtain consistent transit parameters among all models. In such a case of multiple equally good models, we choose the model with the least number of GP regressors in the combination.

%%%%%%%%%%%%%%%%%%%%%%%%%%%%%
%%%%%%%%%% Observation 1 %%%%
%%%%%%%%%%%%%%%%%%%%%%%%%%%%%

\begin{table}
\begin{center}
\caption{BIC calculated using the GP likelihood and log Bayesian evidences (log$_{e}$Z) from \texttt{dynesty} for all possible combinations of GP regressors used to fit Target star light curve alone (top panel of the table labelled `Target LC') using \texttt{New:WLC} and \texttt{New:WLC;No\_Comp}, and to fit the Target divided by the Comparison star light curves (bottom panel `Target/Comparison LC') using \texttt{Conv1:WLC}. The combinations are shown in decreasing order of log$_{e}$Z which is broadly consistent with increasing order of BIC. The best GP regressor combination we choose for the three cases we fit the HAT-P-26 transit white light curves for in Section \ref{sec:WLC} - \texttt{New:WLC}, \texttt{New:WLC;No\_Comp}, and \texttt{Conv1:WLC}, are highlighted in bold below with the corresponding case in brackets. The best fit transit and GP parameters for each of these cases are detailed in Table \ref{tab:WLC_bestfitparams}.}

\label{tab:obs1_bicevi}
\begin{tabular}{ccc}
\hline
\hline
Target LC \\
\hline
GP regressors &  GP BIC & log$_{e}$Z \\
\hline
\textbf{Time, Comp, PSF (\texttt{New:WLC})} & \textbf{-2752.45} & \textbf{1360.13} \\
Comp, Airmass, PSF, CRPA & -2745.64 & 1358.46 \\
Comp, PSF, CRPA & -2751.09 & 1357.8 \\
Time, Comp, PSF, CRPA & -2746.95 & 1357.26 \\
Time, Comp, Airmass, PSF & -2748.32 & 1352.35 \\
Time, Comp, Airmass, PSF, CRPA & -2741.66 & 1352.11 \\
Comp, Airmass, PSF & -2742.03 & 1351.56 \\
Time, Comp, Airmass & -2716.01 & 1349.68 \\
Time, Comp, Airmass, CRPA & -2710.99 & 1347.52 \\
Comp, Airmass & -2724.78 & 1347.45 \\
Time, Comp, CRPA & -2712.1 & 1346.79 \\
Time, Comp & -2720.65 & 1346.04 \\
Comp, Airmass, CRPA & -2715.72 & 1345.09 \\
Comp, CRPA & -2717.5 & 1344.86 \\
Comp, PSF & -2687.77 & 1329.57 \\
Comp & -2630.46 & 1305.13 \\
CRPA & -1920.39 & 961.67 \\
Time, Airmass, CRPA & -1909.55 & 961.37 \\
Airmass, CRPA & -1915.0 & 961.37 \\
Time, CRPA & -1915.05 & 961.33 \\
Time, PSF, CRPA & -1909.54 & 961.24 \\
Time, Airmass, PSF, CRPA & -1907.83 & 960.82 \\
PSF, CRPA & -1914.9 & 960.38 \\
Airmass, PSF, CRPA & -1913.19 & 960.27 \\
\textbf{Time (\texttt{New:WLC;No\_Comp})} & \textbf{-1886.92} & \textbf{947.31} \\
Time, Airmass & -1883.75 & 946.45 \\
Time, PSF & -1887.37 & 946.39 \\
Time, Airmass, PSF & -1881.06 & 945.04 \\
Airmass & -1741.57 & 867.93 \\
Airmass, PSF & -1735.42 & 866.36 \\
PSF & -1163.66 & 584.25 \\ \\
\hline
\hline
Target/Comparison LC \\
\hline
GP regressors &  GP BIC & log$_{e}$Z \\
\hline
Airmass & -2751.16 & 1366.1 \\
Time, CRPA & -2744.1 & 1365.45 \\
\textbf{Time (\texttt{Conv1:WLC})} & \textbf{-2749.47} & \textbf{1363.63} \\
Time, Airmass & -2750.97 & 1363.13 \\
Time, Airmass, CRPA & -2745.71 & 1361.75 \\
Airmass, CRPA & -2750.96 & 1360.75 \\
CRPA & -2728.94 & 1353.64 \\

\end{tabular}
\end{center}
\end{table}

%%%%%%%%%%%%%%%%%%%%%%%%%%%%%
%%%%%%%%%% Observation 2 %%%%
%%%%%%%%%%%%%%%%%%%%%%%%%%%%%

\begin{table}
\caption{Same as Table \ref{tab:obs1_bicevi} but for observation 2.}   
\label{tab:obs2_bicevi}
\begin{tabular}{ccc}
\hline
\hline
Target LC \\
\hline
GP regressors &  GP BIC & log$_{e}$Z \\
\hline
Time, Comp, Airmass, PSF & -7317.6 & 3644.81 \\
\textbf{Time, PSF (\texttt{New:WLC;No\_Comp})} & \textbf{-7305.5} & \textbf{3644.04} \\ 
Comp, Airmass, PSF, CRPA & -7315.53 & 3643.67 \\
Time, Comp, PSF, CRPA & -7307.65 & 3643.45 \\
Time, PSF, CRPA & -7298.95 & 3643.43 \\
Time, Comp, PSF & -7313.89 & 3643.35 \\
Airmass, PSF, CRPA & -7311.87 & 3642.73 \\
Time, Airmass, PSF & -7315.44 & 3642.21 \\
Comp, Airmass, PSF & -7316.14 & 3641.99 \\
Time, Comp, Airmass, PSF, CRPA & -7310.97 & 3641.41 \\
Time, Airmass, PSF, CRPA & -7309.27 & 3641.0 \\
Airmass, PSF & -7262.97 & 3610.49 \\
Time, Comp, CRPA & -7213.6 & 3601.23 \\
Comp, PSF, CRPA & -7220.49 & 3599.57 \\
\textbf{Time, Comp (\texttt{New:WLC})} & \textbf{-7220.01} & \textbf{3599.03} \\
Comp, Airmass, CRPA & -7223.57 & 3598.4 \\
Time, Comp, Airmass & -7220.62 & 3598.38 \\
Time, Comp, Airmass, CRPA & -7217.23 & 3598.13 \\
Comp, Airmass & -7211.17 & 3595.11 \\
Airmass, CRPA & -7202.29 & 3593.22 \\
Time & -7192.99 & 3592.55 \\
Time, Airmass & -7197.75 & 3592.31 \\
Time, CRPA & -7186.79 & 3592.02 \\
Time, Airmass, CRPA & -7195.98 & 3591.63 \\
Comp, CRPA & -7159.79 & 3572.72 \\
Airmass & -7157.87 & 3567.11 \\
PSF, CRPA & -7102.58 & 3546.44 \\
CRPA & -7032.34 & 3515.26 \\
Comp & -6491.33 & 3242.77 \\
Comp, PSF & -6491.87 & 3242.14 \\
PSF & -3717.88 & 1884.53 \\ \\
\hline
\hline
Target/Comparison LC \\
\hline
GP regressors &  GP BIC & log$_{e}$Z \\
\hline
Time, CRPA & -6823.07 & 3409.41 \\
Time, Airmass, CRPA & -6816.64 & 3408.92 \\
\textbf{Time (\texttt{Conv1:WLC})} & \textbf{-6829.35} & \textbf{3407.31} \\
Time, Airmass & -6823.18 & 3401.17 \\
Airmass, CRPA & -6811.13 & 3392.76 \\
CRPA & -6774.79 & 3381.19 \\
Airmass & -6116.87 & 3059.6 \\
\end{tabular}
\end{table}

%%%%%%%%%%%%%%%%%%%%%%%%%%%%%
%%%%%%%%%% Observation 3 %%%%
%%%%%%%%%%%%%%%%%%%%%%%%%%%%%

\begin{table}
\caption{Same as Table \ref{tab:obs1_bicevi} but for observation 3.}   
\label{tab:obs3_bicevi}
\begin{tabular}{ccc}
\hline
\hline
Target LC \\
\hline
GP regressors &  GP BIC & log$_{e}$Z \\
\hline
Time, Comp, PSF & -3885.21 & 1938.94 \\
Comp, CRPA & -3891.73 & 1937.57 \\
Time, Comp, Airmass, CRPA & -3881.97 & 1937.51 \\
\textbf{Time, Comp (\texttt{New:WLC})} & \textbf{-3890.89} & \textbf{1936.67} \\
Time, Comp, CRPA & -3885.91 & 1936.38 \\
Comp, Airmass, PSF, CRPA & -3880.05 & 1936.18 \\
Comp, PSF, CRPA & -3885.92 & 1934.91 \\
Comp, Airmass, CRPA & -3885.92 & 1934.88 \\
Time, Comp, Airmass & -3885.14 & 1933.38 \\
Time, Comp, Airmass, PSF, CRPA & -3874.19 & 1932.66 \\
Time, Comp, Airmass, PSF & -3879.2 & 1932.05 \\
Time, Comp, PSF, CRPA & -3880.22 & 1931.76 \\
Comp & -3867.11 & 1924.29 \\
Comp, Airmass & -3860.11 & 1924.24 \\
Comp, Airmass, PSF & -3855.28 & 1922.76 \\
Comp, PSF & -3861.72 & 1921.71 \\
\textbf{Airmass (\texttt{New:WLC;No\_Comp})} & \textbf{-3098.4} & \textbf{1548.87} \\
Airmass, PSF & -3092.41 & 1548.68 \\
Time, Airmass & -3093.0 & 1546.51 \\
Airmass, PSF, CRPA & -3086.93 & 1546.41 \\
Airmass, CRPA & -3092.73 & 1545.9 \\
Time, Airmass, CRPA & -3086.94 & 1545.77 \\
Time, Airmass, PSF & -3087.04 & 1545.47 \\
Time, Airmass, PSF, CRPA & -3080.92 & 1545.39 \\
Time & -3074.6 & 1537.76 \\
Time, PSF & -3068.33 & 1536.61 \\
Time, CRPA & -3068.48 & 1536.52 \\
Time, PSF, CRPA & -3026.42 & 1535.77 \\
CRPA & -3038.38 & 1520.9 \\
PSF, CRPA & -3032.53 & 1520.04 \\
PSF & -1541.36 & 797.32 \\ \\
\hline
\hline
Target/Comparison LC \\
\hline
GP regressors &  GP BIC & log$_{e}$Z \\
\hline
CRPA & -3914.13 & 1951.6 \\
\textbf{Time (\texttt{Conv1:WLC})} & \textbf{-3910.8} & \textbf{1950.74} \\
Airmass, CRPA & -3908.2 & 1950.73 \\
Time, CRPA & -3908.19 & 1949.33 \\
Time, Airmass & -3904.99 & 1948.04 \\
Time, Airmass, CRPA & -3902.58 & 1946.02 \\
Airmass & -3876.99 & 1929.5 \\
\end{tabular}
\end{table}

%%%%%%%%%%%%%%%%%%%%%%%%%%%%%
%%%%%%%%%% Observation 6 %%%%
%%%%%%%%%%%%%%%%%%%%%%%%%%%%%

\begin{table}
\caption{Same as Table \ref{tab:obs1_bicevi} but for observation 6.}   
\label{tab:obs6_bicevi}
\begin{tabular}{ccc}
\hline
\hline
Target LC \\
\hline
GP regressors &  GP BIC & log$_{e}$Z \\
\hline
\textbf{Time, Comp (\texttt{New:WLC})} & \textbf{-3470.86} & \textbf{1729.74} \\
Time, Comp, Airmass, CRPA & -3463.71 & 1729.59 \\
Time, Comp, Airmass, PSF & -3464.93 & 1729.01 \\
Time, Comp, PSF & -3465.22 & 1728.93 \\
Time, Comp, CRPA & -3465.2 & 1728.82 \\
Time, Comp, Airmass & -3469.46 & 1728.36 \\
Time, PSF & -3458.14 & 1728.31 \\
\textbf{Time (\texttt{New:WLC;No\_Comp})} & \textbf{-3462.02} & \textbf{1728.25} \\
Time, Airmass, PSF, CRPA & -3450.06 & 1727.93 \\
Time, CRPA & -3456.38 & 1727.9 \\
Time, Airmass & -3458.23 & 1727.72 \\
Time, Comp, PSF, CRPA & -3459.56 & 1727.66 \\
Time, Airmass, PSF & -3456.33 & 1726.52 \\
Comp, Airmass, PSF, CRPA & -3464.58 & 1725.28 \\
Airmass, CRPA & -3458.3 & 1724.88 \\
Comp, Airmass & -3465.22 & 1724.86 \\
Time, PSF, CRPA & -3451.36 & 1724.82 \\
Airmass, PSF, CRPA & -3456.38 & 1723.2 \\
Comp, Airmass, CRPA & -3469.22 & 1722.99 \\
Time, Airmass, CRPA & -3451.81 & 1721.35 \\
Time, Comp, Airmass, PSF, CRPA & -3459.28 & 1721.35 \\
Comp, Airmass, PSF & -3464.72 & 1720.7 \\
Comp, PSF, CRPA & -3449.66 & 1720.28 \\
Comp, CRPA & -3455.24 & 1720.15 \\
CRPA & -3440.12 & 1717.28 \\
Airmass, PSF & -3451.3 & 1717.02 \\
PSF, CRPA & -3434.29 & 1716.03 \\
Airmass & -3436.46 & 1713.5 \\
Comp, PSF & -2049.26 & 1067.31 \\
Comp & -2049.62 & 1025.21 \\
PSF & -2003.18 & 1009.93 \\ \\
\hline
\hline
Target/Comparison LC \\
\hline
GP regressors &  GP BIC & log$_{e}$Z \\
\hline
Time, Airmass & -3457.24 & 1728.58 \\
\textbf{Time (\texttt{Conv1:WLC})} & \textbf{-3462.83} & \textbf{1728.26} \\
Time, Airmass, CRPA & -3451.18 & 1728.22 \\
Time, CRPA & -3457.09 & 1724.8 \\
CRPA & -3453.4 & 1724.07 \\
Airmass, CRPA & -3456.81 & 1723.7 \\
Airmass & -3357.8 & 1694.99 \\
\end{tabular}
\end{table}

%%%%%%%%%%%%%%%%%%%%%%%%%%%%%
%%%%%%%%%% Observation 4 %%%%
%%%%%%%%%%%%%%%%%%%%%%%%%%%%%

\begin{table}
\caption{Same as Table \ref{tab:obs1_bicevi} but for observation 4.}   
\label{tab:obs4_bicevi}
\begin{tabular}{ccc}
\hline
\hline
Target LC \\
\hline
GP regressors &  GP BIC & log$_{e}$Z \\
\hline
Time, Comp, Airmass, PSF & -2192.71 & 1082.94 \\
\textbf{Time, Comp, PSF (\texttt{New:WLC})} & \textbf{-2197.27} & \textbf{1082.68} \\
Time, Comp, PSF, CRPA & -2192.41 & 1079.94 \\
Time, Comp, Airmass, PSF, CRPA & -2181.83 & 1076.69 \\
Time, Comp & -2164.5 & 1071.75 \\
Time, Comp, CRPA & -2159.48 & 1071.53 \\
Time, Comp, Airmass & -2159.49 & 1071.45 \\
Time, Comp, Airmass, CRPA & -2154.54 & 1069.72 \\
Comp, Airmass, PSF, CRPA & -2186.96 & 1069.65 \\
Comp, Airmass, CRPA & -2155.06 & 1056.87 \\
Time, Airmass, PSF & -2104.68 & 1048.43 \\
Comp, PSF, CRPA & -2117.88 & 1048.11 \\
Airmass, PSF & -2117.39 & 1045.55 \\
Comp, Airmass, PSF & -2133.95 & 1045.51 \\
Airmass, PSF, CRPA & -2106.45 & 1043.38 \\
\textbf{Time, PSF (\texttt{New:WLC;No\_Comp})} & \textbf{-2098.13} & \textbf{1041.09} \\
Time, PSF, CRPA & -2092.94 & 1040.37 \\
Comp, CRPA & -2093.71 & 1039.24 \\
Time, Airmass, PSF, CRPA & -2098.9 & 1037.5 \\
Comp, Airmass & -2007.12 & 988.35 \\
PSF, CRPA & -1879.64 & 934.87 \\
Time & -1865.92 & 932.78 \\
Time, Airmass, CRPA & -1864.82 & 931.93 \\
Time, Airmass & -1860.93 & 931.8 \\
Time, CRPA & -1862.81 & 931.07 \\
Airmass, CRPA & -1869.76 & 927.73 \\
Airmass & -1788.32 & 892.26 \\
CRPA & -1776.6 & 888.34 \\
Comp, PSF & -957.97 & 479.72 \\
Comp & -903.7 & 455.17 \\
PSF & -727.93 & 440.63 \\ \\
\hline
\hline
Target/Comparison LC \\
\hline
GP regressors &  GP BIC & log$_{e}$Z \\
\hline
Time, Airmass, CRPA & -2081.55 & 1033.03 \\
\textbf{Time, Airmass (\texttt{Conv1:WLC})} & \textbf{-2086.72} & \textbf{1032.28} \\
Airmass, CRPA & -2084.06 & 1030.49 \\
Time & -2064.95 & 1029.94 \\
Time, CRPA & -2059.88 & 1028.8 \\
CRPA & -1757.82 & 886.95 \\
Airmass & -1548.24 & 769.83 \\
\end{tabular}
\end{table}

%%%%%%%%%%%%%%%%%%%%%%%%%%%%%
%%%%%%%%%% Observation 5 %%%%
%%%%%%%%%%%%%%%%%%%%%%%%%%%%%

\begin{table}
\caption{Same as Table \ref{tab:obs1_bicevi} but for observation 5.}   
\label{tab:obs5_bicevi}
\begin{tabular}{ccc}
\hline
\hline
Target LC \\
\hline
GP regressors &  GP BIC & log$_{e}$Z \\
\hline
\textbf{Time, Comp (\texttt{New:WLC})} & \textbf{-1705.44} & \textbf{839.69} \\
Time, Comp, CRPA & -1700.34 & 839.08 \\
Time, Comp, Airmass & -1702.14 & 838.48 \\
Time, Comp, PSF & -1700.46 & 838.41 \\
Time, Comp, Airmass, PSF & -1705.33 & 837.62 \\
Time, Comp, Airmass, PSF, CRPA & -1700.09 & 837.25 \\
Time, Comp, Airmass, CRPA & -1698.0 & 835.85 \\
Time, Comp, PSF, CRPA & -1695.61 & 835.45 \\
Comp, Airmass, PSF, CRPA & -1702.25 & 830.41 \\
Comp, Airmass, CRPA & -1698.15 & 827.49 \\
Time, Airmass, PSF, CRPA & -1645.89 & 806.17 \\
\textbf{Time, Airmass, PSF (\texttt{New:WLC;No\_Comp})} & \textbf{-1650.67} & \textbf{805.84} \\
Airmass, PSF, CRPA & -1646.15 & 805.54 \\
Time, PSF & -1619.75 & 797.54 \\
Comp, CRPA & -1622.41 & 797.52 \\
Time, PSF, CRPA & -1614.88 & 797.38 \\
Comp, PSF, CRPA & -1617.76 & 796.39 \\
Airmass, PSF & -1621.21 & 796.22 \\
Comp, Airmass, PSF & -1616.27 & 794.23 \\
Comp, Airmass & -1566.87 & 779.17 \\
Airmass, CRPA & -1489.8 & 739.88 \\
Time, Airmass & -1492.2 & 739.64 \\
Airmass & -1491.95 & 738.85 \\
Time, Airmass, CRPA & -1487.15 & 736.49 \\
Time & -1477.59 & 735.4 \\
Time, CRPA & -1472.75 & 735.0 \\
PSF, CRPA & -1332.67 & 657.14 \\
CRPA & -1253.85 & 632.24 \\
Comp, PSF & -603.51 & 376.41 \\
Comp & -607.97 & 309.19 \\
PSF & -610.43 & 309.04 \\ \\
\hline
\hline
Target/Comparison LC \\
\hline
GP regressors &  GP BIC & log$_{e}$Z \\
\hline
\textbf{Time, Airmass (\texttt{Conv1:WLC})} & \textbf{-1641.27} & \textbf{807.24} \\
Airmass, CRPA & -1635.99 & 805.33 \\
Time, Airmass, CRPA & -1636.33 & 804.86 \\
Time & -1602.97 & 796.05 \\
Time, CRPA & -1598.17 & 795.92 \\
Airmass & -1462.59 & 745.79 \\
CRPA & -1199.12 & 705.79 \\
\end{tabular}
\end{table}

% If you want to present additional material which would interrupt the flow of the main paper,
% it can be placed in an Appendix which appears after the list of references.

%%%%%%%%%%%%%%%%%%%%%%%%%%%%%%%%%%%%%%%%%%%%%%%%%%

% Don't change these lines
\bsp	% typesetting comment
\label{lastpage}
\end{document}